\documentclass[11 pt]{article}
\usepackage{jheppub} 
\usepackage{orcidlink}
\usepackage{lscape}
\usepackage{amssymb,amsfonts}
\usepackage{xcolor}
\usepackage{caption}
\usepackage{subcaption,longtable,stmaryrd}
\usepackage{slashed}
\usepackage{bigints}
\usepackage{cancel}
\usepackage{multicol}
\usepackage{blindtext}
\usepackage{tikz}
\usepackage{enumitem}
\usepackage[customcolors,shade]{hf-tikz}
\usepackage{graphicx}
\usepackage[bitstream-charter]{mathdesign}
\usepackage{changepage}
\usepackage{cleveref}
\usepackage[most]{tcolorbox}
\usepackage{cancel}
\DeclareSymbolFont{usualmathcal}{OMS}{cmsy}{m}{n}
\DeclareMathAlphabet\mathbfcal{OMS}{cmsy}{b}{n}
\DeclareSymbolFontAlphabet{\mathcal}{usualmathcal}
\DeclareSymbolFont{rmlargesymbols}{OMX}{mdbch}{m}{n}
\DeclareMathSymbol{\rmintop}{\mathop}{rmlargesymbols}{82}
\DeclareMathSymbol{\rmointop}{\mathop}{rmlargesymbols}{72}
\newcommand{\rmint}{\rmintop\nolimits}
\definecolor{mygray}{gray}{0.5}
\newcommand{\Rho}{\mathrm{P}}

{\title{\bf Bootstrapping the spinning two body problem in dynamical  Chern-Simons gravity using worldline QFT}}

\author[a]{Arpan Bhattacharyya \orcidlink{0000-0002-7933-6441},}
\author[b,c]{Debodirna Ghosh \orcidlink{0000-0002-1258-1071},}
\author[a]{Saptaswa Ghosh \orcidlink{0000-0002-0237-7042},} 
\author[a]{ Sounak Pal \orcidlink{0000-0002-2250-0466}}

\affiliation[a]{\it Indian Institute of Technology Gandhinagar,\\ Gujarat-382055, India}
\affiliation[b]{\it Saha Institute of Nuclear Physics,\\ 1/AF, Bidhannagar, Kolkata 700064, India}
\affiliation[c]{Homi Bhabha National Institute, \\Training School Complex, Anushakti Nagar, Mumbai 400094, India}

\emailAdd{abhattacharyya@iitgn.ac.in}
\emailAdd{debodirna.ghosh@saha.ac.in}
\emailAdd{saptaswaghosh@iitgn.ac.in}
\emailAdd{palsounak@iitgn.ac.in}

\abstract{In this paper, we compute the WQFT partition function, specifically the eikonal phase in a black hole scattering event in the dynamical Chern-Simons theory, using the techniques of spinning worldline quantum field theory. We consider the scattering of spinning black holes and highlight the necessary details for the calculation of the partition function. We present the  $\epsilon$-expansion of the essential two-loop integrals using Integration-by-Parts (IBP) reduction and differential equation techniques, which we then utilize to compute the linear-in-order spin eikonal phase up to 3PM. Additionally, we discuss the dependence of the phase on the spin orientations of the black holes.}
\makeatletter
\begin{document}
\maketitle
\flushbottom
\section{Introduction}
The detection of gravitational waves through a network of ground-based detectors provides a new way of “listening” to the Universe in the high-frequency band \cite{LIGOScientific:2014oec,LIGOScientific:2016aoc,LIGOScientific:2016sjg, LIGOScientific:2016vlm,LIGOScientific:2017bnn,LIGOScientific:2019hgc}. Our Universe is full of massive astrophysical objects like black holes, neutron stars, etc. These astrophysical bodies encounter scattering events while moving through space and time. In this process, the bodies are deflected, their spins are altered, and low-frequency gravitational waves are emitted, which can be measured by the earth (space) based detectors. Such tests will become only stronger with the next generation of ground-based and space-based detectors \cite{hyp10}, and the recent pulsar timing arrays \cite{NANOGrav:2023gor,EPTA:2023fyk,NANOGrav:2023wsz,Verbiest:2024nid} which will allow for gravitational wave detection at nano-hertz frequencies. These observations will not only provide information about astrophysical objects but also allow for even more stringent constraints on the modifications to General Relativity. 
 \par
General Relativity is considered the most successful theory describing the classical dynamics of gravity. It explains how massive objects like stars and planets curve spacetime and how this curvature affects the motion of objects and the propagation of light. Despite its success, General Relativity (GR) is considered incomplete because it does not work well at very small scales (high energies), which is referred to as the “UV” (ultraviolet) regime in physics. Thus, General relativity is not a UV complete theory of gravity. At very small scales, the effects of quantum mechanics become significant. Therefore, it is widely expected that there should be a consistent quantum theory of gravity which is UV complete. In this context, Weinberg proposed a novel approach to quantize gravity based on an Effective Field Theory \cite{PhysRev.138.B988}: One should add infinitely many higher derivative operators with Einstein-Hilbert action. These modifications can be the higher curvature modifications of GR. Another way to modify GR is to add extra degrees of freedom in the form of a scalar, gauge field etc, along with the usual graviton degrees of freedom, which have the possibility to explain the phenomenon of dark matter and dark energy \cite{Damour:1992we,Horbatsch:2015bua,Schon:2021pcv,Rainer:1996gw,DeFelice:2011bh}. In principle, one can also invoke additional degrees of freedom (DOFs) along with the higher curvature terms. In this context, the two classes of theories that have received much attention in recent years are the Einstein-Dilaton-Gauss-Bonnet (EdGB) and the dynamical Chern-Simons (dCS) theories \cite{Zwiebach:1985uq, Gross:1986mw,Jackiw:2003pm}. The dilaton-Gauss-Bonnet term appears with a scalar field non-minimally coupled with Gauss-Bonnet curvature and arises from the low energy effective field theory of heterotic string theory \cite{Zwiebach:1985uq, Gross:1986mw}. Whereas, the dCS coupling comes from modifying GR where a scalar field is non-minimally coupled to topological invariant quadratic terms in curvature. Therefore, the action for Chern-Simons (CS) modified gravity theory consists of Einstein-Hilbert action and a new parity-violating, four-dimensional correction. One of the recent interests of this particular model is to make String theory mathematically consistent by introducing such parity-violating corrections \cite{Alexander:2004us,Alexander:2004xd}. In this work, we explore the scattering event of two spinning black holes in the CS modified GR theory. In the case of dCS theory, the scattering events of black holes can be studied, supporting the scale hierarchy of different length scales in which the two scattering black holes are considered to be point particles with extra spin degrees of freedom. It is important to note that for the scattering of non-spinning black holes, the parity-violating CS term does not impose any corrections to the black-hole dynamics. Therefore, it is necessary to consider spinning BHs to observe the contribution of the CP-violating term while computing the classical observables. We explore such an observable in the main sections of this paper.
\par
Several studies related to computing the gravitational waveform in General Relativity have been extensively done in recent years \cite{Blanchet:2013haa, Schafer:2018kuf,Buonanno:1998gg,Blanchet:2004ek,Blanchet:2006gy,Blanchet:2023bwj,Blanchet:2023sbv,Blanchet:2023soy,Warburton:2024xnr, Levi:2018nxp,Wardell:2021fyy}. The LIGO/VIRGO observations of Black holes, Neutron stars, etc. inspirals/mergers require high-precision analytical computations of classical potential and radiation coming from the binary system \cite{Purrer:2019jcp}. Besides the binary inspiral events, there are scattering events along hyperbolic orbits that might be interesting in the context of future space-based detectors. To get an analytical handle on these observables, one should solve Einstein's equation perturbatively. Surprisingly, extracting classical physics from the relativistically quantized theory of gravitons seems more efficient than solving the classical gravitational field equation. It has been discovered that these precise computations are analogous to solving the scattering matrix (S-matrix) in collider physics with advanced quantum field theory (QFT) tools. It is now been proven that the QFT-based approach to handle the precision calculations is significantly more efficient. There are many intrinsically different approaches to solve field equation (for inspiral as well as scattering event): direct amplitude based approaches \cite{Brandhuber:2022qbk,Brandhuber:2023hhy,Kosower:2018adc,DeAngelis:2023lvf,Bjerrum-Bohr:2018xdl,Bjerrum-Bohr:2013bxa,Alessio:2024wmz,Brunello:2024ibk, Bern:2019crd, Bern:2019nnu, Bern:2020buy,Bern:2024adl,Bern:2023ity,Bern:2022kto, Bern:2021yeh,Bern:2021dqo,Bern:2020gjj,Brandhuber:2019qpg,AccettulliHuber:2019jqo,AccettulliHuber:2020oou,Barack:2023oqp,Bjerrum-Bohr:2023iey,Bjerrum-Bohr:2023jau,Bjerrum-Bohr:2022ows,Bjerrum-Bohr:2022blt,Bjerrum-Bohr:2021wwt,Bjerrum-Bohr:2021vuf,Bjerrum-Bohr:2020syg, Bjerrum-Bohr:2019kec, Cristofoli:2019neg,Bjerrum-Bohr:2016hpa,Chen:2024mmm,Alessio:2022kwv,Alessio:2023kgf, ashoke1,ashoke5, Gonzo:2024zxo,Aoude:2023vdk,Aoude:2023dui,Brandhuber:2023hhl,Adamo:2024oxy,Adamo:2023cfp,Adamo:2022ooq,Adamo:2021rfq,Georgoudis:2023eke,Georgoudis:2024pdz,Bini:2024rsy}\footnote{The list is by no means exhaustive. Interested readers are referred to this review \cite{DiVecchia:2023frv} and
references therein for more details. }, and in the latest the  Worldline Quantum Field Theory (WQFT) \cite{Mogull:2020sak,Jakobsen:2021lvp,Jakobsen:2021zvh,Jakobsen:2023ndj,Jakobsen:2023hig,Jakobsen:2022psy,Jakobsen:2023oow,Wang:2022ntx,Klemm:2024wtd,Bhattacharyya:2024aeq, Driesse:2024xad, Adamo:2024oxy, DeAngelis:2023lvf, Cristofoli:2021vyo}. Other approaches also include: solving field equations directly \cite{1978ApJ...224...62K, Damour:1992we,Damour:2016gwp,Damour:2019lcq,Bini:2020rzn,Damour:2022ybd,Bini:2024ijq,DeVittori:2014psa, hyp16}, EFT based approaches \cite{Porto:2007pw,Porto:2008jj,Porto:2012as,Levi:2018nxp,Cheung:2024jpo,Cheung:2023lnj,Cheung:2020gyp,Ivanov:2024sds,Bhattacharyya:2023kbh,Huang:2018pbu,Diedrichs:2023foj,Loebbert:2020aos, Mougiakakos:2021ckm,Riva:2021vnj,Mougiakakos:2022sic,Riva:2022fru,Bernard:2023eul,Porto:2007px,Porto:2017dgs,Dlapa:2024cje}. These complementary approaches give the same results, and the choice depends on the taste of the researcher. In this paper, we work with the novel Worldline quantum field theory (WQFT) formalism \cite{Mogull:2020sak,Jakobsen:2021lvp} to explore the scattering event of two spinning black holes. 
\par
Recently, a duality relation between the expectation value of an operator in WQFT and the S-matrix has been established by Mogull-Plefka-Steinhoff \cite{Mogull:2020sak,Jakobsen:2021lvp}. This formalism is useful in computing the classical observables such as impulse and the time domain waveform of gravitational waves for scalar and spinning particles from black hole scattering encounters, which has been of great interest in the past few years. The formalism is efficient in bypassing certain subtleties, such as the difficulties of finding "super-classical" contributions of the amplitude-based methods. Also, unlike the EFT approach, there is no need to determine a non-observable and gauge-dependent effective potential to solve equations of motion perturbatively. The formalism is also successful in establishing a double-copy relation in WQFTs with bi-adjoint scalars \cite{Shi:2021qsb}, and also finding observables in black hole light bendings, i.e. gravitational lensing phenomena \cite{Bastianelli:2021nbs}. Also, in a recent study \cite{Bhattacharyya:2023kbh}, we have explored different classical observables outside the Einstein-GR regime, i.e. Scalar Tensor theory invoking an extra scalar degree of freedom. To the best of our knowledge, this was the first study of gravitational observables beyond GR using WQFT. One can also use the WQFT formalism by adding higher curvature corrections, e.g., the CS term, or for other scalar-tensor models. \par
A natural extension of the WQFT formalism is to include spin in the theory, i.e., if the scattering objects/particles are spinning, how does it reflect in the worldline action? What are the extra worldline degrees of freedom for spinning bodies? The answer has been nicely given in \cite{Jakobsen:2021lvp,Jakobsen:2021zvh} based on the worldline path integral representation of the Fermionic field (treated as matter in QFT approaches). For GR, the authors show that the spinning scattering system in WQFT enjoys hidden supersymmetry, which turns out to be $\mathcal N=2$ (or spin-1). The symmetry in the system helps to constrain the scattering process in our theory at least up to order-one spin interactions and also find the resemblance with the standard worldline spinning action used in EFT-based approaches. The anti-commuting nature of the worldline vectors  $\psi^{a}$  manifests the supersymmetry associated with the body’s spin tensor ($S^{ab}\sim\bar\psi^{[a}\psi^{b]}$). In WQFT, one surpasses the traditional EFT-based calculation of effective potential and computes only the tree-level diagrams to probe classical observables. In formal amplitude-based approaches, one refers to the classical relativistic scattering from the eikonal phase to investigate both the potential and the radiation zone.  The further extension to the spinning eikonal can also be constructed and used in an economic way using the spinning WQFT formalism. In this work, we use this formalism to analyze the spinning eikonal for the dCS theory.
\par
The study of scattering processes within the Post-Minkowskian (PM) approximation, which involves an expansion in Newton's constant to all orders in velocity, using both EFT-based and amplitude-based approaches, has encountered several technical challenges. One of the major challenges is the integration problem associated with solving Feynman multiloop integrals. Recent progress in Feynman multiloop integral techniques has shown a rich possibility to encounter different algebraic geometry methods such as intersection theory \cite{Mastrolia:2018uzb,Brunello:2023rpq,Brunello:2023fef}, Calabi-Yau n-folds \cite{Frellesvig:2023bbf, Klemm:2024wtd} etc. As one goes higher in PM order, the computation of the multiloop integrals becomes quite tough. Using a connection with modular graph forms \cite{Dorigoni:2022npe} and elliptic curves, one can show the duality between multiloop Feynman graphs and Calabi-Yau two folds (coming from 5 PM integrals). The appearing class of functions differs from multiple polylogarithms \cite{Weinzierl:2007cx} in higher PM calculations, and quadratic combinations appear for elliptic integrals, which can be parametrized in terms of Calabi-Yau two folds. Therefore, there is an extremely beautiful and rich connection with such deep mathematics involved in the BH scattering problems. In this work, we take a digression from the usual GR theories by adding a Chern-Simons type \cite{Yunes:2009hc} term in the bulk action. In connection to Effective one-body (EOB) problems \cite{Buonanno:1998gg} and scattering angle, we calculate the spinning Eikonal up to 3PM, considering the scattering black holes to be spinning. We use multiloop Feynman integral methods to eventually reach our results by handling them analytically. Furthermore, \textit{we comment on the possibility of finding a contribution from adding such a CP-violating term (dCS term) in theory, depending on the spin orientation before the scattering encounters, which represents one of our novel findings in this paper to the best of our knowledge.}\par
Our paper is organised as follows: In Section~(\ref{sec2}), We briefly discuss the WQFT formalism with the effective CP-violating (dCS term) term present in the theory. We also derive the corresponding Feynman rules for the extra spin degrees of freedom for possible vertices coming from the theory. Section~(\ref{sec3}) is devoted mainly to developing the technical tools useful for the computations. We describe the strategies of solving multiloop integrals using Integration-by-parts techniques (IBP) and differential equation solutions in which the IBP-reduced master integrals form a closed algebra between the family of master integrals when differentiated with respect to the kinematic variables. In Section~(\ref{sec4}), we demonstrate the application of the above-mentioned formalism and tools to compute the spinning eikonal, which in turn connects the WQFT partition function via exponentiation. Useful integrals are given in the appendices.\\\\
\textbf{Notations and Conventions}\\
\textbullet $\,\,$ Metric sign convention: ($+,-,-,-$).\\
\textbullet $\,\,$ All computations are done in the unit where, $(c,\hbar)=1$.\\
\textbullet $\,\,$ The impact parameter $b$ is purely spacelike, $b^2=b^\mu b_\mu=-|\boldsymbol{b}|^2$.\\
\textbullet $\,\,$ Black hole velocity parametrization: $v_1=(\gamma,\gamma\beta,0,0)$,  $v_2=(1,0,0,0)$.\\
\textbullet $\,\,$ Planck mass: $m_p=\frac{1}{\sqrt{8\pi G_N}}$, where $G_{N}$ is Newton's constant.\\ 
\textbullet $\,\,$ \textcolor{black}{Scaled delta function: $\hat\delta^{(D)}(\cdots)\equiv (2\pi)^{D}\delta^{(D)}(\cdots)$.}\\
\textbullet $\,\,$ Integration convention: $\int_{k}\equiv \int\frac{d^D k}{(2\pi)^D}\,(4\pi e^{- \gamma_E})^{\epsilon}$.\\ 
\textbullet $\,\,$ $\gamma_E$ is the Euler-Mascheroni constant.\\ 
\section{Chern-Simons gravity and worldline QFT}
\label{sec2}
In this section, we will review the dynamical Chern-Simons (dCS) model and derive all relevant Feynman rules. We also comment on the main differences between the higher curvature EFTs and Einstein GR.
\subsection*{Bulk gravitational action:}
The bulk action for dCS theory is given by,
\begin{align}
    \begin{split}
        S_{\textrm{dCS}}=\rmint d^4 x\sqrt{-g}\Big[-\frac{m_p^2}{2}R+\frac{1}{2}g^{\alpha\beta}\partial_{\alpha}\varphi\partial_{\beta}\varphi-\frac{l_{dCS}^2}{m_p^3}\varphi\, {}^{*}R R\Big]\,.
    \end{split}
\end{align}
Now we expand the bulk gravitational action around flat metric as,
\begin{align}
    \begin{split}
        g_{\mu\nu}=\eta_{\mu\nu}+\frac{h_{\mu\nu}}{m_p},\,\, \varphi\to 0+\varphi,
    \end{split}
\end{align}
Hence, the Chern-Simons term can be expanded in order of fluctuation $h$ as,
\begin{align}
    *RR=\epsilon^{\chi\varepsilon\mu\nu}\Big(\partial_\mu\partial_\beta h_{\nu}^{\,\sigma}-\partial_\mu \partial^\sigma h_{\nu\beta}\Big)\Big(\partial_{\chi}\partial_{\sigma}h_{\delta}^{\,\beta}-\partial_{\chi}\partial^{\beta}h_{\delta\sigma}\Big)+\mathcal{O}(h^3).
\end{align}
Before going to the worldline action for the higher curvature EFTs, we first briefly review the case of General relativity.
\subsection*{SUSY worldline action in General relativity:}
As we will be considering spinning binaries in this paper, following \cite{Jakobsen:2021zvh}, we start with the $\mathcal{N}=2$ supersymmetric worldline action. It  has the following form:
\begin{align}
    \begin{split}
        S_{\mathcal{N}=2}=-\sum_{k=1}^2 
        \int d\tau\, \Big[\frac{1}{2e}g_{\mu\nu}\dot x_k^{\mu}\dot x_k^{\nu}+i\bar \psi_{k,a}\frac{\mathtt{D}\psi^a_k}{\mathtt{D}\tau}+\frac{e}{2}R_{abcd}\,\bar \psi^a_k \psi^b_k \bar \psi^c_k \psi^d_k+\frac{e}{2}m_k^2\Big]
    \end{split}\label{2.4a}
\end{align}
where,
\begin{align}
    \frac{\mathtt{D}\psi ^a_k}{\mathtt{D}\tau}= \dot\psi_k^a+\dot x^\mu\, \omega_{\mu}^{\,\,ab}\psi_{k,b}\,.
\end{align}
Moreover, in a convenient fashion, we can make the following gauge choice $e=1/m_k$ and make a rescaling of the fermions \cite{Jakobsen:2021zvh}:
\begin{align}
    \psi_k^a \to \sqrt{m_k}\,\psi_k^a,\,\,\bar\psi_k^a \to \sqrt{m_k}\,\bar\psi_k^a,
\end{align}
The action in \eqref{2.4a} is invariant under the following asymptotic $\mathcal{N}=2$ supersymmetry transformation.
\begin{align}
    \begin{split}
    &    \delta x^\mu=ie^{\mu}_a\,(\bar\epsilon\psi^a+\epsilon\bar\psi^ a),\,\,\,\delta \psi^a=-\epsilon e^{a}_\mu \dot x^\mu-\delta x^\mu\,{\omega_{\mu}}^{\,\,a}_{\,\,\,\,\,\,\,b}\,\psi^b\\ &
    \textrm{and,}\,\textcolor{black}{\delta}\bar\psi^a=-\bar\epsilon e^{a}_\mu \dot x^\mu-\delta x^\mu\,{\omega_{\mu}}^{\,\,a}_{\,\,\,\,\,\,\,b}\,\bar\psi^b.
    \end{split}
\end{align}
It also has global $U(1)$ symmetry.
\begin{align}
    \delta_{\hat\epsilon}\psi^a=i\hat\epsilon \psi^a,\,\,\delta_{\hat\epsilon}\bar\psi^a=i\hat\epsilon \bar\psi^a,\,\,\delta_{\hat\epsilon}x^\mu=0.
\end{align}
The Noether charges derived from these global transformations give different conservation laws along the worldline \cite{Jakobsen:2021zvh}.\par
In a perturbative framework, one can not distinguish between tetrad indices $(a,b,..)$ and curved indices $(\mu,\nu,..)$. In order to describe the scattering process, one can expand worldline DOF around undeflected trajectories as,
\begin{align}
    \begin{split}
        x_k^\mu\to b_k^\mu+v^\mu_k\tau+z_{k}^{\mu}(\tau),\,\,\, \psi^{a}_k(\tau)\to\Psi_k^{a}+{\psi}_k^a(\tau)\,.
    \end{split}
\end{align}
One can also introduce the constant spin tensor as,
\begin{align}
    \begin{split}
        \mathcal{S}_k^{ab}=-2i\bar \Psi_k^{[a}\Psi_k^{b]}\,.
    \end{split}
\end{align}
From the quadratic part of the action, one can find out the field and worldline propagators,
\begin{align}
    \begin{split}
     &    \begin{minipage}[h]{0.12\linewidth}
	\vspace{4pt}
	\scalebox{1.8}{\includegraphics[width=\linewidth]{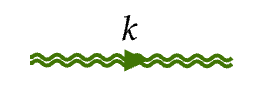}}
\end{minipage}\hspace{1.2 cm}=\frac{iP_{\mu\nu;\rho\sigma}}{k^2+i\varepsilon},\,\begin{minipage}[h]{0.12\linewidth}
	\vspace{4pt}
	\scalebox{1.6}{\includegraphics[width=\linewidth]{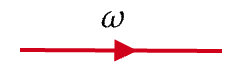}}
\end{minipage}\hspace{1.2 cm}=-\frac{i \eta^{\mu\nu}}{2m_k}\left(\frac{1}{(\omega+i\varepsilon)^2}+\frac{1}{(\omega-i\varepsilon)^2}\right),\\ &
 \begin{minipage}[h]{0.12\linewidth}
	\vspace{4pt}
	\scalebox{1.8}{\includegraphics[width=\linewidth]{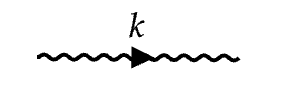}}
\end{minipage}\hspace{1.2 cm}=\frac{i}{k^2+i\varepsilon},\,
        \begin{minipage}[h]{0.12\linewidth}
	\vspace{4pt}
	\scalebox{1.6}{\includegraphics[width=\linewidth]{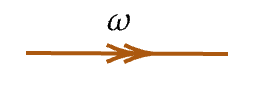}}
\end{minipage}\hspace{1.2 cm}=-\frac{i \eta^{\mu\nu}}{2m_k}\left(\frac{1}{\omega+i\varepsilon}+\frac{1}{\omega-i\varepsilon}\right).
    \end{split}
\end{align}
\textcolor{black}{Note that in this paper, we intend to compute the Eikonal phase, and hence we choose a time-symmetric/Feynman propagator. However, if one wants to compute the observables, then one needs to choose the retarded propagators, which gives the causally sensible classical solutions.} \\\\
One can eventually expand the tetrad and spin connection as,
\begin{align}
    \begin{split}
& e^a_{\,\mu}=\eta^{a\nu}\Big(\eta_{\mu\nu}+\frac{1}{2m_p}h_{\mu\nu}-\frac{1}{8m_p^2}h_{\mu\rho}h^{\rho}_{\,\nu}+\mathcal{O}(1/m_p^3)\Big)\,,\\ &
\omega_{\mu}^{\,\,\,ab}=-\frac{1}{m_p}\partial^{[a}h^{b]}_{\,\,\,\mu}-\frac{1}{2m_p^2}h^{\nu[a}\left(\partial^{b]}h_{\mu \nu}-\partial_\nu h^{b]}_{\,\,\,\,\mu}+\frac{1}{2}\partial_\mu h^{b]}_{\,\,\,\,\nu}\right)+\mathcal{O}(1/m_p^3).
    \end{split}
\end{align}
\subsection*{What happens for modified theories of gravity?}
In some EFT of gravity involving extra scalar degrees of freedom the scalar dipole moment must couple to the scalar field in the worldline Lagrangian. One way, mentioned in \cite{Loutrel:2018ydv}, is to demand the new canonical momenta $\mathcal{P}^\mu$ is now not only a function of graviton degrees of freedom but also a function of the extra degrees of freedom:
\begin{align}
    \begin{split}
        \mathcal{P}_\mu=p_{\mu}+(\textrm{pure gravity contributions})+\frac{\partial \mathcal{P_{\mu}}}{\partial{\varphi}}\Bigg|_{\varphi=0}\,\varphi+\frac{\partial \mathcal{P_{\mu}}}{\partial{\nabla_{\alpha}\varphi}}\Bigg|_{\nabla_\alpha\varphi=0}\,\nabla_{\alpha}\varphi+\cdots\label{2.13k}
    \end{split}
\end{align}
In dCS theory, we have an extra bulk degree of freedom (in the form of a scalar field) along with graviton, and so we need to promote the masses of the black holes as a function of that scalar field $\varphi$, which  does not preserve the SUSY invariance for the following re-scaled fermions as,
\begin{align}
    \psi_k^a \to \sqrt{m_k(\varphi)}\,\psi^a,\,\,\bar\psi_k^a \to \sqrt{m_k(\varphi)}\,\bar\psi^a,
\end{align}
 Extrapolating \cite{Loutrel:2018ydv}, the worldline action takes the form (up-to linear in spin),
\begin{align}
    \begin{split}
        \mathcal{S}_k\subset -\int d\tau \Big[\frac{m_k(\varphi)}{2}g_{\mu\nu}\dot x_k^\mu \dot x_k^\nu+i m_{k}(\varphi)\bar\psi_a^k\, D_{\tau}\psi^a_k-i\,\mathcal{C}_{dCS} \,\dot{x}^{\mu} \nabla_{\alpha}\varphi\,\epsilon_{\mu\,\,\,\rho\,\sigma}^{\,\,\alpha}\bar{\psi}_k^a\,\psi_k^{b}e^{\rho}_a e^{\sigma}_{b}\Big]\label{2.13d}
    \end{split}
\end{align}
where $\mathcal{C}_{dCS}=\frac{\partial \mathcal{P_{\mu}}}{\partial{\nabla_{\alpha}\varphi}}\Bigg|_{\nabla_\alpha\varphi=0}$ is an undetermined coefficient and refers to as scalar-dipole constant and is dependent on the coupling constant(s) of the theory. However, in our computation, we will ignore the finite size corrections. Hence, we can avoid the dipole term from the worldline action.  \par 
Now expanding the mass around $\varphi=0$, we get \footnote{In \cite{Wilson-Gerow:2025xhr}, the author asserts that the dynamical Chern-Simons (dCS) term contributes to the spinless scattering angle by treating the sensitivity parameters as functions of the theory's coupling. However, some  contradictory claims regarding our computation have been made in \cite{Wilson-Gerow:2025xhr} and we like to provide some clarifications to avoid any potential confusion.  First and foremost, we do not compute the scattering angle, contrary to the assertion in \cite{Wilson-Gerow:2025xhr}. Moreover, deriving the scattering angle from the Eikonal phase is nontrivial at the 3PM order and requires careful consideration discussed in Section~\eqref{con}.  Second, as we understand it, the author's approach primarily relies on the on-shell behaviour of the scalar field at far region to establish a link between the sensitivity parameters and the couplings of the underlying theory. While it is true that all parameters in the model are implicitly (or explicitly) dependent on the theory’s coupling, the precise functional relationship between these parameters and the theory itself remains ambiguous. For this reason, we adopt a specific prescription in our analysis, treating these parameters as independent. 
Our focus is on the diagrams where the explicit coupling of the dCS term is directly incorporated (and it can be easily seen that it contributes only for spinning case due its structure), and we defer the investigation of potential indirect effects emerging from the broader theoretical framework to future work. If one wishes to carry out the computation using an alternative prescription, one need only apply the transformation:  
   $ \{s_i, g_i\} \to \{s_i(l_{\textrm{dCS}}), g_i(l_{\textrm{dCS}})\}$
where the sensitivity parameters and couplings are explicitly treated as functions of the dCS coupling and get the appropriate results from the expressions provided in our paper. Also, in the \textit{spinless} part of the eikonal there are contributions form UT-2 polynomial as presented in \eqref{dCSne3waaa} which seems to be absent from the result presented in \cite{Wilson-Gerow:2025xhr}.  },
\begin{align}
    \begin{split}
        m_{k}(\varphi)=m_{k}(0)\Big(1+s_k \varphi +\cdots)
    \end{split}
\end{align}
Upon applying the SUSY transformation, we get, $\delta m_{k}(\varphi)=s_k \partial_{\mu}\varphi \delta x^\mu+\cdots$ and demanding that the sensitivity parameter $s_1$ is very small, we get $\delta m_k(\varphi)\approx \mathcal{O}(s_1)$ and the SUSY is preserved in an approximate sense. However, the breaking of SUSY is not a problem as the action in \eqref{2.13d} eventually recovers the standard spinning worldline action with dynamical mass \cite{Levi:2018nxp}. 
\par
\par
\subsubsection*{Primary differences between GR and modified EFTs:}
For GR, following \cite{Jakobsen:2021zvh}, one can argue that the worldline action is invariant under the SUSY transformation along the whole trajectory of the black holes as well as at the asymptotics, which eventually implies that $p_k^2,\ p_k\cdot \bar\psi_k,\,p_k\cdot \psi_k,\,\textrm{and,}\,\psi_k\cdot \bar\psi_k$ (asymptotic charges) are conserved between initial and final asymptotic states. Now, using those constraints along with the extra condition  $v_k\cdot \Psi_k=v_k\cdot \bar\Psi_k=0$ implies  $p_{k,\mu} \cdot \mathcal{S}_k^{\mu\nu}$  is conserved throughout the scattering event and without loss of generality we can choose it to be zero, which essentially recovers the SSC \cite{1967JMP.....8.1591D,Corinaldesi:1951pb,1964NCim...34..317D}:
\begin{align}
    \textrm{\bf SSC}: \, p_{k,\mu} \cdot \mathcal{S}_k^{\mu\nu}=0,\,\, \textrm{with,}\,\,
     p_k^\mu=m_k v_k^\mu+\mathcal{O}(\mathcal{S}^2).
    \label{SSC17}
\end{align}
\textcolor{black}{For our case, the worldline SUSY invariance is broken (or approximate if we demand that the bulk scalar is varying very slowly with respect to the proper time), and hence, conservation of supercharges is only approximate, and it costs the conservation of $p_k\cdot \mathcal{S}_k^{\mu\nu}(\tau)\,.$  Hence, the SSC (throughout the dynamics of the process) is only approximate, which is indeed true as shown in \cite{Loutrel:2018ydv} \footnote{If one chooses $m_k(\varphi)$ to be renormalized mass such that it satisfies the $D_{\tau} m_{k}(\varphi)=\mathcal{O}(S^4)$, then the SUSY invariance recovers. Interested readers are referred to Appendix (B) of \cite{Loutrel:2018ydv}.  }. However, the theory still enjoys the asymptotic (background) SUSY invariance as in the asymptotic states, there is no gravity. Following \cite{Loutrel:2018ydv}, the SSC as defined in \eqref{SSC17} holds for the asymptotic states i.e, $v_{k,\mu} \cdot \mathcal{S}_k^{\mu\nu}(-\infty)=0$ which we will use in the subsequent computations and the observables or the Eikonal only depends on the asymptotic parameters. So, we can still use the WQFT formalism to compute the scattering observables. Also, we will work in linear order in the spin and small coupling ($l_{dCS}$) limit so that we can replace the canonical momenta $p_k^\mu\to m\,v_k^\mu$. 
}
\par
\subsection*{Main goal and computational procedure:}
In WQFT formalism $h_{\mu\nu}(x),\varphi(x),x^\mu(\tau),\psi^\mu(\tau)$ are promoted to the quantum operators. For this case, the partition function looks like,
\begin{align}
    \begin{split}
        Z_{\textrm{S-WQFT}}\equiv e^{i\chi}=\mathcal{N}&\times\rmint \boldsymbol{\mathcal{D}}[h_{\mu\nu},\varphi]\rmint \prod_{i=1}^n \boldsymbol{\mathcal{D}}[z_i,\psi_i,\bar{\psi}_i,\mathfrak{a}_i,\mathfrak{b}_i,\mathfrak{c}_i]\,,\\ &
        \times \exp\Big[i\Big(S_{EH}+S_{\textrm{dCS}}+\sum_i (S_{\textrm{sp}}^{i}+S^{i}_{\textrm{ghost}})\Big)\Big],
    \end{split}
\end{align}
where, $\chi$ is the eikonal phase and the extra ``ghost'' term comes from the metric dependent worldline measure,
\begin{align}
    \begin{split}
       \boldsymbol{ \mathcal{D}}[x]&=D[x]\prod_{0\le \sigma\le T}\sqrt{-g[x(\sigma)]}\,,\\ &
        =D[x]\underbrace{\rmint \boldsymbol{\mathcal{D}}[\mathfrak{a},\mathfrak{b},\mathfrak{c}]\exp \Big[-i\rmint d\tau\Big(\frac{1}{2}g_{\mu\nu}(\mathfrak{a}^\mu\mathfrak{a}^\mu+\mathfrak{b}^\nu\mathfrak{c}^\nu)\big)\Big]}_{\textrm{ghost term}}\,.
    \end{split}
\end{align}
In the classical computations, the ghost field does not appear and can be ignored. Now the question is, what does the WQFT partition function do with scattering amplitude? It has been shown that at the classical limit, the eikonal phase is given by \cite{Amati:1987wq,Amati:1990xe},
\begin{align}
    e^{i\chi}=\frac{1}{4m_1 m_2}\int  \frac{d^4 q}{(2\pi)^4} \, \hat\delta(q\cdot v_1)\hat\delta(q\cdot v_2) e^{iq\cdot b}\,\langle \,\textrm{final}\,|\mathbb{S}|\,\textrm{initial}\,\rangle\xrightarrow[]{} Z_{\textrm{WQFT}}\label{2.18a}
\end{align}
So, the WQFT partition function is equivalent to computing the eikonal phase, which is nothing but the $2\to2$ scattering amplitude, Fourier transformed into impact parameter space.
Now, we have all the ingredients to compute the Feynman rules. Our primary goal in this paper is to compute the eikonal phase for the spinning binaries, especially to investigate the contribution due to the extra degrees of freedom (scalar field) present in the theory and see how the dCS modifies the phase. 
We will mainly concentrate on the spin part of the worldline action as the spinless part (from the scalar field) has already been done in \cite{Bhattacharyya:2024aeq}. For simplicity, we only focus on the terms where spins are linear. Now, to read off the interaction vertices from the supersymmetric worldline action, write the bulk fields in Fourier space.
\begin{align}
    \begin{split}
        \chi(x_i)=\sum_{n-0}^{\infty}\frac{i^n}{n!}\rmint_{k,\omega_1,\cdots \omega_n}e^{ik\cdot b_i}e^{i(k\cdot v_i+\sum_{j=1}^{n}\omega_j)\tau}\Big(\prod_{j=1}^n k\cdot z_i(-\omega_j)\Big)\chi(-k)
    \end{split}
\end{align}
and the worldline fields can be written in energy space as,
\begin{align}
    \begin{split}
        z_{i}^{\mu}(\tau)=\rmint_{\omega} e^{i\omega \tau}z_i^\mu{(-\omega)},\,\,\,\,\,\,\breve{\psi}_i^\mu(\tau)=\rmint_{\omega}e^{i\omega \tau}\breve{\psi}_i^\mu(-\omega)\,.
    \end{split}
\end{align}
The linear spin term in the worldline action takes the following form,
\begin{align}
    \begin{split}
        S_{sp}\sim& -i\rmint d\tau \Big(m_k(0)+\frac{m_k s_k}{m_p} \varphi+\frac{m_k g_k }{m_p^2}\varphi^2\Big)\Big(\bar\psi_{k,a}\dot \psi_k^a+\dot x_k^\mu \omega_{\mu}^{\,\,ab}\bar \psi_{k,a} \psi_{k,b}\Big)\\ &+(\textrm{dCS contribution to the worldline action})\,.\label{2.13}
    \end{split}
\end{align}
Note that in \eqref{2.13} at leading order ($\mathcal{O}(\varphi^0)$) we can add a total derivative term $-1/2\, d_{\tau}(\bar\psi^a\,\psi_a)$ which preserves the SUSY transformation. But when we go to the linear (or higher order in $\varphi$), this addition will be important so that we can have both $\varphi-\psi$ and $\varphi-\bar\psi$ vertices. Again, as mentioned earlier, the SUSY invariance is approximate, and we restrict ourselves to the linear order of $\varphi\,.$ 
Keeping terms that are linear in $\varphi$ we have the following,
\begin{align}
    \begin{split}
        S_{sp}\subset -i\frac{m_i s_i}{2 m_p}\int d\tau\,\varphi\, \,\,\Big(\bar\psi _a \dot \psi^a-\dot{\bar\psi}_a \psi^a\Big)\,. \label{2.19}
    \end{split}
\end{align}
\\We can now compute the Feynman rules in $\mathcal{O}(z^n,{\psi}^n,\chi^n)$. We will mainly focus on the diagrams in which we have an external $\varphi$ field.\\\\
\vspace{-1.3 cm}
\subsection*{Feynman rules:}
 We list down the extra important vertices that we need in our computations. Other relevant vertices are derived in \cite{Jakobsen:2021zvh} for graviton-worldline coupling, and vertices due to the extra scalar/scalar-graviton are derived in \cite{Bhattacharyya:2024aeq}.
\begin{align}
    \begin{split}
      &  \mathcal{V}_1:   \begin{minipage}[h]{0.12\linewidth}	
\scalebox{1.5}{\includegraphics[width=\linewidth]{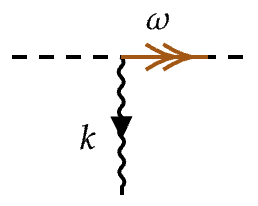}}
\end{minipage}\hspace{1 cm}
\equiv 
\hspace{0 cm}i\frac{m_a s_a}{m_p} \,\bar\Psi^a_{\eta}\,\omega\,\hat\delta(\omega+k\cdot v_a) e^{ik\cdot b_a}\,,\\ &
\mathcal{V}_2:   \begin{minipage}[h]{0.12\linewidth}	
\scalebox{1.5}{\includegraphics[width=\linewidth]{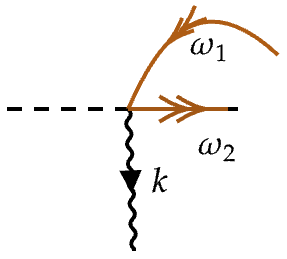}}
\end{minipage}\hspace{1 cm}
\equiv 
\hspace{0 cm}i\frac{m_a s_a}{2 m_p}\hat\delta(k\cdot v_a-\omega_1+\omega_2 )\,(\omega_2+\omega_1) e^{ik\cdot b_a}\,,\\ &
\mathcal{V}_3:   \begin{minipage}[h]{0.12\linewidth}	
\scalebox{1.3}{\includegraphics[width=\linewidth]{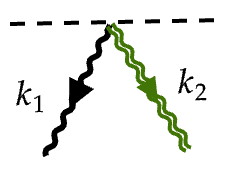}}
\end{minipage}\hspace{1 cm}
\equiv 
\hspace{0 cm}-i\frac{m_a s_a}{2m_p^2}e^{i(k_1+k_2)\cdot b}\,\hat\delta(k_1\cdot v_a+k_2\cdot v_a)\Big(v_a^\mu v_a^\nu+i (k_2\cdot \mathcal{S}_a)^{(\mu}v_a^{\nu)}\Big)\,,
\\ &
\hspace{-0.3 cm}\mathcal{V}_4:   \begin{minipage}[h]{0.12\linewidth}	
\scalebox{1.6}{\includegraphics[width=\linewidth]{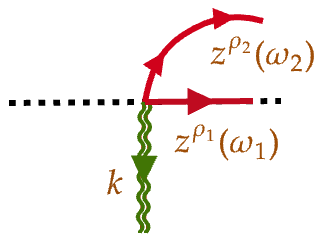}}
\end{minipage}\hspace{1 cm}
\equiv \, i\frac{m_a }{m_p} \,e^{ik\cdot b_a}\,\hat\delta(k\cdot v_a+\omega_1+\omega_2) \Bigg(\frac{1}{2} k_{\rho_1} k_{\rho_2} v_a^\mu v_a^\nu +\omega_1 k_{\rho_2}v_a^{(\mu}\delta^{\nu)}_{\rho_1}+\omega_2 k_{\rho_1}v_a^{(\mu}\delta^{\nu)}_{\rho_2}+\omega_1\omega_2 \delta^{(\mu}_{\rho_1}\delta^{\nu)}_{\rho_2}\\&
\hspace{4 cm}+\left.\frac{i}{2}\Bigg(\left(\omega _1 \,k\cdot S^{(\mu }\delta ^{\nu )}_{\rho _1}k_{\rho _2}+\omega _2\,k \cdot S^{(\mu }\delta ^{\nu )}_{\rho _2}k_{\rho _1}\right)+\frac{1}{2}k\cdot S^{(\eta }\delta ^{\eta )}_{(\nu }v_{a,\mu )}k_{\rho _1}k_{\rho _2}\Bigg)\Bigg)\right]
    \end{split}
\end{align}
\section{Bootstrapping Post-Minkowskian (PM) physics from Post-Newtonian (PN) data: the idea, technology and example}
\label{sec3}
In this section, we briefly discuss the idea of bootstrapping of Post-Minkowskian (relativistic) observables from Post-Newtonian (non-relativistic) data \cite{Dlapa:2023hsl,Henn:2014qga},  which primarily based on differential equation technique for evaluating multiloop Feynman integrals \cite{Kotikov:1990kg,Gehrmann:1999as} and also illustrate the method with a specific two loop example relevant for our computations. In the upcoming sections, we widely encounter two loop integrals, which can be solved using IBP relations. In this section, we will present the results for the master integrals that we will use for our computations of the eikonal phase in the subsequent section. \\\\
\textbf{Identifying the integrands:} We encounter a general family of n-loop integrals that has the following form, which is essentially equivalent to computing $2\to 2$ scattering amplitude,
\begin{align}
           &\langle \,\textrm{final}\,|\mathbb{S}|\,\textrm{initial}\,\rangle\sim\mathbfcal{M}^{a_1\cdots a_n;\pm\cdots \pm}_{\alpha_1\cdots \alpha_n;\beta_1\cdots \beta_m}(|q|,\gamma)=\Bigg(\prod_{i=1}^n\int_{\ell_i}\frac{\hat\delta(\ell_i\cdot v_{a_i})}{(\pm \ell_i\cdot v_{\cancel{a}_i}+i\varepsilon)^{\alpha_i}}\Bigg)\frac{1}{D_1^{\beta _1 }D_2^{\beta_2}\cdots D_m^{\beta_m}}\,,
\end{align}
where $a_i$s are particle index label with $\cancel{1}=2,\,\cancel{2}=1$. $D_i$s are the graviton (scalar) propagators along with irreducible scalar products that form the numerators of the loops integrals and have the following form,
$$D_i=P_i^2,\,\,P_i=\tilde\Upsilon_{ij}\,l_j+\tilde\Upsilon_i\, q,\,\,\textrm{with},\,\,\tilde\Upsilon_{ij},\tilde\Upsilon_i\in(0,\pm 1)\,\,\forall\,\,\, 1\le i,j\le m$$
where $q$ is the external momentum and $v_{1,2}$ are the velocities of the two astrophysical objects. Furthermore, one would notice that each loop integral is supported by a Dirac-$\delta$ function, $\hat\delta(l_i\cdot v_{a_i})$. At this point, to utilize the multiloop collider physics techniques \cite{Weinzierl:2022eaz}, it is convenient to represent the $\delta$-functions by reverse-unitarity (sometimes called velocity-cut),
\begin{align}
    \begin{split}
       \frac{i}{(-1)^{s+1}} \hat\delta^{(s)}(D_i)=\frac{1}{(D_i+i\varepsilon)^{s+1}}-\frac{1}{(D_i-i\varepsilon)^{s+1}}
    \end{split}
\end{align}                 
which allows us to identify the $\delta$-functions as cut propagators and one can freely apply the Integration-by-parts (IBP) technique to evaluate the integrals.  \\\\
\textbf{IBP reduction:}  After identifying the integrands, the next task is to find the IBP identities, first introduced in\cite{Chetyrkin:1981qh}, which help to identify the master integrals in a given family of Feynman integrals. IBP identities can be obtained by demanding that under dimensional regularisation, loop integrals of any total derivative identically vanish \cite{Weinzierl:2022eaz},
\begin{align}
    \int_{l_i}\frac{\partial}{\partial l_k^\mu}\Bigg(\frac{\eta^{\mu}}{(\pm l_i\cdot v_{\slashed{a}_i}+i\varepsilon)^{\alpha_i}\cdots D_1^{\beta_1}\,D_2^{\beta_2}\cdots \slashed{D}_1^{\gamma_1}\slashed{D}_2^{\gamma_2}\cdots}\Bigg)=0\,,\label{3.3a}
\end{align}
where, $l_k$s are loop momenta, $\eta^\mu$ is the linear combination of loop momenta and external momenta and $\slashed{D}$ are the `cut' propagators coming from the $\delta$-functions. The IBP reduction procedure generates a large number of homogeneous linear systems of equations of form,
\begin{align}
    \sum_k \gamma_k(|q|,d,\gamma)\mathbfcal{M}^{(\pm\pm)i}_{\alpha_1+\kappa_{k,1},\alpha_2+\kappa_{k,2},\cdots}(|q|,\gamma,d)=0\,.
\end{align}
However, all the equations are not independent but rather related by various symmetry relations. Ultimately, one needs to solve the equations in unique sectors. Fortunately, there are several publicly available packages for doing this task \textbf{LiteRed} \cite{Lee:2012cn,Lee:2013mka}, \textbf{Kira} \cite{Maierhofer:2017gsa}, \textbf{Reduze 2} \cite{vonManteuffel:2012np} and \textbf{FIRE6} \cite{Smirnov:2019qkx}. For our computations, we extensively use \textbf{LiteRed} . \\\\
\textbf{Master integrals and differential equations:}
One of the main tasks of the IBP procedure is to expand a general Feynman integral as a linear combination of master integrals. For example, let us take a look at the two-loop case. A general two-loop integral has the following form,
\begin{align}
    \begin{split}
        \mathbfcal{M}^{\pm,\pm,\textrm{2-loop}}_{\alpha_1,\alpha_2;\beta_1\cdots \beta_{5}}=\int_{l_{1,2}}\Bigg(\frac{\hat\delta^{(\gamma_1)}(\slashed D_1)\hat\delta^{(\gamma_2)}(\slashed D_2)}{(\pm D_1+i\varepsilon)^{\alpha_1}(\pm D_2+i\varepsilon)^{\alpha_2}}\Bigg)\frac{1}{D_3^{\beta_1}\cdots D_7^{\beta_5}}\label{3.4a}
    \end{split}
\end{align}
where, $\{D_i,\{\slashed D_k\}\}\in\Big(\ell_{1}\cdot v_1,\,\,\ell_{2}\cdot v_2,\,,\,\ell_{1}^2,\,\, \ell_{2}^2,\,\, (\ell_{1}+\ell_{2}-q)^2,\,\, (\ell_{1}-q)^2,\,\, (\ell_{2}-q)^2,\,\, \{\ell_{1}\cdot v_2,\,\, \ell_{2}\cdot v_1\}\Big) $. The set $\{D_i,\slashed D_k\}$ s are called basis function set and the number of independent basis functions (scalar-product) can be determined by noting the number of loop momenta and external momenta: $N_{\textrm{basis}}=\frac{L(L+1)}{2}+LE$, where $L$ and $E$ are the number of loop integrals and external momenta respectively. For the two-loop case, we have $2$ loop momenta and $3$ external momenta, hence the number of basis functions is, indeed, $N_{\textrm{basis}}^{2-\textrm{loop}}=9$. Once we have the basis function, we solve the IBP relation using \textbf{LiteRed}; after solving, one gets the unique master integral for this particular family of integrals. Now using built-in command \textbf{IBPReduce} \cite{Lee:2012cn,Lee:2013mka}, one can write the general two-loop integrals as,
\begin{align}
    \begin{split}
         \mathbfcal{M}^{\textrm{2-loop}}=\sum_{n=1}^{N_{m}} a_{n}(\gamma,d)\,\mathbfcal{M}^{\textrm{master}}_n(\gamma, d)\,.
    \end{split}
\end{align}
We face 16 independent master integrals (excluding the different signs of $i\varepsilon$-prescription, which can be derived further using appropriate field redefinition and invoking symmetry relations) after solving 17 unique sectors using \textbf{LiteRed}.\\\\
\textbf{Solving master integrals using differential equation:} In general, the master integrals do not factorize into the kinematic variable $\gamma$ and dimensional regularisation parameter $\epsilon$. In this case, it is difficult to find any closed-form expression. So, one needs to find the solutions order by order in $\epsilon$. For this purpose, the method of differential equations is very useful \cite{Remiddi:1997ny,Kotikov:1991pm, Henn:2013pwa}. We first define a column vector consisting of relevant master integrals and redefine the kinematic variable $\gamma$ to $x$ as $\gamma\to \frac{x^2+1}{2x}$,
\begin{align}
    \begin{split}
        \vec f(x,\epsilon)=\{\mathbfcal{M}_1^{\textrm{master}},\cdots, \mathbfcal{M}_{N_m}^{\textrm{master}} \}
    \end{split}
\end{align}
which satisfies the following differential equation,
\begin{align}
      \partial_x \vec f(x,\epsilon)=A(x,\epsilon)
     \vec f(x,\epsilon)\,.\label{3.7}
\end{align}
For all practical purposes, the differential equation is quite hard to solve. But it is convenient to find a suitable transformation \cite{Henn:2014qga},
\begin{align}
    \begin{split}
        \vec f(x,\epsilon)=\mathbb{T}(x,\epsilon)\vec{g}(x,\epsilon),
    \end{split}
\end{align}
such that the differential system can be brought to the $\epsilon$-form, which is easier to solve. Now, the system of differential equations has the form,
\begin{align}
    \begin{split}
        \partial_x \vec g(x,\epsilon)=\mathbb{T}^{-1}(A \mathbb{T}-\partial_x \mathbb{T})\vec g(x,\epsilon)\equiv \epsilon\, \mathbb{S}(x) \vec g(x,\epsilon)\,.\label{2.14a}
    \end{split}
\end{align}
The goal is to find the basis transformation matrix $\mathbb{T}(x,\epsilon)$ such that $\mathbb{T}(A \mathbb{T}-\partial_x \mathbb{T})=\epsilon \,\mathbb{S}(x)$. Note that $\epsilon$ is now completely factorized from the matrix $\mathbb{S}(x)\,.$, It was first conjectured by Henn \cite{Henn:2013pwa}. The $\mathbf {\epsilon}$-\textbf{factorization} can be systematically done using Lee's algorithm \cite{Lee:2014ioa} and is (semi-) automatized in the following packages \textbf{Fuchsia} \cite{Gituliar:2017vzm}, \textbf{epsilon} \cite{Prausa:2017ltv}, \textbf{Libra} \cite{Lee:2020zfb} and \textbf{CANONICA} \cite{Meyer:2017joq} and the matrix equation in $\bf{\epsilon}$-form has a general solution in terms of path-ordered exponential (Bootstrap equation), which can be solved systematically using \textbf{Libra} \cite{Lee:2020zfb},
\begin{align}
    \begin{split}
        \vec g(x,\epsilon)=\underbrace{\Bigg(\mathbfcal{P}e^{\epsilon\int_{x_0}^{x}S(x')\,dx'}\Bigg)}_{\mathbb{B}(x,x_0)}\vec g(x_0,\epsilon).\label{3.10a}
    \end{split}
\end{align}
Once we have \eqref{3.10a}, we need to use the boundary condition to fix $\vec g(x_0,\epsilon)\,.$ For relativistic two-body problem it is easy to find the solution of integration at `\textit{soft}' (near-static) limit which is at $\gamma\to 1^{+}\,(x\to 1^{-})$, which is nothing but the solution of the master integrals in Post-Newtonian (non-relativistic) regime: \textit{this is nothing but bootstrapping the complete relativistic integrals (PM-integrals) using the non-relativistic integrals (PN-integrals) \cite{Dlapa:2023hsl}.}

\par

\noindent
For illustration, we show the details of computation in the \textbf{Sectors:\{0,0,0,0,1,1,1\}\&\{1,1,0,0,1,1,1\}}, where we face the following {\bf four} master integrals (the original basis is sometimes called the Laporta basis \cite{Laporta:2000dsw}). We take the following vector \footnote{For the sake of convenience, we will henceforth drop the subscript "master" from the $\mathbfcal{M}$'s\, and the $|q|$ dependence can be recovered from dimensional analysis, as $|q|$ being a dimensionful scale.},
\begin{align}
\begin{split}
    \vec f(\gamma,\epsilon)=&\{\mathbfcal{M}_{0,0,0,0,1,1,1,\slashed 1,\slashed 1},\mathbfcal{M}_{0,0,0,0,2,1,1,\slashed 1,\slashed 1},\mathbfcal{M}_{0,0,0,0,1,2,1,\slashed 1,\slashed 1}, \mathbfcal{M}_{1,1,0,0,1,1,1,\slashed 1,\slashed 1}\}^{T},
    \end{split}
\end{align}
which satisfies, $\partial_x \vec f(x,\epsilon)=A(x,\epsilon)\vec f(x,\epsilon)$ with,
\begin{align}
      A(x,\epsilon)=  \left(
\begin{array}{cccc}
 \frac{3 \left(x^4+6 x^2+1\right) \epsilon -\left(x^2+1\right)^2}{x \left(x^4-1\right)} & -\frac{x^2-1}{2 \left(x^3+x\right)} & \frac{2 x (2 \epsilon +1)}{\epsilon -x^4 \epsilon } & 0 \\
 \frac{6 \left(x^2-1\right) \epsilon ^2}{x^3+x} & \frac{x^4 (\epsilon +1)+x^2 (6 \epsilon +2)+\epsilon +1}{x-x^5} & \frac{4 x (2 \epsilon +1)}{x^4-1} & 0 \\
 -\frac{6 \left(x^2-1\right) \epsilon ^2}{x^3+x} & \frac{\epsilon -x^2 \epsilon }{x^3+x} & \frac{\left(x^2-1\right) (2 \epsilon +1)}{x^3+x} & 0 \\
 0 & -\frac{4}{x^2-1} & 0 & \frac{2 \left(x^2+1\right)}{x-x^3} \\
\end{array}
\right)\,.\label{3.12}
\end{align}
The matrix equation can be brought to $\epsilon$-form (canonical form) using suitable basis transformation using the publicly available packages mentioned earlier. In the canonical basis/Uniform Transcendental (UT) basis, the differential equation takes the form, $\partial_x \vec g(x,\epsilon)=\epsilon \mathbb{S}(x)\vec g(x,\epsilon)$, where $\mathbb{S}(x)$ and the transformation matrix $\mathbb{T}(x,\epsilon)$ are given by,
\begin{align}
\begin{split}
   &\mathbb{T}(x,\epsilon)= \left(
\begin{array}{cccc}
 -\frac{7 x}{x^2-1} & \frac{x}{2-2 x^2} & \frac{3 x}{x^2-1} & 0 \\
 -\frac{6 x \epsilon }{x^2-1} & -\frac{5 x \epsilon }{x^2-1} & \frac{6 x \epsilon }{x^2-1} & 0 \\
 \frac{12 \left(x^2-1\right) \epsilon ^2}{2 x \epsilon +x} & \frac{2 \left(x^2-1\right) \epsilon ^2}{2 x \epsilon +x} & -\frac{12 x \epsilon ^2}{2 \epsilon +1} & 0 \\
 0 & 0 & 0 & \frac{2 x^2}{\left(x^2-1\right)^2} \\
\end{array}
\right),\quad\textrm{and}\\ &
\mathbb{S}(x)=
\left(
\begin{array}{cccc}
 \frac{9 \left(2 x^2+1\right)}{2 \left(x^3-x\right)} & \frac{2 x^2+3}{4 \left(x^3-x\right)} & -\frac{6 x}{x^2-1} & 0 \\
 -\frac{3 \left(2 x^2+5\right)}{x^3-x} & -\frac{6 x^2+5}{2 \left(x^3-x\right)} & \frac{12 x}{x^2-1} & 0 \\
 \frac{6}{x} & -\frac{1}{3 x} & -\frac{2}{x} & 0 \\
 \frac{12}{x} & \frac{10}{x} & -\frac{12}{x} & 0 \\
\end{array}
\right)=\sum_{i=1}^3\frac{\mathbb{S}_i}{x-x_i},\,(x_i=1,-1,0)
\end{split}
\end{align}
where the matrix residues take the form,
\begin{align}
\mathbb{S}_0=\left(\begin{array}{cccc}
         -  \frac{9}{2}  & -\frac{3}{4}& 0 &0  \\
            15 & \frac{5}{2} & 0 &0\\
            6&-\frac{1}{3}&-2&0\\
            12 & 10 & -12 &0 
     \end{array}
     \right)\,, \quad    
    \mathbb{S}_{\pm 1}=\left(\begin{array}{cccc}
           \frac{27}{4}  & \frac{5}{8}& -3 &0  \\
            -\frac{21}{2} & -\frac{11}{4} & 6 &0\\
            0&0&0&0\\
            0 & 0 & 0 &0 
     \end{array}
     \right)\,.  \label{2.17} 
\end{align}
In this fuchsified form, the equation looks like this,
\begin{align}
    \begin{split}
        \partial_x\vec g(x,\epsilon)=\epsilon \Bigg(\frac{S_0}{x}+\frac{S_1}{x+1}+\frac{S_{-1}}{x-1}\Bigg)\vec g(x,\epsilon)\label{3.15}
    \end{split}
\end{align}
which has the solution in terms of a path ordered exponential as mentioned in \eqref{3.10a}. Finally, in  the Laporta basis, the formal solution takes the form,
\begin{align}
    \begin{split}
        \vec f(x,\epsilon)=\mathbb{T}(x,\epsilon)\,\Bigg(\mathbfcal{P}e^{\epsilon\int_{1^{-}}^x \,\mathbb{S}(x')\,dx'}\Bigg)\,\mathbb{T}^{-1}(1^{-},\epsilon) \,\vec f(1^{-},\epsilon)\,.\label{3.16}
    \end{split}
\end{align}
Taking the boundary limit is a bit tricky, and one needs to take the limit very carefully.First we expand the $\mathbb T^{-1}(x_0=1-v_{\infty})$ matrix  and  also expand the boundary vector $\vec f(x_0,\epsilon)$ around $v_\infty=0$ using the method of regions. For generic two-loop integrals, we will have both the contribution from potential and the radiation region \cite{Dlapa:2023hsl}. \textit{ However, at 3PM, the full conservative dynamics of the boundary integrals are captured by the contribution from the potential region only \cite{Dlapa:2023hsl}.}\\ \\
\textbf{\textit{Potential region:}} In momentum space, the potential region is characterized by the following loop momenta scaling: $l_{i}\sim |\boldsymbol{q}|(v_{\infty},1)$ and, the velocity of black holes are parametrized as \cite{Dlapa:2023hsl}, 
\begin{align}
    \begin{split}
        v_2=(1,0,0,0),\,\, v_1=(1,v_{\infty},0,0).\label{2.23a}
    \end{split}
\end{align}
Using the above-mentioned scaling, the integral is reduced to,
\begin{align}
    \begin{split}
     \mathbfcal{M}\Big|_{v_{\infty}\to 0}^{\textrm{pot.}}\sim    v_{\infty}^{-\sum n_i}\Bigg(\prod_{i}\int_{\vec \ell_i}\frac{1}{(\pm\vec \ell_i\cdot \boldsymbol{n}+i\varepsilon)^{n_i}}\Bigg)\frac{1}{\boldsymbol {D_1}^{m_1}\cdots\boldsymbol{D_k}^{m_k}},\,\boldsymbol D\equiv \textrm{spatial part of the squared propagators.}
    \end{split}
\end{align}
For our case, we are interested in the following two loop integrals in potential regions.
\begin{align}
    \begin{split}
\mathbfcal{M}^{\textrm{2-loop},(\mp)}_{n_1,n_2;m_1,..,m_5}\Big|_{v_{\infty}\to 0}^{\textrm{pot.}}&\sim\int_{\ell_{1},\ell_{2}}\frac{\hat\delta(\ell_{1}^0)\hat\delta(\ell_{2}^0-v_{\infty}\ell_{2}^{(x)})}{(\ell_{1}^0-v_{\infty}\ell_{1}^{(x)}+i\varepsilon)^{n_1}(\ell_{2}^0+i\varepsilon)^{n_2}}\frac{1}{\boldsymbol{D_1}^{m_1}\cdots\boldsymbol{D_5}^{m_5} }+\mathcal{O}(v_{\infty}^{-1})\\ &
=v_{\infty}^{-n_1-n_2}\int_{\boldsymbol{\ell_1},\boldsymbol{\ell_2}}\frac{1}{(-\boldsymbol{\ell_1}\cdot \boldsymbol{n}+i\varepsilon)^{n_1}(\boldsymbol{\ell_2}\cdot \boldsymbol{n}+i\varepsilon)^{n_2}}\frac{1}{\boldsymbol{D_1}^{m_1}\cdots\boldsymbol{D_5}^{m_5} }+\mathcal{O}(v_{\infty}^{1-n_1-n_2})\,.
    \end{split}
\end{align}
The details of the computation of potential mode master integrals are given in Appendix~(\ref{AppB}).\par

\noindent
Now, the solution \eqref{3.16} to the differential equation to the differential equation can be presented in terms of multiple
polylogarithms (MPLs) \cite{Remiddi:1999ew,goncha},
\begin{align}
    G(a_1,a_2,\cdots a_n;z)=\int_{0}^z \frac{dt}{t-a_1}G(a_2,\cdots,a_n;t),\,\, G(\,;z)=1,\,\,\forall a_i,z\in \mathbb{C}\,.
\end{align}
If all $a_i's$ are zero, 
\begin{align}
    G(\vec 0_n;z)=\frac{1}{n!}\,\log^{n}(z)
\end{align}
and, for equal $a_i=a$ one can obtain,
\begin{align}
    G(\vec a_n,z)=\frac{1}{n! }\log^{n}\Big(1-\frac{z}{a}\Big),\,\,\forall a\ne 0\,.
\end{align}
Apart from ordinary logarithms, MPLs also contain classical polylogarithms, which are defined as,
\begin{align}
    \begin{split}
        G(\vec 0_{n-1},1;z)=- \textrm{{\bf Li}}_2(z)\,.
    \end{split}
\end{align}
Now, after accumulating all the facts mentioned above and suitably regularizing the logarithmic divergences, we get the solutions for the Laporta master integrals that are needed for our subsequent computations, and we give the matrix with epsilon form, $\mathbb{S}(x)$, in the Appendix~\eqref{Ca} \footnote{{We match all the listed master integrals in \eqref{MasterM} with \cite{Jakobsen:2022fcj}. At leading order (upto which the result of the integrals given in \cite{Jakobsen:2022fcj}), the results have perfect agreement with our results except $\mathbfcal{M}_{0,0;1,1,2,1,1}$ }. However, we have crosschecked our result by explicitly matching it with the boundary value. S.G would like to thank Gustav Jacobsen for confirming the same. }.
\begin{adjustwidth}{-2cm}{2cm}
\begin{tcolorbox}[enhanced, height=21.1cm, width=20.4 cm, colback=brown!2!white, colframe=black!2000!brown, title=\textit{Master Integrals}, breakable]
\begin{align*}
&
(-q^2)^{2\epsilon}\mathbfcal{M}_{0,0;0,0,1,1,1}=-\frac{x \log (x)}{32 \pi ^2 \left(x^2-1\right) \epsilon }+\frac{3x \left(\log^2(x)+\textbf{Li}_2(1-x^2)\right)}{32 \pi ^2 \left(x^2-1\right)}-\frac{ x \epsilon}{192 \pi ^2 \left(x^2-1\right)}  \Bigg[-7 \pi ^2 \log(x)\\ &\hspace{1.4 cm}+216 \boldsymbol{\mathscr{J}}(\{-1,-1,0\},x)-216 \boldsymbol{\mathscr{J}}(\{-1,0,0\},x)+216 \boldsymbol{\mathscr{J}}(\{-1,1,0\},x)-216 \boldsymbol{\mathscr{J}}(\{0,-1,0\},x)+120 \boldsymbol{\mathscr{J}}(\{0,0,0\},x)\\ &\hspace{1.6 cm}-216 \boldsymbol{\mathscr{J}}(\{0,1,0\},x)+216 \boldsymbol{\mathscr{J}}(\{1,-1,0\},x)-216 \boldsymbol{\mathscr{J}}(\{1,0,0\},x)+216 \boldsymbol{\mathscr{J}}(\{1,1,0\},x)\Bigg]+\mathcal{O}(\epsilon^2)\,,\\ &
(-q^2)^{2\epsilon+1}\mathbfcal{M}_{0,0;0,0,1,2,1}=-\frac{x^2+1}{64 \pi ^2 x}+\frac{x^2+(1-x^2) \log (x)+1}{32 \pi ^2 x}\epsilon+\frac{\epsilon^2}{384 \pi ^2 x}\Big[120 \left(x^2-1\right) (\textbf{Li}_2(1-x)-\textbf{Li}_2(-x))\\ &\hspace{2.3 cm}-3 \left(8+\pi ^2\right) x^2+12 \log (x) \left(\left(9 x^2-1\right) \log (x)+\left(x^2-1\right) (2-10 \log (x+1))\right)+17 \pi ^2-24\Big]\,,\\ &\
(-q^2)^{2\epsilon+1}\mathbfcal{M}_{0,0;0,0,2,1,1}=-\frac{x \log (x)}{16 \pi ^2 \left(x^2-1\right)}-\epsilon  \frac{x   \left(\log^2(x)+\textbf{Li}_2(1-x^2)\right)}{16 \pi ^2 \left(x^2-1\right)}     \tikzmarkend{ma1}+\mathcal{O}(\epsilon^2)\,,\\ &
     (-q^2)^{2\epsilon+1} \textcolor{black}{\mathbfcal{M}^{(++)}_{1,1;0,0,1,1,1}}=\frac{x^2}{8 \pi ^2 \left(x^2-1\right)^2 \epsilon ^2}-\frac{x^2 \left(\pi ^2-6 \log ^2(x)\right)}{48 \left(\pi ^2 \left(x^2-1\right)^2\right)}+\frac{x^2 \epsilon }{24 \pi ^2 \left(x^2-1\right)^2}\Big[ (-12 (\boldsymbol{\mathscr{J}}(\{0,-1,0\},x)-\boldsymbol{\mathscr{J}}(\{0,0,0\},x)\\ &\hspace{4 cm}+\boldsymbol{\mathscr{J}}(\{0,1,0\},x))-32 \zeta (3))\Big]+O\left(\epsilon ^2\right)=  (-q^2)^{2\epsilon+1}\frac{1}{2}\mathbfcal{M}^{(+-)}_{1,1;0,0,1,1,1}\,,
 \\     &(-q^2)^{1+2\epsilon}\mathbfcal{M}^{(+\pm)}_{0,0;1,1,0,1,1}=\frac{1}{(4\pi)^{3-2\epsilon}}\frac{\Gamma^4(1/2-\epsilon)\Gamma^2(1/2+\epsilon)}{\Gamma^2(1-2\epsilon)},\\ &
(-q^2)^{2+2\epsilon}\mathbfcal{M}^{(+\pm)}_{0,0;1,1,1,1,1}=\frac{x \log (x)}{16 \pi ^2 (x^2-1) \epsilon }+\frac{\Big(\log^2(x)+\textbf{Li}_2(1-x^2)\Big)}{16\pi^2(x^2-1)}-\frac{\epsilon }{96 \pi ^2 \left(x^2-1\right)} \Bigg(24 x (3 \boldsymbol{\mathscr{J}}(\{-1,-1,0\},x)\\&\hspace{2.4 cm} -3 \boldsymbol{\mathscr{J}}(\{-1,0,0\},x)+3 \boldsymbol{\mathscr{J}}(\{-1,1,0\},x)-3 \boldsymbol{\mathscr{J}}(\{0,-1,0\},x)+\boldsymbol{\mathscr{J}}(\{0,0,0\},x)-3 \boldsymbol{\mathscr{J}}(\{0,1,0\},x)\\ & \hspace{2.4 cm}+3 \boldsymbol{\mathscr{J}}(\{1,-1,0\},x)-3 \boldsymbol{\mathscr{J}}(\{1,0,0\},x)+3 \boldsymbol{\mathscr{J}}(\{1,1,0\},x))-5 \pi^2 x \log (x)\Bigg)+\mathcal{O}(\epsilon^2)\,,\\ &
(-q^2)^{3+2\epsilon}\mathbfcal{M}^{(+\pm)}_{0,0;1,1,2,1,1}=\frac{x \left(x^4-4 x^2 \log (x)-1\right)}{16 \pi ^2 \left(x^2-1\right)^3 \epsilon }+\frac{x}{16 \pi ^2 \left(x^2-1\right)^3}\Bigg[5 \left(x^4-1\right)-2 \left(x^4+4 x^2+1\right) \log (x)\\ & \hspace{1.2 cm}+4x^2 \Big(\log^2 x+\textbf{Li}_2(1-x^2)\Big)\Bigg]+\epsilon\frac{x}{96 \pi ^2 \left(x^2-1\right)^3}\Big[288 x^2\Bigg( \boldsymbol{\mathscr{J}}(\{-1,-1,0\},x)- \boldsymbol{\mathscr{J}}(\{-1,0,0\},x)+ \boldsymbol{\mathscr{J}}(\{-1,1,0\},x)\\ &\hspace{1.2 cm}- \boldsymbol{\mathscr{J}}(\{0,-1,0\},x)+ \frac{1}{3} \boldsymbol{\mathscr{J}}(\{0,0,0\},x)- \boldsymbol{\mathscr{J}}(\{0,1,0\},x)+ \boldsymbol{\mathscr{J}}(\{1,-1,0\},x)- \boldsymbol{\mathscr{J}}(\{1,0,0\},x)+ \boldsymbol{\mathscr{J}}(\{1,1,0\},x)\Bigg)\\ &\hspace{1.2 cm}+24 \left(x^4+12 x^2+1\right) \textbf{Li}_2(1-x)-24 \left(x^4+12 x^2+1\right) \textbf{Li}_2(-x)+54 \left(x^4-1\right)+36 x^4 \log ^2(x)-24 x^4 \log (x) \log (x+1)\\ &\hspace{1.2 cm}-96 x^4 \log (x)+\pi ^2 \left(3 x^2 \left(x^2-8\right)-7\right)+144 x^2 \log ^2(x)-288 x^2 \log (x) \log (x+1)-20 \pi ^2 x^2 \log (x)\\ &\hspace{1.2 cm}+24 x^2 \log (x)-12 \log ^2(x)-24 \log (x) \log (x+1)-96 \log (x)\Big]+\mathcal{O}(\epsilon^2)\,,\\ &
 (-q^2)^{2\epsilon} \textcolor{black}{\mathbfcal{M}^{(+\pm)}_{0,0;0,1,1,0,1}}=0,\,\,
 (-q^2)^{2\epsilon} \textcolor{black}{\mathbfcal{M}^{(+\pm)}_{0,1;0,0,1,1,1}}=-\frac{i\,x}{32\pi \epsilon (x^2-1)}+\mathcal{O}(\epsilon^0)
\end{align*}\label{MasterM}
\end{tcolorbox}
\end{adjustwidth}
\vspace{-3 cm}
\newpage
The iterative integral in \eqref{MasterM} is defined as,
\begin{align}
    \boldsymbol{\mathscr{J}}(\{i,j,k\},x)\equiv \int_{1^-}^x \frac{dt}{t-i}\int_{1^-}^{t}\frac{dt'}{t'-j}\int_{1^-}^{t'}\frac{dt''}{t''-k} \boldsymbol{\mathscr{J}}(\{\},t'');~~~~~~\boldsymbol{\mathscr{J}}(\{\},t'')=1,\label{3.43k}
\end{align}
and, the relevant 3-point iterative UT integrals are defined in Appendix~\eqref{D1}\footnote{Note that the integrals defined in \eqref{3.43k} are not exactly Goncharov MPLs (which is defined for iterative integrals with lower limit `$0$') because here the lower limit is `$1$'. However one can write \eqref{3.43k} as a difference of Goncharov MPLs.}. 
In \eqref{MasterM} we extensively use the following identity,
\begin{align}
    \textbf{Li}_2(-x)-\textbf{Li}_2(1-x)=-\frac{\pi^2}{12}-\log(x)\log(1+x)-\frac{1}{2}\textbf{Li}_2(1-x^2),
\end{align}
which can be derived from the fundamental identities of dilogarithm, namely \textit{Reflection formula and Abel's duplication formula}  \cite{goncha}. From now on, we make the $(++)$ sign of the last master integral of \eqref{MasterM} implicit for all computations. All the derivations are done by implicitly making the $i\varepsilon$ prescription positive for all linear propagators using proper field redefinition.\par
Now, we provide a bit of detail regarding the computation of master integrals. Let's focus on the following $\mathbfcal{M}_{0,1;0,0,1,1,1}$ and $\mathbfcal{M}_{0,0;0,1,0,1,1}$ that we do not include in the differential system mentioned in \eqref{3.12}. Fortunately, the first integral can be found separately using the differential equation method as it closes itself, and the second one identically vanishes because of the tadpole nature of $\ell_{1}$ loop integral,
\begin{align}
    \begin{split}
     &   \frac{d}{d\gamma}\mathbfcal{M}_{0, 1; 0, 0, 1, 1, 1, \slashed 1, \slashed 1}(\gamma,\epsilon)=-\frac{\gamma}{\gamma^2-1}\mathbfcal{M}_{0, 1; 0, 0, 1, 1, 1, \slashed 1, \slashed 1}(\gamma,\epsilon)+\frac{1}{\gamma^2-1}{\mathbfcal{M}_{0, 0; 0, 0, 1, 1, 1, \slashed 2, \slashed 1}}\,.
     \label{3.29}
    \end{split}
\end{align}
It appears that the master $\mathbfcal{M}_{0, 1; 0, 0, 1, 1, 1, \slashed 1, \slashed 1}(\gamma,\epsilon)$ does not close, but we now sketch that the master $\mathbfcal{M}_{0, 0; 0, 0, 1, 1, 1, \slashed 2, \slashed 1}$, that appears in the second term in \eqref{3.29} does not contribute and can be proved by noting that the integral has a flip symmetry $\ell_i\to -\ell_i,\,q\to -q$.
\begin{align}
\begin{split}
    \mathbfcal{M}_{0, 0; 0, 0, 1, 1, 1, \slashed 2, \slashed 1}(|q|,\gamma)&\propto \int_{\ell_{1},\ell_{2}}\frac{\hat\delta(\ell_{2}\cdot v_1)}{\ell_{1}^2 \ell_{2}^2 (\ell_{1}+\ell_{2}-q)^2}\Bigg(\frac{1}{(\ell_{1}\cdot v_2+i\varepsilon)^2}-\frac{1}{(\ell_{1}\cdot v_2-i\varepsilon)^2}\Bigg)\,,\\ &\xrightarrow[l_i\to -\ell_{1}]{q\to -q}\int_{\ell_{1},\ell_{2}}\frac{\hat\delta(\ell_{2}\cdot v_1)}{\ell_{1}^2 \ell_{2}^2 (\ell_{1}+\ell_{2}-q)^2}\Bigg(\frac{1}{(\ell_{1}\cdot v_2-i\varepsilon)^2}-\frac{1}{(\ell_{1}\cdot v_2+i\varepsilon)^2}\Bigg)\,,\\ &
    =-  \mathbfcal{M}_{0, 0; 0, 0, 1, 1, 1, \slashed 2, \slashed 1}\implies   \mathbfcal{M}_{0, 0; 0, 0, 1, 1, 1, \slashed 2, \slashed 1}\to 0\,.
    \end{split}
\end{align}
Therefore, the solution can be written straightforwardly and is given by,
\begin{align}
    \begin{split}
     &   \mathbfcal{M}_{0, 1; 0, 0, 1, 1, 1}(\gamma,\epsilon)=\frac{C_1(\epsilon,\gamma_0)}{\sqrt{\gamma^2-1}},
     \\ &
   \mathbfcal{M}_{0,0;0,1,0,1,1}(\gamma,\epsilon)  =0. \label{MasterC}
    \end{split}
\end{align}
To fix the boundary data, we evaluate the integrals in the following limit,
\begin{align}
    \begin{split}
   \lim_{\gamma\to 1^{++}}     \sqrt{\gamma^2-1} \,\mathbfcal{M}_{0, 1; 0, 0, 1, 1, 1, \slashed 1, \slashed 1}^{(++)}(\gamma,\epsilon)&=\lim_{\gamma\to 1^{+}}\sqrt{\gamma^2-1}\int_{\ell_{1},\ell_{2}}\frac{{\hat\delta(\ell_{1}^{(0)})\hat\delta(\gamma \ell_{2}^{0}-\gamma\beta \ell_{2}^{(x)})}}{(\ell_{2}^{(0)}+i\varepsilon)(\ell_{1}+\ell_{2}-q)^2(\ell_{1}-q)^2(\ell_{2}-q)^2}\,,\\ &
   =-\int_{\boldsymbol{\ell_1},\boldsymbol{\ell_2}}\frac{1}{(\boldsymbol{\ell_2}\cdot \boldsymbol{n}+ i \delta)\,\boldsymbol{\ell_1}^2\,\boldsymbol{\ell_2}^2 \,(\boldsymbol{\ell_1}+\boldsymbol{\ell_2}-\boldsymbol{q})^2}\,,\\ &
   =\frac{i 4^{2 \epsilon -3} \pi ^{2 \epsilon -\frac{5}{2}} \Gamma \left(\frac{1}{2}-2 \epsilon \right) \Gamma (-\epsilon ) \Gamma \left(2 \epsilon +\frac{1}{2}\right)
   \Gamma^2(\frac{1}{2}-\epsilon)}{\Gamma \left(\frac{1}{2}-3 \epsilon \right) \Gamma(1-2\epsilon)}\frac{1}{(-q^2)^{1/2+2\epsilon}}\,.\label{3.37h}
    \end{split}
\end{align}
The appearance of negative sign in \eqref{3.37h} can be justified by noticing that $\mathbfcal{M}_{0,1;0,0,1,1,1}^{(+)}(+i\varepsilon)=-\mathbfcal{M}_{0,1;0,0,1,1,1}^{(-)}(-i\varepsilon)$, which can be achieved by momentum flip $\ell_i\to -\ell_i,\,q\to -q$. 
Now, we can extract the constant $C_1$ mentioned in \eqref{MasterC} and is given by,
\begin{align}
    \begin{split}
        C_1(\gamma_0\to 1^{+},\epsilon)=\frac{i\times 4^{2 \epsilon -3} \pi ^{2 \epsilon -\frac{5}{2}} \Gamma \left(\frac{1}{2}-2 \epsilon \right) \Gamma (-\epsilon ) \Gamma \left(2 \epsilon +\frac{1}{2}\right)
   \Gamma^2(\frac{1}{2}-\epsilon)}{\Gamma \left(\frac{1}{2}-3 \epsilon \right) \Gamma(1-2\epsilon)}\,.
    \end{split}
\end{align}
Therefore, the master integral is given by,
\begin{align}
    \begin{split}
       (-q^2)^{1/2+2\epsilon} \mathbfcal{M}_{0, 1; 0, 0, 1, 1, 1}^{(++)}(|q|,\gamma,\epsilon)&=\frac{i\times 4^{2 \epsilon -3} \pi ^{2 \epsilon -\frac{5}{2}}\Gamma \left(\frac{1}{2}-2 \epsilon \right) \Gamma (-\epsilon ) \Gamma \left(2 \epsilon +\frac{1}{2}\right)
   \Gamma^2(\frac{1}{2}-\epsilon)}{\sqrt{\gamma^2-1}\Gamma \left(\frac{1}{2}-3 \epsilon \right) \Gamma(1-2\epsilon)}\,,\\ &
       =-\frac{i}{64\pi \epsilon\sqrt{\gamma^2-1}}+\mathcal{O}(\epsilon^0)=  (-q^2)^{1/2+2\epsilon} \mathbfcal{M}_{0, 1; 0, 0, 1, 1, 1}^{(+-)}\,.
    \end{split}
\end{align}
\par 
At two loops we encounter another family of integrals defined in \eqref{3.4a} with basis functions,
 $\{D_i,\slashed D_k\}\in\Big(\ell_{1}\cdot v_1,\,\,\ell_{2}\cdot v_1,\,,\,\ell_{1}^2,\,\, \ell_{2}^2,\,\, (\ell_{1}+\ell_{2}-q)^2,\,\, (\ell_{1}-q)^2,\,\, (\ell_{2}-q)^2,\,\, \ell_{1}\cdot v_2,\,\, \ell_{2}\cdot v_2\Big) $, where the last two entries are the cut propagators. They are just `potential region' counterparts of the $\mathbfcal{M}$-type master integrals discussed earlier. After carrying out the  IBP reduction, we found 8 master integrals, and they are completely factorized in $\epsilon$ and $\gamma$, and we do not need differential equation techniques to solve them. Of these, we will need only two for our subsequent computations, and we list the results for those two below:
 \begin{align}
     \begin{split}
& (-q^2)^{2\epsilon}\mathbfcal{L}_{0,0;0,0,1,1,1}\equiv \frac{1}{(4\pi)^{3-2\epsilon}}\frac{\Gamma^3(1/2-\epsilon)\Gamma(2\epsilon)}{\Gamma(3/2-3\epsilon)}\,,\\ &
(-q^2)^{2\epsilon+1}\mathbfcal{L}^{(++)}_{1,1;0,0,1,1,1}=(-q^2)^{2\epsilon+1}2\mathbfcal{L}^{(+-)}_{1,1;0,0,1,1,1}\equiv \frac{1}{(4\pi)^{2-2\epsilon}}\frac{\Gamma^3(-\epsilon)\Gamma(1+2\epsilon)}{3(\gamma^2-1)\Gamma(-3\epsilon)}\,. \label{MasterL}
\end{split}
 \end{align}\\
At one loop, we encounter the following family of integral (we always look for the case when $\gamma=1$),
\begin{align}
    \begin{split}
        \mathbfcal{G}^{\pm,\textrm{1-loop}}_{\alpha_1;\beta_1 \beta_{2}}=\int_{l_{1,2}}\frac{\hat\delta^{(\gamma_1)}(\slashed D_1)}{(\pm D_1+i\varepsilon)^{\alpha_1}}\frac{1}{D_2^{\beta_1}D_3^{\beta_2}}\label{3.41a}
    \end{split}
\end{align}
with $\{D_i\,\lceil\slashed D_k\rceil\}\in \Big(\ell_{1}\cdot v_1,\,\ell_{1}^2,\,(\ell_{1}-q)^2,\,\lceil \ell_{1}\cdot v_2\rceil\Big)$. After IBP reduction we found two master integrals: ($\mathbfcal{G}_{0;1,1},\,\mathbfcal{G}_{1;1,1}$). We will require only one of them, and it is given by,
\begin{align}
    \begin{split}
        &(-q^2)^{\epsilon+1/2}\mathbfcal{G}_{0;1,1}=\frac{2^{2 \epsilon -3} \pi ^{\epsilon -\frac{3}{2}} \Gamma \left(\frac{1}{2}-\epsilon \right)^2 \Gamma \left(\epsilon +\frac{1}{2}\right)}{\Gamma (1-2 \epsilon )}\,. \label{3.4aa}
    \end{split}
\end{align}
\section{Computation of eikonal phase}
\label{sec4}
In this section, we provide the detailed computation of the eikonal phase upto the third Post-Minkowskian (PM) order. We mainly focus on the scalar field and dCS corrections to the phase as the contribution of the Einstein-Hilbert term has already been investigated earlier \cite{Jakobsen:2022fcj, Jakobsen:2021zvh}. We restrict our subsequent computation upto linear order in spin, leaving the study beyond that order for future investigation.  \\\\ 
\textbf{Spinning eikonal at 2PM:}
In this subsection, we compute the eikonal phase at 2PM order by considering only the scalar-graviton interaction terms by noting that spin only appears when there are scalar-graviton vertices or scalar-fermion vertices in the worldline. The spinless diagrams involving scalar and scalar-graviton have been done in \cite{Bhattacharyya:2024aeq}.\\ \\
\textbullet $\,\,$ We start with the diagram of the following topology where we have a scalar-superfield vertex,
\begin{align}
    \begin{split}
         \begin{minipage}[h]{0.12\linewidth}
	\vspace{4pt}
	\scalebox{1.8}{\includegraphics[width=\linewidth]{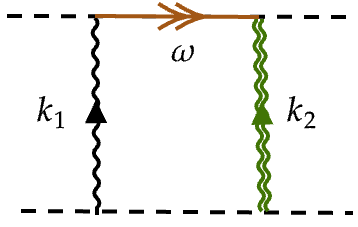}}
 \end{minipage}&
    \end{split}
\end{align}
The corresponding amplitude is given by
\begin{align}
    \begin{split}
      i \chi\Big|_{\mathcal{S}^1}= -i\frac{s_1 s_2 m_1 m_2^2}{8 m _p^4} \int e^{-i(k_1+k_2)\cdot b}\frac{\hat\delta(\omega-k_1\cdot v_1)\hat\delta(\omega+k_2\cdot v_1)\hat\delta(k_1\cdot v_2)\hat\delta(k_2\cdot \textcolor{black}{v_2})}{\omega\,k_1^2\, k_2^2}\mathcal{N}
    \end{split}
\end{align}
where the numerator takes the form:
\begin{align}
    \begin{split}
        \mathcal{N}=\omega \,\bar\Psi_{1\eta}\,\Psi_1^{\sigma}\eta^{\eta\delta}k_{2[\delta}\delta_{\sigma]}^{(\mu}v_1^{\nu)}P_{\mu\nu;\alpha\beta}v_2^\alpha v_2^\beta\,.
    \end{split}
\end{align}
Now, using the spin-supplementary condition (\ref{SSC17}) and integrating over the worldline energy we get,
\begin{align}
    \begin{split}
      i  \chi\Big|_{\mathcal{S}^1}&= \frac{\gamma s_1 s_2 m_1 m_2^2}{32 m _p^4} v_{2\eta}\mathcal{S}_1^{\eta\delta}\int e^{-i(k_1+k_2)\cdot b}\frac{\hat\delta(k_1\cdot v_1+k_2\cdot v_1)\hat\delta(k_1\cdot v_2)\hat\delta(k_2\cdot v_2)\,k_{2\delta}}{k_1^2 k_2^2}\,,\\ &
        =\frac{\gamma s_1 s_2 m_1 m_2^2}{32 m _p^4}v_{2\eta}\mathcal{S}_1^{\eta\delta}\int_{q} e^{-iq\cdot b}\hat\delta(q\cdot v_1)\hat\delta(q\cdot v_2)\int_{k_1}\frac{\hat\delta(k_1\cdot v_2)\,(q-k_1)_\delta}{k_1^2(k_1-q)^2}\,.
    \end{split}
\end{align}
Now, using Passarino-Veltman (PV) reduction and inserting the vertex factors, we get,\\

\begin{align}
    \begin{split} 
      i  \chi\Big|_{\mathcal{S}^1}&= \frac{\gamma s_1 s_2 m_1 m_2^2}{512 m _p^4}v_{2\eta}\mathcal{S}_1^{\eta i}\int e^{-iq\cdot b}\frac{\hat\delta(q\cdot v_1)\hat\delta(q\cdot v_2)\, q_i}{(-q^2)^{1/2}}\,,\\ &
        =-i\frac{ s_1 s_2 m_1 m_2^2}{1024 \pi m _p^4|\boldsymbol{b}|^2}\Big(\frac{1+x^2}{1-x^2}\Big)\Big(\hat b\cdot \mathcal{S}_1\cdot v_2\Big)\,.
        \end{split}
\end{align}
Here we have replaced $\gamma$ by $\frac{x^2+1}{2 x}.$ All the subsequent expressions for the eikonal phase will be expressed in terms of $x\,.$ For the $q$ integral, we have used \eqref{int} of Appendix~(\ref{D}). In all the subsequent computations, whenever we will encounter this kind of vector integral, we will use the result mentioned in \eqref{int}.\\\\
\textbullet $\,\,$ Another possible diagram in this order can be obtained by replacing the graviton line with the scalar line in the previous diagram.
\begin{align}
    \begin{split}
         \begin{minipage}[h]{0.12\linewidth}
	\vspace{4pt}
	\scalebox{1.8}{\includegraphics[width=\linewidth]{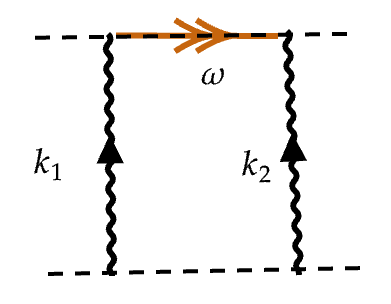}}
 \end{minipage}&
    \end{split}
\end{align}
The corresponding phase takes the following form,
\begin{align}
    \begin{split}
        i\chi\sim\int e^{-i(k_1+k_2)\cdot b}\frac{\hat\delta(\omega-k_1\cdot v_1)\hat\delta(\omega+k_2\cdot v_1)\hat\delta(k_1\cdot v_2)\hat\delta(k_2\cdot v_2)}{\omega\,k_1^2\,k_2^2}\mathcal{N}
    \end{split}
\end{align}
where the numerator takes the form,
\begin{align}
    \begin{split}
        \mathcal{N}=\omega^2\, \bar\Psi_{\eta}\Psi_{\delta}\,\eta^{\eta\delta}\,.
    \end{split}
\end{align}
Now, using the fact that another diagram with opposite spinor flow will eventually add up to this diagram such that we will eventually have terms that are antisymmetric in of $(\eta,\delta)$ and therefore, this diagram will not contribute to the eikonal phase.\textit{This structure eventually ensures that only the scalar line along with worldline propagators does not contribute to the spinning eikonal.}
\\\\\\
\textbullet $\,\,$ Now we show the detailed computation of the following 2PM diagram with $\mathbb{U}$-type topology. Note that we only focus on the spinning part of the diagram as the spinless part has been done in \cite{Bhattacharyya:2024aeq}.
\begin{align}
    \begin{split}
         \begin{minipage}[h]{0.12\linewidth}
	\vspace{4pt}
	\scalebox{1.7}{\includegraphics[width=\linewidth]{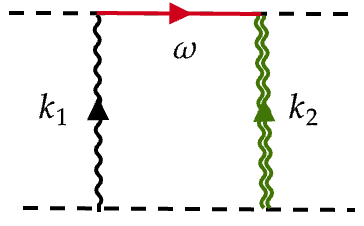}}
 \end{minipage}&
    \end{split}
\end{align}
The contribution to the eikonal phase from this diagram takes the following form,
\begin{align}
    \begin{split}
       i \chi\Big|_{\mathcal{S}^1} =-i\frac{s_1 s_2 m_1 m_2^2}{16m_p^4} \int e^{-i(k_1+k_2)\cdot b}\frac{\hat\delta(k_1\cdot v_1+k_2\cdot v_1)\hat\delta(k_1\cdot v_2)\hat\delta(k_2\cdot v_2)}{\omega^2 \,k_1^2\, k_2^2}\mathcal{N}\,,
    \end{split}
\end{align}
where the numerator $\mathcal{N}$ has the following form,
\begin{align}
    \begin{split}
        \mathcal{N}=&i\gamma (k_2\cdot \mathcal{S}_1\cdot v_2)\Big[k_1\cdot k_2-2 (k_1\cdot v_1)^2-2\, k_1\cdot v_1\,k_2\cdot v_1\Big]-4i\gamma \,(k_2\cdot \mathcal S_2\cdot v_1)(k_1\cdot v_1)^2+i\gamma (k_2\cdot \mathcal{S}_2\cdot k_1)(k_1\cdot v_1)\\ &
        -\frac{i}{2}(k_2\cdot \mathcal{S}_1\cdot k_1)(k_1\cdot v_1)\,.
    \end{split}
\end{align}
Now, redefining $k_1+k_2\to q$ and ignoring the tadpoles, we will get \footnote{We will always throw away the tadpoles before performing the IBP reduction. From now on, we will not mention it explicitly. It should be understood.},
\begin{align}
    \begin{split}
      i  \chi\Big|_{\mathcal{S}^1}\sim \int e^{-iq\cdot b}\hat\delta(q\cdot v_1)\hat\delta(q\cdot v_2)&\int \frac{\hat\delta(k_1\cdot v_2)}{(k_1\cdot v_1)^2\,k_1^2\,(k_1-q)^2}\Bigg[\frac{\gamma}{2}q^2\Big(q\cdot \mathcal{S}_1\cdot v_2-k_1\cdot \mathcal{S}_1\cdot v_2\Big)-4\gamma\Big(q\cdot \mathcal{S}_2\cdot v_1-k_1\cdot \mathcal S_2\cdot v_1\Big)\\ &
    \times (k_1\cdot v_1)^2+\gamma(q\cdot \mathcal S_2\cdot k_1)(k_1\cdot v_1) -\frac{1}{2}(q\cdot \mathcal{S}_1\cdot k_1)(k_1\cdot v_1)   \Bigg]\,.\label{4.9}
    \end{split}
\end{align}
To proceed further, we need to evaluate the following tensor integrals:
\begin{align}
    \begin{split}
        \theta_1^{\eta}=\int_{k_1}\frac{\hat\delta(k_1\cdot v_2)\,k_1^\eta}{(k_1\cdot v_1)^2\,k_1^2\,(k_1-q)^2}&=\frac{v_1^\eta-\gamma v_2^\eta}{1-\gamma^2}\int_{k_1}\frac{\hat\delta(k_1\cdot v_2)}{(k_1\cdot v_1)\,k_1^2\,(k_1-q)^2}+\frac{q^\eta}{2}\int_{k_1}\frac{\hat\delta(k_1\cdot v_2)}{(k_1\cdot v_1)^2\,k_1^2\,(k_1-q)^2}\,,\\ &
        =\frac{v_1^\eta-\gamma v_2^\eta}{1-\gamma^2} \,\mathbfcal{G}_{1;1,1}+\frac{q^\eta}{2}\frac{8 \epsilon } {(\gamma^2 -1) q^2}\mathbfcal{G}_{0;1,1}\,,\\ &
        \xrightarrow[]{\epsilon\to 0}\frac{v_1^\eta-\gamma v_2^\eta}{1-\gamma^2} \,\mathbfcal{G}_{1;1,1},\\ &
        \hspace{-4.8 cm}
        \theta_2^\eta =\int_{k_1}\frac{\hat\delta(k_1\cdot v_2)\,k_1^\eta}{(k_1\cdot v_1)\,k_1^2\,(k_1-q)^2}=\frac{v_1^\eta-\gamma v_2^\eta}{1-\gamma^2}\mathbfcal{G}_{0;1,1}+\frac{q^\eta}{2} \mathbfcal{G}_{1;1,1},\,\,\\&  \hspace{-4.8 cm}\theta_3^\eta=\int_{k_1}\frac{\hat\delta(k_1\cdot v_2)\,k_1^\eta}{k_1^2\,(k_1-q)^2}=\frac{q^\eta}{2}\mathbfcal{G}_{0;1,1}\,.
    \end{split}\label{4.10}
\end{align}
 In \eqref{4.10}, we have used the IBP reduction using a suitable choice of basis functions and write the full integral in terms of two master integrals: ($\mathbfcal{G}_{1;1,1},\,\mathbfcal{G}_{0;1,1}$). These are defined in \eqref{3.41a}. 
Now, using \eqref{4.10} as well as the anti-symmetric nature of the spin tensor and the spin supplementary condition {\eqref{SSC17}}, we get,\\
\begin{align}
    \begin{split}
    i    \chi\Big|_{\mathcal{S}^1}=&\frac{s_1 s_2 m_1 m_2^2}{16m_p^4} \int e^{-iq\cdot b}\hat\delta(q\cdot v_1)\hat\delta(q\cdot v_2)\Bigg(\frac{\gamma}{1-\gamma^2}\,(q\cdot \mathcal{S}_2\cdot v_1)\,+\frac{\gamma}{2(1-\gamma^2)}(q\cdot \mathcal{S}_1\cdot v_2)\Bigg)\mathbfcal{G}_{0;1,1}\,,\\&
    =i\frac{ s_1 s_2 m_1 m_2^2\,x^2(1+x^2)}{64\, \pi\, |\boldsymbol{b}|^2 m_p^4\left(1-x^2\right)^3} \Big(\hat b\cdot \mathcal{S}_2\cdot v_1 +\frac{1}{2}\hat{b}\cdot \mathcal{S}_1\cdot v_2\Big)\,.
    \end{split}
\end{align}
As we can see, although the intermediate step is tedious, the result is remarkably simple and depends on only one master integral $\mathbfcal{G}_{0;1,1}$ defined in \eqref{3.4aa} \footnote{Note that we have only shown the regularization independent finite part after doing an expansion in small $\epsilon\,.$ We will follow this same strategy for all the diagrams.}. We will use a similar methodology to evaluate the rest of the diagrams (at 3PM also).\\\\
\textbullet $\,\,$ Another scalar-graviton worldline interaction vertex having a $\mathbb{V}$-type topology will contribute to the 2PM eikonal phase. Again, for this case, the spinless part has been done in \cite{Bhattacharyya:2024aeq}\,.
\begin{align}
    \begin{split}
         \begin{minipage}[h]{0.12\linewidth}
	\vspace{4pt}
	\scalebox{1.7}{\includegraphics[width=\linewidth]{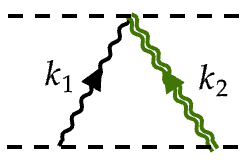}}
 \end{minipage}&
    \end{split}
\end{align}
The eikonal phase contribution is given by,
\begin{align}
    \begin{split}
        i\,\chi\Big|_{\mathcal{S}^1} & =-i\frac{s_1 s_2 m_1 m_2^2}{8m_p^4} \int e^{-i(k_1+k_2)\cdot b}\frac{\hat\delta(k_1\cdot v_1+k_2\cdot v_1)\hat\delta(k_1\cdot v_2)\hat\delta(k_2\cdot v_2)}{k_1^2 k_2^2}\\ &
        \hspace{3 cm}\times\Big(v_1^\mu v_1^\nu-i(k_2\cdot \mathcal{S}_1)^{(\mu}v_1^{\nu)}\Big)P_{\mu\nu;\alpha\beta}\Big(v_2^\alpha v_2^\beta+i (k_2\cdot \mathcal{S}_2)^{(\alpha}v_2^{\beta)}\Big)\,,\\ &
        =\frac{ \gamma s_1 s_2 m_1 m_2^2}{8m_p^4} \int e^{-iq \cdot b}\hat\delta(q\cdot v_1)\hat\delta(q\cdot v_2)\int \frac{\hat\delta(k_1\cdot v_2)}{k_1^2 (k_1-q)^2}(-k_1\cdot 
    \mathcal{S}_2\cdot v_1+k_1\cdot \mathcal{S}_1\cdot v_2+q\cdot \mathcal{S}_2\cdot v_1-q\cdot \mathcal{S}_1\cdot v_2),\\ &
    =\frac{ \gamma s_1 s_2 m_1 m_2^2}{16m_p^4} \int e^{-iq\cdot b}\hat\delta(q\cdot v_1)\hat\delta(q\cdot v_2)\Big((q\cdot \mathcal S_2\cdot v_1)-(q\cdot \mathcal S_1\cdot v_2)\Big)\,\mathbfcal{G}_{0;1,1}\,,\\&
    =i\frac{ s_1 s_2 m_1 m_2^2\,(1+x^2)}{256\,\pi\,  |b|^2 m_p^4\left(1-x^2\right)} \Big(\hat b\cdot \mathcal{S}_2\cdot v_1 -\hat{b}\cdot \mathcal{S}_1\cdot v_2\Big)\,.
    \end{split}
\end{align}
Hence the final form of the eikonal phase is,\\
\begin{align}
    \begin{split}
i\chi\Big|_{\mathcal{S}^{1}}=  i\frac{ s_1 s_2 m_1 m_2^2\,(1+x^2)}{256\,\pi\,  |\boldsymbol{b}|^2 m_p^4\left(1-x^2\right)} \Big(\hat b\cdot \mathcal{S}_2\cdot v_1 -\hat{b}\cdot \mathcal{S}_1\cdot v_2\Big)\,.
    \end{split}
\end{align}
At this point, we would like to remind the readers that one also needs to add the contributions from 2PM diagrams obtained by interchanging the worldlines one and two in all the diagrams mentioned above. We will not show them explicitly as the contributions from those diagrams will be similar to those we have already computed and can be obtained easily by exchanging labels one and two. \\\\
Finally adding all the $\mathcal{O}(\mathcal{S}^1)$ at 2PM contributions we have,\\
\begin{tcolorbox}[enhanced,  height= 3 cm, width=15 cm, colback=brown!2!white, colframe=black!2000!brown, title=\textit{Spinning Eikonal at 2PM}, breakable]
\begin{align*}   \chi\Big|_{\mathcal{S}^1}^{\textrm{2PM}}=&\frac{m_1m_2^2}{m_p^4}\Bigg[\Rho_{01}\Big(\hat b\cdot \mathcal{S}_{1}\cdot v_2\Big)+\Rho_{02}\Big(\hat b\cdot \mathcal{S}_2\cdot v_1 \Big)\Bigg]+(1\leftrightarrow 2)\,,
\end{align*}
\end{tcolorbox}
where, $$\Rho_{01}=\frac{s_1\,s_2\,\left(x^2+1\right) \left(5 x^4-18 x^2+5\right)}{1024 \pi  |\boldsymbol{b}|^2  \left(x^2-1\right)^3}\,,\quad \Rho_{02}=-\frac{s_1\,s_2\,\left(x^2+1\right)^3}{256 \pi  |\boldsymbol{b}|^2  \left(x^2-1\right)^3}\,.$$
\subsection*{Spinning eikonal at 3PM}
Now, we will compute the contributions to the eikonal phase at 3PM order. We will focus on the correction to the eikonal phase due to the extra scalar degree of freedom. \\

\textbullet $\,\,$ First, we will compute the Feynman diagrams coming purely from the scalar part. We start with the first comb-type diagram. 
\begin{align}
    \begin{split}
   \hspace{-4 cm}     \begin{minipage}[h]{0.12\linewidth}
	\vspace{4pt}
	\scalebox{4.3}{\includegraphics[width=\linewidth]{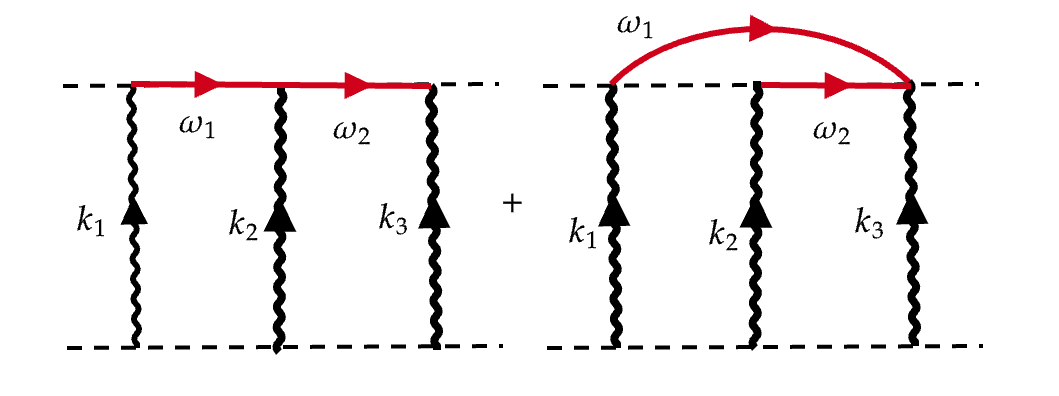}}
\end{minipage}
    \end{split}
\end{align}
The contribution to the eikonal phase looks like,
\begin{align}
    \begin{split}
       i\, \chi\Big|_{\mathcal{S}^0}&=-i \frac{s_1^3 s_2^3 m_1 m_2^3}{64 m_p^6} \rmint_{k_i,\omega_i}\hat\delta(k_1\cdot v_2)\hat\delta(k_2\cdot v_2)\hat\delta(k_3\cdot v_2)\hat\delta(-k_1\cdot v_1+\omega_1)\hat\delta(-k_2\cdot v_1+\omega_2-\omega_1)\hat\delta(\omega_2+k_3\cdot v_1)\\ &
       \hspace{2 cm} \times\frac{ \mathcal{N}}{\prod_i k_i^2}\left(\sum_{(\pm)} \frac{1}{(\pm\omega_1+i\varepsilon)^2(\pm\omega_2+i\varepsilon)^2}\right)\,e^{-i(k_1+k_2+k_3)\cdot b}
    \end{split}
\end{align}
Now relabeling the momenta $k_1\to \ell_1,\,k_2\to q-\ell_1-\ell_2\,,k_3\to\ell_2$ we can write the numerator as,\footnote{We have performed all the tensor contractions used in this paper using \textbf{FeynCalc} \cite{Mertig:1990an, Shtabovenko:2020gxv,Shtabovenko:2023idz}\,.},
\begin{align}
    \begin{split}
        \mathcal{N}=&4(\ell_1\cdot v_1)(\ell_2\cdot v_1)\left(\ell_1\cdot v_1+\ell_2\cdot v_1\right)^2
    \end{split}
\end{align}
\noindent
Now, taking care of the $\pm i\varepsilon$ prescription and then doing IBP reduction using \textbf{LiteRed} \cite{Lee:2012cn,Lee:2013mka}, one can find the eikonal phase in terms of two master integrals and is given by, 
\begin{align}
    \begin{split}
      &  i\chi\Big|_{\mathcal{S}^0}= -i \frac{s_1^3 s_2^3 m_1 m_2^3}{64 m_p^6}  \int_{q}e^{-iq\cdot b}\hat\delta(q\cdot v_1)\hat\delta(q\cdot v_2)\Bigg[  \boldsymbol{\tilde h_1}\mathbfcal{L}_{0,0;0,0,1,1,1}(q) -\boldsymbol{\boldsymbol{\tilde h_2}}\,q^2\mathbfcal{L}^{(+-)}_{1,1;0,0,1,1,1}(q)\Bigg]\,.
    \end{split}
\end{align}
where, $\boldsymbol{\tilde h_1}$ and $\boldsymbol{\tilde h_2}$ are defined in Appendix~(\ref{AppE})\,.
Then using \eqref{MasterL}, we finally get,\\
\begin{align}
    \begin{split} \label{eq1}
\chi\Big|_{\mathcal{S}^0}=\frac{ m_1 m_2^3}{m_p^6}\Bigg(\frac{s_1^3 s_2^3 x}{256 \pi ^3 |\boldsymbol{b}|^2 \left(x^2-1\right)}\Bigg) ,
    \end{split}
\end{align}
where $\gamma_E$ is the Euler-Mascheroni constant. For the $q$ integral, we have used \eqref{int1} of Appendix~(\ref{D}). In all the subsequent computations, whenever we will encounter this kind of scalar integral, we will use the result mentioned in \eqref{int1}.
The second diagram gives the following contribution,
\begin{align}
    \begin{split}
      i\chi\Big|_{\mathcal{S}^0}=-i \frac{s_1^3 s_2^3 m_1 m_2^3}{256 m_p^6} \int e^{-iq\cdot b}\hat\delta(q\cdot v_1)\hat\delta(q\cdot v_2)\int \sum_{(\pm)}\frac{\hat\delta(\ell_1\cdot v_2)\hat\delta(\ell_2\cdot v_2)\,\mathcal{N}}{(\pm\ell_1\cdot v_1+i\varepsilon)^2\,(\pm\ell_2\cdot v_1+i\varepsilon)^2\,\ell_1^2\,\ell_2^2\,(\ell_1+
        \ell_2-q)^2}
    \end{split}
\end{align}
where the numerator has the following form,
\begin{align}
    \begin{split}
        \mathcal{N}=&\frac{1}{8}\Bigg[(\ell_1-q)^2(\ell_2-q)^2+4q^2 (\ell_1\cdot v_1)(\ell_2\cdot v_1)\Bigg]\,.
    \end{split}
\end{align}
Therefore, the eikonal phase is given by,
\begin{align}
    \begin{split} 
    i\chi\Big|_{\mathcal{S}^0}=-i \frac{s_1^3 s_2^3 m_1 m_2^3}{64 m_p^6}\int e^{-iq\cdot b}\hat\delta(q\cdot v_1)\hat\delta(q\cdot v_2)\Bigg[\boldsymbol{\tilde h}_3\,\mathbfcal{L}_{0,0;0,0,1,1,1}+\frac{1}{4}\boldsymbol{\tilde h}_4\,q^2\,\mathbfcal{L}^{(++)}_{1,1;0,0,1,1,1}\Bigg]
    \end{split}
\end{align}
where, $ \boldsymbol{\tilde h_3},  \boldsymbol{\tilde h_4}$ are defined in Appendix~(\ref{AppE}).
Using \eqref{MasterL}, we finally get,\\
\begin{align}
    \begin{split} \label{eq2}
 \chi\Big|_{\mathcal{S}^0}=\frac{m_1 m_2^3}{m_p^6}\Bigg(\frac{ s_1^3 s_2^3 x^3 \left(x^4+1\right) (6 \log (|\boldsymbol{b}|)+5 \gamma_E +\log \left(\frac{\pi }{16}\right))}{512 \pi ^3 |\boldsymbol{b}|^2 \left(x^2-1\right)^5}\Bigg)\,.
    \end{split}
\end{align}
Then, the total contribution to the eikonal phase coming from these two diagrams after summing (\ref{eq1}) and (\ref{eq2}) is the following, 
\begin{align}
    \begin{split} \label{eqt1}
 \chi\Big|_{\mathcal{S}^0}=\frac{m_1 m_2^3}{m_p^6}\Delta_{\mathfrak{01}}^{\textrm{poly}}\,,
    \end{split}
\end{align}
where, $$\Delta_{\mathfrak{01}}^{\textrm{poly}}=\Bigg(\frac{s_1^3 s_2^3 x \left(\left(x^6+x^2\right) \log \left(\frac{\pi  |\boldsymbol{b}|^6}{16}\right)+2 x^8+(5 \gamma_E -8) x^6+12 x^4+(5 \gamma_E -8) x^2+2\right)}{512 \pi ^3 |\boldsymbol{b}|^2 \left(x^2-1\right)^5}\Bigg)\,.$$

\textbullet $\,\,$ Apart from the previous one, we have the following two comb-type diagrams where one of the scalar propagators is replaced by the graviton propagator. 
\begin{align}
    \begin{split}
   \hspace{-4 cm}     \begin{minipage}[h]{0.12\linewidth}
	\vspace{4pt}
	\scalebox{4.3}{\includegraphics[width=\linewidth]{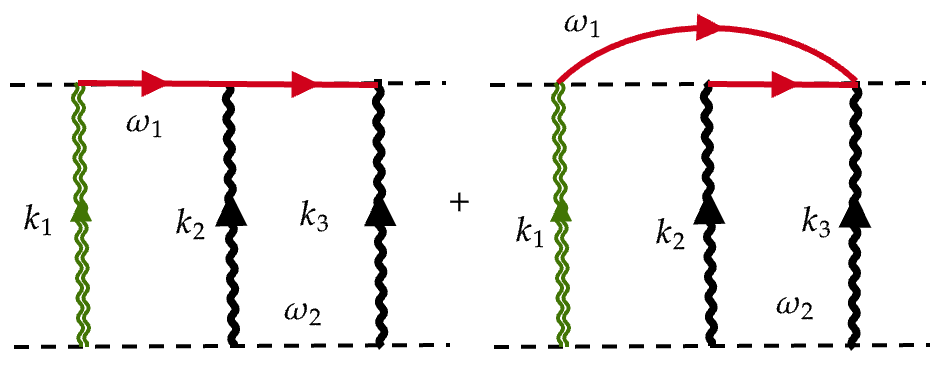}}
\end{minipage}
\end{split}
\end{align}
We proceed as before. The non-spinning part of the eikonal phase is given by, 
\begin{align}
    \begin{split} \label{eq3}
      i\chi\Big|_{\mathcal{S}^0}=-i\frac{m_1m_2^3 s_1^2 s_2^2}{32 m_p^6}\int e^{-iq\cdot b}\hat\delta(q\cdot v_1)\hat\delta(q\cdot v_2)\,\Bigg[ \boldsymbol{\tilde h}_5 \,\mathbfcal{L}_{0,0;0,0,1,1,1}-\frac{1}{2}\boldsymbol{\tilde h}_6 \,q^2\,\mathbfcal{L}^{(+-)}_{1,1;0,0,1,1,1}\Bigg]
    \end{split}
\end{align}
$ \boldsymbol{\tilde h_5},  \boldsymbol{\tilde h_6}$ are defined in Appendix~(\ref{AppE}).
Similarly, the spinning part is given by, 
\begin{align}
    \begin{split}  \label{eq4}
      i\chi\Big|_{\mathcal{S}^1}=-i\frac{m_1m_2^3 s_1^2 s_2^2}{32 m_p^6}\int e^{-iq\cdot b}\hat\delta(q\cdot v_1)\hat\delta(q\cdot v_2)\,\Bigg[ &i\,\Bigg(\boldsymbol{\tilde h}_7 \,\mathbfcal{L}_{0,0;0,0,1,1,1}-\boldsymbol{\tilde h}_8 \,q^2\,\mathbfcal{L}^{(+-)}_{1,1;0,0,1,1,1}\Bigg)(q\cdot \mathcal{S}_1\cdot v_2)+\\&i\,\Bigg(\boldsymbol{\tilde h}_9\,\mathbfcal{L}_{0,0;0,0,1,1,1}-\frac{1}{2}\boldsymbol{\tilde h}_{10}\,q^2\,\mathbfcal{L}^{(+-)}_{1,1;0,0,1,1,1}\Bigg)(q\cdot \mathcal{S}_2\cdot v_1)\Bigg]\,.
    \end{split}
\end{align}

Now, we compute the contribution to the eikonal phase from the second diagram.
The spinless part of the eikonal phase is given by,
\begin{align}
    \begin{split}\label{eq5}
        i\chi\Big|_{\mathcal{S}^0}=-i\frac{m_1m_2^3 s_1^2 s_2^2}{32 m_p^6}\int e^{-iq\cdot b}\hat\delta(q\cdot v_1)\hat\delta(q\cdot v_2)\,\Bigg[\boldsymbol{\tilde h}_{11}\,\mathbfcal{L}_{0,0;0,0,1,1,1}+\frac{1}{4}\boldsymbol{\tilde h}_{12}q^2\,\mathbfcal{L}^{(++)}_{1,1;0,0,1,1,1}\,\Bigg]\,.
    \end{split}
\end{align}
Similarly, the spinning part is given by, 
\begin{align}
    \begin{split} \label{eq6}
      i\chi\Big|_{\mathcal{S}^1}=-i\frac{m_1m_2^3 s_1^2 s_2^2}{32 m_p^6}\int e^{-iq\cdot b}\hat\delta(q\cdot v_1)\hat\delta(q\cdot v_2)\,\Bigg[ &i\,\Bigg(\boldsymbol{\tilde h}_{13} \,\mathbfcal{L}_{0,0;0,0,1,1,1}+ \frac{1}{4}\boldsymbol{\tilde h}_{14} \,q^2\,\mathbfcal{L}^{(++)}_{1,1;0,0,1,1,1}\Bigg)(q\cdot \mathcal{S}_1\cdot v_2)+\\ &i\,\Bigg(\boldsymbol{\tilde h}_{15}\,\mathbfcal{L}_{0,0;0,0,1,1,1}+ \frac{1}{4}\boldsymbol{\tilde h}_{16}\,q^2\,\mathbfcal{L}^{(++)}_{1,1;0,0,1,1,1}\Bigg)(q\cdot \mathcal{S}_2\cdot v_1)\Bigg]\,.
    \end{split}
\end{align}
Again all the $ \boldsymbol{\tilde h_i}'s\,, i=7,\cdots 16\,.$ are defined in Appendix~(\ref{AppE}). Note that, at $\mathcal{O}(\mathcal{S})$, we encounter tensor integrals. We first have to perform a Veltman-Passarino Reduction (PV) to write it in terms of a set of scalar integrals. Then, we do the IBP reduction for each of those scalar integrals. We will use this same strategy throughout this paper.  \par
Finally, the total contribution at $\mathcal{O}(\mathcal{S}^0)$ to the eikonal phase coming from these two diagrams after summing (\ref{eq3}) and (\ref{eq5}) is the following, 
\begin{align}
    \begin{split}  \label{eqt2}
   \chi\Big|_{\mathcal{S}^0} =\frac{m_1m_2^3}{m_p^6}\Delta_{\mathfrak{02}}^{\textrm{poly}}\,,
 \end{split}
\end{align}
where, 
\begin{align}
    \begin{split} 
\Delta_{\mathfrak{02}}^{\textrm{poly}}= &\Bigg(\frac{s_1^2s_2^2}{12288 \pi ^3 |\boldsymbol{b}|^2 x \left(x^2-1\right)^7}\Bigg)\Bigg[6 \left(-x^{12}+10 x^{10}+x^8+44 x^6+x^4+10 x^2-1\right) x^2 \log \left(\pi  |\boldsymbol{b}|^6\right)\\&\hspace{4.5cm}-9 x^{16}+x^{14} (-36 \gamma_E +56+\log (4096))+4 x^{12} (82 \gamma_E  -25-46 \log (2))\\&\hspace{4.5cm}-4 x^{10} (7 \gamma_E -6+35 \log (2))+2 x^8 (696 \gamma_E  -35-456 \log (2))\\&\hspace{4.5cm}-4 x^6 (7 \gamma_E  -6+35 \log (2))+4 x^4 (82 \gamma_E  -25-46 \log (2))\\&\hspace{4.5cm}+x^2 (-36 \gamma_E  +56+\log (4096))-9\Bigg]
     \end{split}
\end{align}
and the total contribution at $\mathcal{O}(\mathcal{S}^1)$ to the eikonal phase coming from these two diagrams after summing (\ref{eq4}) and (\ref{eq6}) is the following, 
\begin{align}
    \begin{split}   \label{eqt3}
     \chi\Big|_{\mathcal{S}^1}=\frac{m_1m_2^3}{m_p^6}\Bigg[\Delta_{\mathfrak{03}}^{\textrm{poly}}(\hat{b}\cdot \mathcal{S}_1\cdot v_2)+\Delta_{\mathfrak{04}}^{\textrm{poly}}(\hat{b}\cdot \mathcal{S}_2\cdot v_1)\Bigg]\,,
     \end{split}
\end{align}
where, 
\begin{align}
    \begin{split}
    \Delta_{\mathfrak{03}}^{\textrm{poly}}= \Bigg(\frac{s_1^2s_2^2\,x} {3072 \pi ^3 |\boldsymbol{b}|^3 \left(x^2-1\right)^5 }\Bigg)&\Bigg[-6 \left(x^8+6 x^7-2 x^6-26 x^5-6 x^4-26 x^3-2 x^2+6 x+1\right) \log (|\boldsymbol{b}|)\\&\hspace{0 cm}+5 x^8+x^8 \log (64)-x^8 \log (\pi )+x^7 \log (4096)-6 x^7 \log (\pi )-6 x^6\\ &+2 x^6 \log \left(\frac{\pi }{64}\right)-81 x^5+10 x^5 \log (4)-26 x^5 \log \left(\frac{16}{\pi }\right)+2 x^4\\ &-6 x^4 \log (64)+6 x^4 \log (\pi )-81 x^3+10 x^3 \log (4)-26 x^3 \log \left(\frac{16}{\pi }\right)-6 x^2\\ &+2 x^2 \log \left(\frac{\pi }{64}\right)-4 \gamma_E  \left(x^8+9 x^7-2 x^6-35 x^5-6 x^4-35 x^3-2 x^2+9 x+1\right)\\ &+x \log \left(\frac{4096}{\pi ^6}\right)+5+\log (64)-\log (\pi )\Bigg]\,,
        \end{split}
\end{align}
\begin{align}
    \begin{split}    \Delta_{\mathfrak{04}}^{\textrm{poly}}=\Bigg(\frac{s_1^2s_2^2\,x^2(1+x^2)} {768  \pi ^3 |\boldsymbol{b}|^3 \left(x^2-1\right)^7 }\Bigg)&\Bigg[12 \left(x^8+x^6-8 x^4+x^2+1\right) \log (|\boldsymbol{b}|)+13 x^8+x^8 \log \left(\frac{\pi ^2}{64}\right)-159 x^6\\ &-12 x^6 \log (2)+2 x^6 \log (\pi )+392 x^4+68 x^4 \log (2)-16 x^4 \log (\pi )\\ &-159 x^2-12 x^2 \log (2)+2 x^2 \log (\pi )+\gamma_{E}  \left(11 x^8+8 x^6-78 x^4+8 x^2+11\right)\\ &+13+\log \left(\frac{\pi ^2}{64}\right)\Bigg]\,.
    \end{split}
\end{align}
\\
\textbullet $\,\,$ We have another two comb-type diagrams with two graviton line 
\begin{align}
    \begin{split}
   \hspace{-4 cm}     \begin{minipage}[h]{0.12\linewidth}
	\vspace{4pt}
	\scalebox{4.3}{\includegraphics[width=\linewidth]{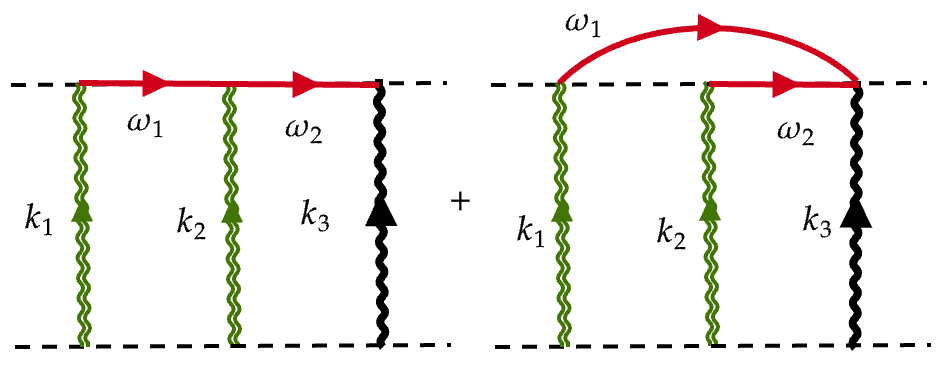}}
\end{minipage}
\end{split}
\end{align}


For the first diagram, we have, 

\begin{align}
    \begin{split} \label{eq8}
        i\chi\Big|_{\mathcal{S}^0} =i\frac{m_1 m_2^3 s_1 s_2}{64 m_p^6}\int e^{-iq\cdot b}\hat\delta(q\cdot v_1)\hat\delta(q\cdot v_2)\Bigg( \boldsymbol{\tilde h}_{17}\mathbfcal{L}_{0,0;0,0,1,1,1}-\frac{1}{2}\boldsymbol{\tilde h}_{18}q^2\,\mathbfcal{L}^{(+-)}_{1,1;0,0,1,1,1}\Bigg)\,,
    \end{split}
\end{align}
and 
\begin{align}
    \begin{split} \label{eq9}
        i\chi\Big|_{\mathcal{S}^1} =i\frac{m_1 m_2^3 s_1 s_2}{64 m_p^6}\int e^{-iq\cdot b}\hat\delta(q\cdot v_1)\hat\delta(q\cdot v_2)&\Bigg[i\Bigg( \boldsymbol{\tilde h}_{19}\mathbfcal{L}_{0,0;0,0,1,1,1}-\frac{1}{2}\boldsymbol{\tilde h}_{20}q^2\,\mathbfcal{L}^{(+-)}_{1,1;0,0,1,1,1}\Bigg)(q\cdot \mathcal{S}_1\cdot v_2)\\&i\,\Bigg( \boldsymbol{\tilde h}_{21}\mathbfcal{L}_{0,0;0,0,1,1,1}-\frac{1}{2}\boldsymbol{\tilde h}_{22}q^2\,\mathbfcal{L}^{(+-)}_{1,1;0,0,1,1,1}\Bigg)(q\cdot \mathcal{S}_2\cdot v_1)\Bigg]\,,
    \end{split}
\end{align}

We now compute the eikonal phase for the second diagram.
The spinless eikonal phase is given by,
\begin{align}
    \begin{split} \label{eq10}
        \chi\Big|_{\mathcal{S}^0} =-i\frac{m_1 m_2^3 s_1 s_2}{64 m_p^6} \int e^{-iq\cdot b}\hat\delta(q\cdot v_1)\,\hat\delta(q\cdot v_2)\,\Bigg[\boldsymbol{\tilde h}_{23} \,\mathbfcal{L}_{0,0;0,0,1,1,1}+\frac{1}{4}\,\boldsymbol{\tilde h}_{24}\,q^2\,\mathbfcal{L}^{(++)}_{1,1;0,0,1,1,1}\Bigg]
    \end{split}
\end{align}
The spinning eikonal phase is given by,
\begin{align}
    \begin{split} \label{eq11}
        i\chi\Big|_{\mathcal{S}^1} =-i\frac{m_1 m_2^3 s_1 s_2}{64 m_p^6}\int e^{-iq\cdot b}\hat\delta(q\cdot v_1)\hat\delta(q\cdot v_2)&\Bigg[i\Bigg( \boldsymbol{\tilde h}_{25}\mathbfcal{L}_{0,0;0,0,1,1,1}+\frac{1}{4}\boldsymbol{\tilde h}_{26}q^2\,\mathbfcal{L}^{(++)}_{1,1;0,0,1,1,1}\Bigg)(q\cdot \mathcal{S}_1\cdot v_2)\\&i\,\Bigg( \boldsymbol{\tilde h}_{27}\mathbfcal{L}_{0,0;0,0,1,1,1}+\frac{1}{4}\boldsymbol{\tilde h}_{28}q^2\,\mathbfcal{L}^{(++)}_{1,1;0,0,1,1,1}\Bigg)(q\cdot \mathcal{S}_2\cdot v_1)\Bigg]\,,
    \end{split}
\end{align}
Again, all the coefficients $\boldsymbol{\tilde h}'s$ are defined in Appendix ~(\ref{AppE}).
Finally, the total contribution at $\mathcal{O}(\mathcal{S}^0)$ to the eikonal phase coming from these two diagrams after summing (\ref{eq8}) and (\ref{eq10}) is the following, 
\begin{align}
    \begin{split}  \label{eqt4}
     \chi\Big|_{\mathcal{S}^0}=\frac{m_1m_2^3}{m_p^6}\Delta_{\mathfrak{05}}^{\textrm{poly}}
     \end{split}
\end{align}
where,
\begin{align}
    \begin{split}
\Delta_{\mathfrak{05}}^{\textrm{poly}}=\frac{s_1s_2}{24576 \pi ^3 |\boldsymbol{b}|^2 x \left(x^2-1\right)^5}\Bigg[&24 \left(x^8+x^4\right) \log \left(\pi  |\boldsymbol{b}|^6\right)+x^2 \Big(-5 x^6 (37+20 \log (2))-4 x^4 (\log (4)-47)\\ &-5 x^2 (37+20 \log (2))+x^8 \left(x^2 (\log (16)-43)+134+\log (16)\right)\\ &+134+\log (16)\Big)+2 \gamma_{E}  \left(x^{12}+x^{10}+59 x^8-2 x^6+59 x^4+x^2+1\right)\\ &-43+\log (16)\Bigg]\,.
        \end{split}
\end{align}
The total contribution at $\mathcal{O}(\mathcal{S}^1)$ to the eikonal phase coming from these two diagrams after summing (\ref{eq9}) and (\ref{eq11}) is the following,
\begin{align}
    \begin{split}  \label{eqt5}
     \chi\Big|_{\mathcal{S}^1}=\frac{m_1m_2^3}{m_p^6}\Bigg[\Delta_{\mathfrak{06}}^{\textrm{poly}}(\hat{b}\cdot \mathcal{S}_1\cdot v_2)+\Delta_{\mathfrak{07}}^{\textrm{poly}}(\hat{b}\cdot \mathcal{S}_2\cdot v_1)\Bigg]\,,
     \end{split}
\end{align}
where, 
\begin{align}
    \begin{split}   &\Delta_{\mathfrak{06}}^{\textrm{poly}}=\frac{s_1s_2}{147456 \pi ^3 |\boldsymbol{b}|^3  \left(x^2-1\right)^7}\Bigg[3\alpha_0-3\alpha_1(x^2+x^{12})+3\alpha_2x^{14}-3\alpha_3(x^6+x^{8})+3\alpha_4(x^4+x^{10})\\&\hspace{5.2cm}-13824 x^{11}+41472 x^9-55296 x^7+41472 x^5-13824 x^3\Bigg]\,,\\
& \Delta_{\mathfrak{07}}^{\textrm{poly}}=-\frac{s_1 s_2}{24576 \pi ^3 |\boldsymbol{b}|^3 \left(x^2-1\right)^7 }\Bigg[\hat{\alpha}_0+8\hat{\alpha}_1(x^5+x^{9})+\hat{\alpha}_0x^{14}+\alpha_2(x^2+x^{12})+\hat{\alpha}_3(x^4+x^{10})\\&\hspace{5.2cm}+\hat{\alpha}_4(x^6+x^{8})+8\hat{\alpha}_5(x^3+x^{11})-360 x^{13}-360 x\\&\hspace{5.2cm}-16 x^7 (18 \log (|\boldsymbol{b}|)+14 \gamma_E -161-14 \log (2)+3 \log (\pi ))\Bigg]
\end{split}
\end{align}
with, 
\begin{align}
    \begin{split} 
   & \alpha_0= 8 \log \left(\frac{\pi  |\boldsymbol{b}|^6}{16}\right)+72 \log ^2(b)+24 \left(5 \gamma_E -3+\log \left(\frac{\pi }{16}\right)\right) \log (|\boldsymbol{b}|)+50 \gamma_E ^2+\pi ^2-19 \gamma_E +89+\\&\hspace{1cm} 2 \log ^2(\pi )+48 \log (2)+16 \log (2) \log (4)+\log (4)+20 \gamma_E  \log \left(\frac{\pi }{16}\right)-16 \log (2) \log (\pi )-12 \log (\pi )\,,\\&
   \alpha_1=24 \log (|\boldsymbol{b}|) (3 \log (b)+5 \gamma_E +24)+4 \log \left(\frac{\pi }{16}\right) (6 \log (b)+5 \gamma_E )+\pi ^2+\gamma_E  (473+50 \gamma )+573+32 \log ^2(2)\\&\hspace{1cm}-2 (199+8 \log (\pi )) \log (2)+2 \log (\pi ) (48+\log (\pi ))\,,\\
    &\alpha_2=24 \left(\log \left(\frac{\pi  |\boldsymbol{b}|^3}{16}\right)+5 \gamma_E -1\right) \log (|\boldsymbol{b}|)+50 \gamma_E ^2+\pi ^2+89+\log (4) \left(9+8 \log \left(\frac{4}{\pi }\right)\right)+2 (\log (\pi )-2) \log (\pi )\\&\hspace{1cm}+\gamma_E  (-19-80 \log (2)+20 \log (\pi ))\,,\nonumber
     \end{split}
\end{align}
 \begin{align}
    \begin{split} 
& \alpha_3=-504 \log \left(\frac{\pi  |\boldsymbol{b}|^6}{16}\right)+24 \log (|\boldsymbol{b}|) (3 \log (b)+5 \gamma_E -3)+4 \log \left(\frac{\pi }{16}\right) (6 \log (|\boldsymbol{b}|)+5 \gamma_E)+\gamma_E (50 \gamma_E -2583)+\pi ^2\\&\hspace{1cm}+2245+(21+16 \log (2)) \log (4)+2 \log (\pi ) (-6-8 \log (2)+\log (\pi ))\,,\\&
\alpha_4=-20 \log \left(\frac{\pi  |\boldsymbol{b}|^6}{16}\right)+36 \left(6 \log (|\boldsymbol{b}|) (3 \log (|\boldsymbol{b}|)+5 \gamma_E -3)+\log \left(\frac{\pi }{16}\right) (6 \log (|\boldsymbol{b}|)+5 \gamma_E )\right)+450 \gamma_E ^2+9 \pi ^2-651 \gamma_E \\&\hspace{1cm}+2697+(205+144 \log (2)) \log (4)+18 \log (\pi ) (-6-8 \log (2)+\log (\pi ))\,,\\&\hat{\alpha}_0=11 \log \left(\frac{\pi  |\boldsymbol{b}|^6}{16}\right)+65 \gamma_E +54+10 \log (4)\,,\\&\hat{\alpha}_1=72 \log ^2(|\boldsymbol{b}|)+12 (10 \gamma_E-7-8 \log (2)+2 \log (\pi )) \log (|\boldsymbol{b}|)+50 \gamma_E ^2+\pi ^2-271+4 \log (2) (13+\log (256))+\\&\hspace{1cm}2 \left(\log \left(\frac{\pi }{256}\right)-7\right) \log (\pi )-4 \gamma_E  (18+20 \log (2)-5 \log (\pi))\,,\\&\hat{\alpha}_2=32 \log \left(\frac{\pi }{16}\right) (6 \log (|\boldsymbol{b}|)+5 \gamma_E)+6 \log (|\boldsymbol{b}|) (96 \log (|\boldsymbol{b}|)+160 \gamma_E -85)+400 \gamma_E ^2+8 \pi ^2-435 \gamma_E +684+\\&\hspace{1cm}16 \left(\log ^2(16)+\log ^2(\pi )+\log (2) (20-8 \log (\pi ))\right)-85 \log (\pi )\,,\\&\hat{\alpha}_3=-275 \log \left(\frac{\pi  |\boldsymbol{b}|^6}{16}\right)-64 \left(6 \log (|\boldsymbol{b}|) (3 \log (|\boldsymbol{b}|)+5 \gamma_E -3)+\log \left(\frac{\pi }{16}\right) (6 \log (|\boldsymbol{b}|)+5 \gamma_E )\right)-\\&\hspace{1cm}\gamma_E  (429+800 \gamma_E )-4 \left(4 \pi ^2+398+\log (2) (199+128 \log (2))+8 \log (\pi ) (-6-8 \log (2)+\log (\pi ))\right)\,,\\&\hat{\alpha}_4=-131 \log \left(\frac{\pi |\boldsymbol{b}|^6}{16}\right)-160 \left(6 \log (|\boldsymbol{b}|) (3 \log (|\boldsymbol{b}|)+5 \gamma_E -3)+\log \left(\frac{\pi }{16}\right) (6 \log (|\boldsymbol{b}|)+5 \gamma_E )\right)+\\&\hspace{1cm}\gamma_E  (1759-2000 \gamma_E)-40 \pi ^2+1814-4 \log (2) (473+320 \log (2))-80 \log (\pi ) (-6-8 \log (2)+\log (\pi ))\,,\\&
\hat{\alpha}_5=\log \left(\frac{\pi |\boldsymbol{b}|^6}{4}\right)+6 \gamma_E +163\,.
    \end{split}
\end{align}
\textbullet $\,\,$ Next we compute the following two diagrams. 
\begin{align}
    \begin{split}
  \hspace{-4 cm}      \begin{minipage}[h]{0.12\linewidth}
	\vspace{4pt}
	\scalebox{4.2}{\includegraphics[width=\linewidth]{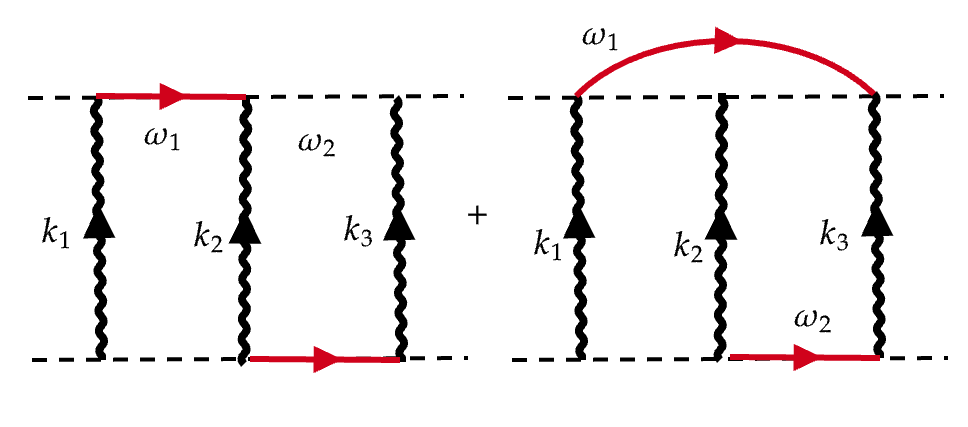}}
\end{minipage}\hspace{1 cm}&
    \end{split}
\end{align}
The amplitude for the first diagram is given by,
\begin{align}
    \begin{split}
        i\chi&=  i\,\frac{s_1^3 s_2^3 m_1^2 m_2^2} {128 m_p^6}\int_q e^{-iq\cdot b}\hat\delta(q\cdot v_1)\hat\delta(q\cdot v_2)\int_{\ell_{1},\ell_{2}}\left(\sum_{(\pm)}\frac{1}{(\pm\omega_1+i\varepsilon)^2(\pm\omega_2+i\varepsilon)^2}\right)\frac{\hat\delta(\ell_{1}\cdot v_2)\hat\delta(\ell_{2}\cdot v_1)}{\ell_{1}^2 \ell_{2}^2(q-\ell_{1}-\ell_{2})^2}\\&\hspace{5cm}\times\Big((q-\ell_{1}-\ell_{2})\cdot \ell_{1}\Big)\Big( (q-\ell_{1}-\ell_{2})\cdot \ell_{2}\Big)\Bigg|_{\omega_1\to \ell_1\cdot v_1,\omega_2\to\ell_2\cdot v_2}\,.\\ &
    \end{split}
\end{align}
After IBP reduction, we are left with the following integral,
\begin{align}
    \begin{split}
       i\, \chi\Big|_{\mathcal{S}^0} &=
         i\,\frac{s_1^3 s_2^3 m_1^2 m_2^2} {128 m_p^6}\int_q e^{-iq\cdot b}\hat\delta(q\cdot v_1)\hat\delta(q\cdot v_2)\Big[\boldsymbol{\tilde h}_{29}\mathbfcal{M}_{0,0;0,0,1,1,1}+\boldsymbol{\tilde h}_{30}\,q^2\,\mathbfcal{M}_{0,0;0,0,2,1,1}  +\boldsymbol{\tilde h}_{31}\,q^2\,\mathbfcal{M}_{0,0;0,0,1,2,1}\\ &\hspace{7 CM}-\frac{1}{2}\boldsymbol{\tilde h}_{32} q^2\,\mathbfcal{M}^{(++)}_{1,1;0,0,1,1,1} \Big]\,.
    \end{split}
\end{align}
The contribution to the phase from the second diagram has the following form,
\begin{align}
    \begin{split}
       i \chi\Big|_{\mathcal{S}^0} =i\frac{s_1^3 s_2^3 m_1^2 m_2^2} {128 m_p^6} \int_{k_i}e^{-i(k_1+k_2+k_3)\cdot b}\hat\delta(k_1\cdot v_2)\hat\delta(k_2\cdot v_1)\hat\delta(k_1\cdot v_1+k_3\cdot v_1)\hat\delta(k_2\cdot v_2+k_3\cdot v_2)\\ &\hspace{6 cm}        \times\frac{\mathcal{N}}{(k_1\cdot v_1+i\varepsilon)^2(k_3\cdot v_2+i\varepsilon)^2\,k_1^2 \,k_2^2\,k_3^2} 
    \end{split}
\end{align}
Now, doing the following variable transformation, $k_1\to \ell_1-q,\,k_2\to (\ell_2-q)\,,k_3\to (q-\ell_1-\ell_2)$, we will get,
\begin{align}
    \begin{split}
   \hspace{-0.5 cm}   i\,  \chi \Big|_{\mathcal{S}^0}=i\frac{s_1^3 s_2^3 m_1^2 m_2^2} {128 m_p^6}  \int e^{i q\cdot b}\hat\delta(q\cdot v_1)\hat\delta(q\cdot v_2)\int_{\ell_1,\ell_2}\sum_{\mp}\frac{\hat\delta(\ell_1\cdot v_2)\hat\delta(\ell_2\cdot v_1)\mathcal{N}}{(\pm\ell_1\cdot v_1+i\varepsilon)^2(\mp\ell_2\cdot v_2+i\varepsilon)^2\,(\ell_1-q)^2(\ell_2-q)^2(\ell_1+\ell_2-q)^2},
    \end{split}
\end{align}
where, the numerator $\mathcal{N}$ takes the form (ignoring the tadpoles),
\begin{align}
    \begin{split}
        \mathcal{N}=\frac{1}{8}\ell_1^2\,\ell_2^2-\frac{1}{16}(\ell_1^4+\ell_2^4)+\frac{q^4}{16}\,.
    \end{split}
\end{align}
Therefore, the contribution to the eikonal phase from the second diagram is given by,
\begin{align}
    \begin{split}
       i\, \chi\Big|_{\mathcal{S}^0}&=i\frac{s_1^3 s_2^3 m_1^2 m_2^2} {128 m_p^6}  \int e^{iq\cdot b}\hat\delta(q\cdot v_1)\hat\delta(q\cdot v_2)\Bigg[\boldsymbol{\tilde h}_{33}\,\mathbfcal{M}_{0,0;0,0,1,1,1}+\boldsymbol{\tilde h}_{34}\,q^2 \mathbfcal{M}_{0,0;0,0,2,1,1}+\boldsymbol{\tilde h}_{35}\,q^2 \mathbfcal{M}_{0,0;0,0,1,2,1}\\ &\hspace{6 cm}+\frac{1}{4}\boldsymbol{\tilde h}_{36}\,q^2 \mathbfcal{M}^{(+-)}_{1,1;0,0,1,1,1}\Bigg]\,.
    \end{split}
\end{align}
We finally get the total contribution to the eikonal phase from these two diagrams at $\mathcal{O}(\mathcal{S}^0)$, 
\begin{align}
\begin{split} \label{eqt6}
 \chi\Big|_{\mathcal{S}^0} =\frac{m_1^2m_2^2}{m_p^6}\Bigg(\Delta_{\mathfrak{08}}^{\textrm{poly}}+\Delta_{\mathfrak{01}}^{\textrm{log}}\log (x)\Bigg)
\end{split}
\end{align}
where,
\begin{align}
\begin{split}
&\Delta_{\mathfrak{08}}^{\textrm{poly}}= -\frac{ s_1^3 s_2^3 x^4 \left(x^2+1\right) (258 \log (|\boldsymbol{b}|)+215 \gamma_E -12-172 \log (2)+43 \log (\pi ))}{6144 \pi ^3 |\boldsymbol{b}|^2 \left(x^2-1\right)^5 }\,,\\&
\Delta_{\mathfrak{01}}^{\textrm{log}}=-\frac{ s_1^3 s_2^3 x^4}{512 \pi ^3 |\boldsymbol{b}|^2 \left(x^2-1\right)^4 }\,.
\end{split}
\end{align}
We have used the results for the master integrals mentioned in \eqref{MasterM}. 
\\\\
\textbullet $\,\,$ Now, we move on to calculating  Feynman topology with two scalar lines and one graviton line, which appear at 3PM. In these diagrams, the spin of the black holes appears because of the presence of graviton. 
\begin{align}
    \begin{split}
    \hspace{-4 cm}    \begin{minipage}[h]{0.12\linewidth}
	\vspace{4pt}
	\scalebox{4.2}{\includegraphics[width=\linewidth]{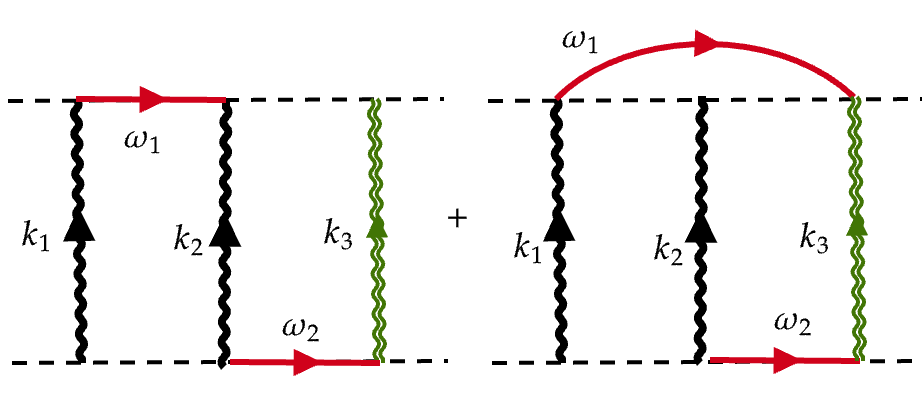}}
\end{minipage}\hspace{1 cm}
    \end{split}
\end{align}
The corresponding phase is given by,
\begin{align}
    \begin{split}
        i\chi&=i\,\frac{s_1^2 s_2^2 m_1^2 m_2^2} {128 m_p^6} \int_{k_i,\omega_i}e^{-i(k_1+k_2+k_3)\cdot b}\hat\delta(k_1\cdot v_2)\hat\delta(k_3\cdot v_1)\hat\delta(\omega_1-k_1\cdot v_1)\hat\delta(\omega_1+k_2\cdot v_1)\\ &
\hspace{2 cm}\times \hat\delta(\omega_2+k_2\cdot v_2)\hat\delta(\omega_2-k_3\cdot v_2)\,\frac{\mathcal{N}}{k_1^2 \,k_2^2\, k_3^2}\left(\sum_{(\pm)}\frac{1}{(\pm\omega_1+i\varepsilon)^2(\pm\omega_2+i\varepsilon)^2}\right)
    \end{split}
\end{align}
Now redefining $k_1\to \ell_{1}, k_3\to \ell_{2}, q=\ell_{1}+k_2+\ell_{2}$, we can write the numerator up to linear order in spin in the following way ($\mathcal{N}= \mathcal{N}|_{\mathcal{S}^0}+ \mathcal{N}|_{\mathcal{S}^1}$).
\begin{align}
    \begin{split}
    &   \mathcal{N}\Big|_{\mathcal{S}^0}= -\frac{\gamma^2(d-2)-2}{d-2}(\ell_2-q)^2(\ell_2\cdot v_2)^2\,,\\ &
     \mathcal{N}\Big|_{\mathcal{S}^1}=i\gamma \Big(\ell_2\cdot \mathcal{S}_1\cdot v_2\Big)(\ell_2\cdot v_2)^2(\ell_2-q)^2\,.
    \end{split}
\end{align}
The spinless part of the eikonal phase is given by,
\begin{align}
    \begin{split}
        i\chi\Big|_{\mathcal{S}^0}&
        =-i\,\frac{s_1^2 s_2^2 m_1^2 m_2^2} {128 m_p^6} \frac{\gamma^2(d-2)-2}{d-2}\int e^{-iq\cdot b}\hat\delta(q\cdot v_1)\hat\delta(q\cdot v_2)\Bigg(\boldsymbol{\tilde h}_{37}\mathbfcal{M}_{0,0;0,0,1,1,1}+\boldsymbol{\tilde h}_{38}\,q^2\mathbfcal{M}_{0,0;0,0,2,1,1}\\ &\hspace{11 cm}+\boldsymbol{\tilde h}_{39}\,q^2\mathbfcal{M}_{0,0;0,0,1,2,1}\Bigg)\,.
    \end{split}
\end{align}
At $\mathcal{O}(\mathcal{S})$ we get, 
\begin{align}
    \begin{split}
      i\chi\Big|_{\mathcal{S}^1}  = i\,\frac{s_1^2 s_2^2 m_1^2 m_2^2} {128 m_p^6} \int_{q}e^{-i q\cdot b}\hat\delta(q\cdot v_1)\,\hat\delta(q\cdot v_2)
      &\Bigg[i\gamma\Bigg(\boldsymbol{\tilde h}_{40}\mathbfcal{M}_{0,0;0,0,1,1,1}+\boldsymbol{\tilde h}_{41} q^2\,\mathbfcal{M}_{0,0;0,0,2,1,1}\\&+\boldsymbol{\tilde h}_{42}q^2\,\mathbfcal{M}_{0,0;0,0,    1,2,1}\Bigg)\Bigg]\Big(q\cdot \mathcal{S}_1\cdot v_2\Big)\,,\\&
    \end{split}
\end{align}\\
The contribution from the second diagram to the eikonal phase is given by,
\begin{align}
    \begin{split}
i\chi\Big|_{\mathcal{S}^0}=-\frac{i m_1^2 m_2^2 s_1^2 s_2^2}{64 m_p^6}\,\int e^{-i q \cdot b}\, \hat\delta(q\cdot v_1)\hat\delta(q\cdot v_2)&\Big(\boldsymbol{\tilde h}_{43}\,\,\mathbfcal{M}_{0,0;0,0,1,1,1}+\boldsymbol{\tilde h}_{44}\,\,q^2\mathbfcal{M}_{0,0;0,0,2,1,1}\\&+\boldsymbol{\tilde h}_{45}\,\,q^2\mathbfcal{M}_{0,0;0,0,1,2,1}+\frac{1}{4}\boldsymbol{\tilde h}_{46}\,\,q^2\mathbfcal{M}^{(+-)}_{1,1;0,0,1,1,1}\Big)\,,
   \end{split}
\end{align}
Similarly,
\begin{align}
    \begin{split}
       i\, \chi\Big|_{\mathcal{S}^1}=-i \frac{s_1^2 s_2^2 m_1^2 m_2^2}{64 m_p^6} \int_{q}e^{i q\cdot b}\hat\delta(q\cdot v_1)\,\hat\delta(q\cdot v_2)&\Bigg[i\,\Bigg(\boldsymbol{\tilde h}_{47}\,\,\mathbfcal{M}_{0,0;0,0,1,1,1}+\boldsymbol{\tilde h}_{48}\,\,q^2\mathbfcal{M}_{0,0;0,0,2,1,1}\\&+\boldsymbol{\tilde h}_{49}\,\,q^2\mathbfcal{M}_{0,0;0,0,1,2,1}+\frac{1}{4}\boldsymbol{\tilde h}_{50}\,\,q^2\mathbfcal{M}^{(+-)}_{1,1;0,0,1,1,1}\Bigg)\big(q\cdot\mathcal{S}_1\cdot v_2\big)+\\&i\,\Bigg(\boldsymbol{\tilde h}_{51}\,\,\mathbfcal{M}_{0,0;0,0,1,1,1}+\boldsymbol{\tilde h}_{52}\,\,q^2\mathbfcal{M}_{0,0;0,0,2,1,1}\\&+\boldsymbol{\tilde h}_{53}\,\,q^2\mathbfcal{M}_{0,0;0,0,1,2,1}+\frac{1}{4}\boldsymbol{\tilde h}_{54}\,\,q^2\mathbfcal{M}^{(+-)}_{1,1;0,0,1,1,1}\Bigg)\big(q\cdot\mathcal{S}_2\cdot v_1\big)\Bigg]
    \end{split}
\end{align}
We finally get the total contribution to the eikonal phase from these two diagrams at $\mathcal{O}(\mathcal{S}^0)$, 
\begin{align}
\begin{split} \label{eqt7}
 \chi\Big|_{\mathcal{S}^0} =\frac{m_1^2m_2^2}{m_p^6}\Bigg(\Delta_{\mathfrak{09}}^{\textrm{poly}}+\Delta_{\mathfrak{02}}^{\textrm{log}}\log (x)\Bigg)
\end{split}
\end{align}
where, 
\begin{align}
\begin{split}
&\Delta_{\mathfrak{09}}^{\textrm{poly}}=-\frac{s_1^2 s_2^2}{4096 \pi ^3 |\boldsymbol{b}|^2 \left(x^2-1\right)^5}\Bigg(\left(x^2+1\right) \big(-12 \left(11 x^4-6 x^2+11\right) x^2 \log (|\boldsymbol{b}|)+5 x^8-22 x^6 \left(5 \gamma_E +2+\log \left(\frac{\pi }{16}\right)\right)\\&\hspace{5cm}+6 x^4 (10 \gamma_E +9-8 \log (2)+2 \log (\pi ))-22 x^2 \left(5 \gamma_E +2+\log \left(\frac{\pi }{16}\right)\right)+5\big)\Bigg)\,,\\&
\Delta_{\mathfrak{02}}^{\textrm{log}}=-\frac{s_1^2 s_2^2 \left(x^4+6 x^2+1\right)}{512 \pi ^3 |\boldsymbol{b}|^2 \left(x^2-1\right)^2}\,.
\end{split}
\end{align}
Also the total contribution at $\mathcal{O}(\mathcal{S}^1)$ is, 
\begin{align}
\begin{split}
\chi\Big|_{\mathcal{S}^1} =\frac{m_1^2m_2^2}{m_p^6}&\Bigg[\Big(\Delta_{\mathfrak{10}}^{\textrm{poly}}+\Delta_{\mathfrak{03}}^{\textrm{log}}\log (x)+\Delta_{\mathfrak{01}}^{\textrm{Polylog}}\left(\textbf{Li}_2(1-x^2)+\log^2(x)\right)+\boldsymbol{\mathfrak{U}}_{\mathfrak{1}}^{\textrm{LI}_3}\Big\{54\, \textbf{Li}_3\left(x^2\right)-216\, \textbf{Li}_3\left(\frac{x+1}{1-x}\right)
\\ &+432\, \textbf{Li}_3(1-x)+216\, \textbf{Li}_3\left(\frac{x+1}{x-1}\right)+432\, \textbf{Li}_3(x+1)+216 \Big(-\textbf{Li}_2\left(\frac{x+1}{1-x}\right)-\textbf{Li}_2(1-x)\\ &+\textbf{Li}_2\left(\frac{x+1}{x-1}\right)+\textbf{Li}_2(x+1)\Big) \log \left(\frac{2}{x+1}-1\right)-54 \left(\pi ^2-2 \log ^2(x)\right) \log \left(1-x^2\right)-20 \log ^3(x)\\ &+108 i \pi  \log ^2(1-x)+108 i \pi  \log ^2(x+1)-11 \pi ^2 \log (x)-432 \zeta (3)\Big\}\Big)\Big(\hat{b}\cdot \mathcal{S}_1\cdot v_2\Big)+\\&\Big(\Delta_{\mathfrak{11}}^{\textrm{poly}}+\Delta_{\mathfrak{04}}^{\textrm{log}}\log (x)+\Delta_{\mathfrak{02}}^{\textrm{Polylog}}\left(\textbf{Li}_2(1-x^2)+\log^2(x)\right)+\boldsymbol{\mathfrak{U}}_{\mathfrak{2}}^{\textrm{LI}_3}\Big\{54\, \textbf{Li}_3\left(x^2\right)-216\, \textbf{Li}_3\left(\frac{x+1}{1-x}\right)
\\ &+432\, \textbf{Li}_3(1-x)+216\, \textbf{Li}_3\left(\frac{x+1}{x-1}\right)+432\, \textbf{Li}_3(x+1)+216 \Big(-\textbf{Li}_2\left(\frac{x+1}{1-x}\right)-\textbf{Li}_2(1-x)\\ &+\textbf{Li}_2\left(\frac{x+1}{x-1}\right)+\textbf{Li}_2(x+1)\Big) \log \left(\frac{2}{x+1}-1\right)-54 \left(\pi ^2-2 \log ^2(x)\right) \log \left(1-x^2\right)-20 \log ^3(x)\\ &+108 i \pi  \log ^2(1-x)+108 i \pi  \log ^2(x+1)-11 \pi ^2 \log (x)-432 \zeta (3)\Big\}\Big)\Big)\Big(\hat{b}\cdot \mathcal{S}_2\cdot v_1\Big)\Bigg]
\end{split}
\end{align}
where,
\begin{align}
\begin{split}
\Delta_{\mathfrak{10}}^{\textrm{poly}}=\frac{xs_1^2 s_2^2}{12288 \pi ^3 |\boldsymbol{b}|^3 \left(x^2-1\right)^5 }\Bigg[&-12 \left(12 x^6+77 x^5+72 x^4+154 x^3+72 x^2+77 x+12\right) x \log(|\boldsymbol{b}|)\\ &-13 x^8-120 \gamma_E  x^7+60 x^7+24 x^7 \log \left(\frac{16}{\pi }\right)-770 \gamma_E  x^6+398 x^6+154 x^6 \log \left(\frac{16}{\pi }\right)\\ &-720 \gamma_E  x^5+372 x^5+144 x^5 \log \left(\frac{16}{\pi }\right)-1540 \gamma_E  x^4+630 x^4+308 x^4 \log \left(\frac{16}{\pi }\right)\\ &-720 \gamma_E  x^3+372 x^3+144 x^3 \log \left(\frac{16}{\pi }\right)-770 \gamma_E  x^2+398 x^2+154 x^2 \log \left(\frac{16}{\pi }\right)\\ &-120 \gamma_E  x+60 x+24 x \log \left(\frac{16}{\pi }\right)-13\Bigg]\,,
\\ &\hspace{-5 cm}
\Delta_{\mathfrak{3}}^{\textrm{log}}=\frac{1}{18432 \pi ^3 |\boldsymbol{b}|^3 \left(x^2-1\right)^6 \left(x^2+1\right) }\Bigg[-48(1+x^{12})+(-24 x^8-24 x^4)\beta_1+2(x^5+x^7)\beta_2+3(x^3+x^9)\beta_3+\\ & \hspace{1cm}(x+x^{11})\beta_4-24x^6\beta_5+6(x^2+x^{10})\beta_6\Bigg]\,,\nonumber
 \end{split}
\end{align}
\begin{align}
\begin{split}
&\Delta_{\mathfrak{01}}^{\textrm{Polylog}}=\frac{x^2s_1^2 s_2^2}{6144 \pi ^3 |\boldsymbol{b}|^3 \left(x^2-1\right)^6 \left(x^2+1\right) }\Bigg[72 \left(x^{10}+24 x^9+3 x^8-10 x^6-48 x^5-10 x^4+3 x^2+24 x+1\right) \log(|\boldsymbol{b}|)\\ &\hspace{6.5cm}+9 x^{10}-12 x^{10} \log \left(\frac{16}{\pi }\right)-1236 x^9-288 x^9 \log \left(\frac{16}{\pi }\right)+9 x^8\\ &\hspace{6.5cm}-36 x^8 \log \left(\frac{16}{\pi }\right)-392 x^7+222 x^6+120 x^6 \log \left(\frac{16}{\pi }\right)+3224 x^5\\ &\hspace{6.5cm}+576 x^5 \log \left(\frac{16}{\pi }\right)+222 x^4+120 x^4 \log \left(\frac{16}{\pi }\right)-392 x^3+9 x^2\\ &\hspace{6.5cm}-36 x^2 \log \left(\frac{16}{\pi }\right)+60 \gamma_E  (x^{10}+24 x^9+3 x^8-10 x^6-48 x^5-10 x^4\\ &\hspace{6.5cm}+3 x^2+24 x+1)-1236 x-288 x \log \left(\frac{16}{\pi }\right)+9-12 \log \left(\frac{16}{\pi }\right)\Bigg]\,,\\
&\boldsymbol{\mathfrak{U}_1}^{\textrm{LI}_3}=\frac{ \pi ^2 s_1^2 s_2^2 x}{9216 \pi ^5 |\boldsymbol{b}|^3 \left(x^2-1\right)^6}\Bigg[x^8+24 x^7+2 x^6-24 x^5-12 x^4-24 x^3+2 x^2+24 x+1\Bigg],
\\ &
\Delta_{\mathfrak{04}}^{\textrm{log}}=\frac{1}{36864 \pi ^3 |\boldsymbol{b}|^3 (x^2-1)^6 \left(x^2+1\right) }\Bigg[96(1+x^{12})+48(x^3+x^{9})\delta_1+3(x^2+x^{10})\delta_2+48(x^5+x^7)\delta_3\\ &\hspace{6.5cm}+4(x^4+x^8)\delta_4+x^6\delta_5\Bigg]\,,\\&
\Delta_{\mathfrak{11}}^{\textrm{poly}}= \frac{x^2 s_1^2 s_2^2}{6144 \pi ^3 |\boldsymbol{b}|^3 \left(x^2-1\right)^5 }\Bigg[12 (42 x^4-36 x^3+89 x^2-36 x+42) x \log(\boldsymbol{b})-12 x^6+420 \gamma_E  x^5-323 x^5\\ &\hspace{5 cm}-84 x^5 \log \left(\frac{16}{\pi }\right)-360 \gamma_E  x^4+108 x^4+72 x^4 \log \left(\frac{16}{\pi }\right)+890 \gamma_E  x^3-558 x^3\\ &\hspace{
       5 cm}-178 x^3 \log \left(\frac{16}{\pi }\right)-360 \gamma_E  x^2+108 x^2+72 x^2 \log \left(\frac{16}{\pi }\right)+420 \gamma_E  x-323 x\\ &\hspace{
       5 cm}-84 x \log \left(\frac{16}{\pi }\right)-12\Bigg]\,,\\ &    \Delta_{\mathfrak{02}}^{\textrm{Polylog}}=\frac{x^3 s_1^2 s_2^2}{768 \pi ^3 |\boldsymbol{b}|^3 (x^2-1)^6 \left(x^2+1\right) }\Bigg[18 \left(12 x^8-7 x^6-12 x^5-38 x^4-12 x^3-7 x^2+12\right) \log (|\boldsymbol{b}|)-150 x^8\\ &\hspace{5 cm}-36 x^8 \log \left(\frac{16}{\pi }\right)-15 x^7+30 x^6+21 x^6 \log \left(\frac{16}{\pi }\right)+39 x^5+36 x^5 \log \left(\frac{16}{\pi }\right)\\& \hspace{5 cm}+554 x^4+114 x^4 \log \left(\frac{16}{\pi }\right)+39 x^3+36 x^3 \log \left(\frac{16}{\pi }\right)+30 x^2+21 x^2 \log \left(\frac{16}{\pi }\right)\\ &\hspace{4.7 cm}+15 \gamma_E  \left(12 x^8-7 x^6-12 x^5-38 x^4-12 x^3-7 x^2+12\right)-15 x-150-36 \log \left(\frac{16}{\pi }\right)\Bigg]\,,\\ &   \boldsymbol{\mathfrak{U}}_{\mathfrak{2}}^{\textrm{LI}_3}=\frac{s_1^2 s_2^2 x^2}{4608 \pi ^3 |\boldsymbol{b}|^3 \left(x^2-1\right)^6}\Bigg[12 x^6-19 x^4-12 x^3-19 x^2+12\Bigg]
    \end{split}
\end{align}
with,
\begin{align}
    \begin{split}
       & \beta_1=90 \log (|\boldsymbol{b}|)+75 \gamma_E -26-15 \log \left(\frac{16}{\pi }\right),\\ &
       \beta_2=-1080 \log ^2(|\boldsymbol{b}|)-1800 \gamma_E  \log (|\boldsymbol{b}|)+666 \log (|\boldsymbol{b}|)+3 \log \left(\frac{\pi }{16}\right) \left(-120 \log (|\boldsymbol{b}|)+37+10 \log \left(\frac{16}{\pi }\right)\right)\\ &\hspace{1 cm}-750 \gamma_E ^2-25 \pi ^2+555 \gamma_E +216+300 \gamma_E  \log \left(\frac{16}{\pi }\right),\\ &
       \beta_3=216 \log ^2(|\boldsymbol{b}|)+18 (20 \gamma_E +1-16 \log (2)+4 \log (\pi )) \log (|\boldsymbol{b}|)+5 \pi ^2+176+15 \gamma_E  (10 \gamma_E +1-16 \log (2)\\ &\hspace{1 cm}+4 \log (\pi ))+3 \log \left(\frac{16}{\pi }\right) \left(\log \left(\frac{256}{\pi ^2}\right)-1\right)\,,\\ &
       \beta_4=216 \log ^2(|\boldsymbol{b}|)+18 (20 \gamma_E +3-16 \log (2)+4 \log (\pi )) \log (|\boldsymbol{b}|)+5 \pi ^2+144+15 \gamma_E  (10 \gamma_E +3-16 \log (2)\\ &\hspace{1 cm}+4 \log (\pi ))+3 \log \left(\frac{16}{\pi }\right) \left(\log \left(\frac{256}{\pi ^2}\right)-3\right)\,,\\ &
       \beta_5=432 \log ^2(|\boldsymbol{b}|)+6 (120 \gamma_E -137-96 \log (2)+24 \log (\pi )) \log (|\boldsymbol{b}|)+10 \pi ^2+151+12 \log ^2\left(\frac{\pi }{16}\right)+548 \log (2)\\ &\hspace{1 cm}-137 \log (\pi )+5 \gamma_E  (60 \gamma_E -137-96 \log (2)+24 \log (\pi ))\,,\\ &
       \beta_6=2 \Big(432 \log ^2(|\boldsymbol{b}|)+6 (120 \gamma_E -103-96 \log (2)+24 \log (\pi )) \log (|\boldsymbol{b}|)+10 \pi ^2+12 \log ^2\left(\frac{\pi }{16}\right)\\ &\hspace{1 cm}+103 (1+\log (16)-\log (\pi ))+5 \gamma_E  (60 \gamma_E -103-96 \log (2)+24 \log (\pi ))\Big)
    \end{split}
\end{align}
and,
\begin{align}
    \begin{split}
        &\delta_1=30 \log(|\boldsymbol{b}|)+25 \gamma_E +6-5 \log \left(\frac{16}{\pi }\right),\\ &
        \delta_2=-192 \log(|\boldsymbol{b}|) (18 \log(|\boldsymbol{b}|)+30 \gamma_E -25)+96 \log \left(\frac{16}{\pi }\right) \left(12 \log(|\boldsymbol{b}|)+10 \gamma_E +\log \left(\frac{\pi }{16}\right)\right)\\ &\hspace{0.7 cm}-800 \gamma_E  (3 \gamma_E -5)-80 \pi ^2-1199-800 \log \left(\frac{16}{\pi }\right)\,,\\ &
        \delta_3=216 \log ^2(|\boldsymbol{b}|)+6 (60 \gamma_E -13-48 \log (2)+12 \log (\pi )) \log(|\boldsymbol{b}|)+5 \left(\gamma_E  (30 \gamma_E -13)+\pi ^2-14\right)\\ &
        \hspace{0.7 cm}+\log \left(\frac{\pi }{16}\right) (60 \gamma_E -13-24 \log (2)+6 \log (\pi ))\,,\\ &
        \delta_4=84 \log \left(\frac{\pi }{16}\right) (6 \log(|\boldsymbol{b}|)+5 \gamma_E )+72 \log(|\boldsymbol{b}|) (21 \log(|\boldsymbol{b}|)+35 \gamma_E -10)+5 \left(30 \gamma_E  (7 \gamma_E -4)+7 \pi ^2-54\right)\\ &\hspace{0.7 cm}+6 \log \left(\frac{16}{\pi }\right) (20+28 \log (2)-7 \log (\pi ))\,,\\ &
        \delta_5=96 \left(19 \log \left(\frac{\pi }{16}\right) (6 \log(|\boldsymbol{b}|)+5 \gamma_E )+6 \log(|\boldsymbol{b}|) (57 \log(|\boldsymbol{b}|)+95 \gamma_E -91)\right)+760 \pi ^2+240 \gamma_E  (95 \gamma_E -182)\\ &\hspace{0.7 cm}+9642+14592 \log ^2(2)+7296 \log (8)+3264 \log \left(\frac{16}{\pi }\right)+912 \left(\log \left(\frac{\pi }{256}\right)-6\right) \log (\pi )\,.
    \end{split}
\end{align}
\textbullet $\,\,$ We now compute the 3PM diagrams with a graviton propagator in the middle. The topology has the following form,
\begin{align}
    \begin{split}
    \hspace{-4 cm}     \begin{minipage}[h]{0.12\linewidth}
	\vspace{4pt}
	\scalebox{4.2}{\includegraphics[width=\linewidth]{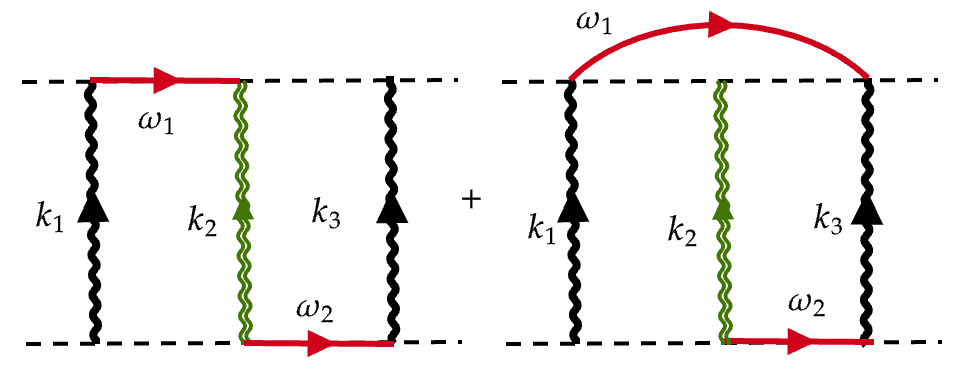}}
 \end{minipage}
    \end{split}
\end{align}
 At $\mathcal{O}(\mathcal{S}^0)$ we get the total contribution from these two diagrams after doing the IBP reduction, 
\begin{align}
    \begin{split}
\chi\Big|_{\mathcal{S}^0}=-\frac{ m_1^2 m_2^2 s_1^2 s_2^2}{256 m_p^6}\,\int e^{-i q \cdot b}\, \hat\delta(q\cdot v_1)\hat\delta(q\cdot v_2)&\Big(\boldsymbol{\tilde h}_{55}\,\,\mathbfcal{M}_{0,0;0,0,1,1,1}+\boldsymbol{\tilde h}_{56}\,\,q^2\mathbfcal{M}_{0,0;0,0,2,1,1}\\&+\boldsymbol{\tilde h}_{57}\,\,q^2\mathbfcal{M}_{0,0;0,0,1,2,1}+\boldsymbol{\tilde h}_{58}\,\,q^2\mathbfcal{M}^{(++)}_{1,1;0,0,1,1,1}\Big)\,,
   \end{split}
\end{align}
and proceeding as before we get at $\mathcal{O}(\mathcal{S}^1)$, 
\begin{align}
    \begin{split}
\chi\Big|_{\mathcal{S}^1}=-\frac{ m_1^2 m_2^2 s_1^2 s_2^2}{64m_p^6}\,\int e^{-i q \cdot b}\, \hat\delta(q\cdot v_1)\hat\delta(q\cdot v_2)&\Bigg[i\,\Big(\boldsymbol{\tilde h}_{59}\,\,\mathbfcal{M}_{0,0;0,0,1,1,1}+\boldsymbol{\tilde h}_{60}\,\,q^2\mathbfcal{M}_{0,0;0,0,2,1,1}\\&+\boldsymbol{\tilde h}_{61}\,\,q^2\mathbfcal{M}_{0,0;0,0,1,2,1}-\frac{1}{2}\boldsymbol{\tilde h}_{62}\,\,q^2\mathbfcal{M}^{(++)}_{1,1;0,0,1,1,1}\\&-\frac{1}{4}\boldsymbol{\tilde h}_{63}\,\,q^2\mathbfcal{M}^{(+-)}_{1,1;0,0,1,1,1}\Big)\big(q\cdot\mathcal{S}_1\cdot v_2\big)\\&+i\,\Big(\boldsymbol{\tilde h}_{64}\,\,\mathbfcal{M}_{0,0;0,0,1,1,1}+\boldsymbol{\tilde h}_{65}\,\,q^2\mathbfcal{M}_{0,0;0,0,2,1,1}\\&+\boldsymbol{\tilde h}_{66}\,\,q^2\mathbfcal{M}_{0,0;0,0,1,2,1}-\frac{1}{2}\boldsymbol{\tilde h}_{67}\,\,q^2\mathbfcal{M}_{1,1;0,0,1,1,1}\\&-\frac{1}{4}\boldsymbol{\tilde h}_{68}\,\,q^2\mathbfcal{M}^{(+-)}_{1,1;0,0,1,1,1}\Big)\big(q\cdot\mathcal{S}_2\cdot v_1\big)\Bigg]\,.
 \end{split}
\end{align}
Finally, using \eqref{MasterM} we get, \\
\begin{align}
    \begin{split} \label{eqt7}
        \chi\Big|_{\mathcal{S}^0}=\frac{m_1^2m_2^2}{m_p^6}&\Bigg[\Big(\Delta_{\mathfrak{12}}^{\textrm{poly}}+\Delta_{\mathfrak{05}}^{\textrm{log}}\log (x)\Bigg] 
        \end{split}
\end{align}
where, 
\begin{align}
    \begin{split}
   & \Delta_{\mathfrak{12}}^{\textrm{poly}}=-\frac{s_1^2s_2^2 (1+x^2)}{8192 \pi ^3 |\boldsymbol{b}|^2 \left(x^2-1\right)^5}\Bigg[48 \left(3 x^4-2 x^2+3\right) x^2 \log (|\boldsymbol{b}|)+x^8+24 x^6 \left(5 \gamma_E +1+\log \left(\frac{\pi }{16}\right)\right)\\&\hspace{5cm}-2 x^4 (40 \gamma_E +17-32 \log (2)+8 \log (\pi ))+24 x^2 \left(5 \gamma_E +1+\log \left(\frac{\pi }{16}\right)\right)+1\Bigg]\,,\\&
   \Delta_{\mathfrak{05}}^{\textrm{log}}=-\frac{ s_1^2 s_2^2 \left(3 x^4+22 x^2+3\right)}{6144 \pi ^3 |\boldsymbol{b}|^2 \left(x^2-1\right)^2 }\,.
      \end{split}
\end{align}
The total contribution at $\mathcal{O}(\mathcal{S}^1)$ is, 
\begin{align}
\begin{split} \label{eqt8}
\chi\Big|_{\mathcal{S}^1} =\frac{m_1^2m_2^2}{m_p^6}&\Bigg[\Big(\Delta_{\mathfrak{13}}^{\textrm{poly}}+\Delta_{\mathfrak{06}}^{\textrm{log}}\log (x)\Big)\Big(\hat{b}\cdot \mathcal{S}_1\cdot v_2\Big)+\Big(\Delta_{\mathfrak{14}}^{\textrm{poly}}+\Delta_{\mathfrak{07}}^{\textrm{log}}\log (x)\Big)\Big(\hat{b}\cdot \mathcal{S}_2\cdot v_1\Big)\Bigg]
\end{split}
\end{align}
where,
\begin{align}
\begin{split}
&\Delta_{\mathfrak{13}}^{\textrm{poly}}=\frac{s_1^2s_2^2\,x}{3072 \pi ^3 |\boldsymbol{b}|^3 \left(x^2-1\right)^5}\Bigg[132 \left(x^3+x\right)^2 \log (|\boldsymbol{b}|)+3 x^8-6 x^7+2 x^6 (55 \gamma_E -38-44 \log (2)+11 \log (\pi ))\\&\hspace{5cm}-18 x^5+22 x^4 (10 \gamma_E -5-8 \log (2)+2 \log (\pi ))-18 x^3\\&\hspace{5cm}+2 x^2 (55 \gamma_E -38-44 \log (2)+11 \log (\pi ))-6 x+3\Bigg]\,,\\&
\Delta_{\mathfrak{14}}^{\textrm{poly}}=-\frac{s_1^2s_1^2 x}{6144 \pi ^3 |\boldsymbol{b}|^3 \left(x^2-1\right)^5 }\Bigg[48 \left(48 x^4+55 x^2+48\right) x^2 \log (|\boldsymbol{b}|)+15 x^8+2 x^6 (960 \gamma_E -359-768 \log (2)\\&\hspace{5cm}+192 \log (\pi ))-24 x^5+10 x^4 (220 \gamma_ E -49-176 \log (2)+44 \log (\pi ))-24 x^3\\&\hspace{5cm}+2 x^2 (960 \gamma_E -359-768 \log (2)+192 \log (\pi ))+15\Bigg]\,,\\&
\Delta_{\mathfrak{6}}^{\textrm{log}}=\frac{3s_1^2 s_2^2 x^3 \left(x^4-1\right)}{256 \pi ^3 |\boldsymbol{b}|^3 \left(x^2-1\right)^5 }\,,\Delta_{\mathfrak{7}}^{\textrm{log}}=-\frac{s_1^2 s_2^2 x \left(x^8-10 x^6+x^4\right)}{128 \pi ^3 |\boldsymbol{b}|^3 \left(x^2-1\right)^6 \left(x^2+1\right) }\,.
\end{split}
\end{align}
\textbullet $\,\,$ Now we consider the following diagram. It comes from the 3-point interaction vertex between the graviton and the kinetic term of the scalar field. 
\begin{align}
    \begin{split}
         \begin{minipage}[h]{0.12\linewidth}
	\vspace{4pt}
	\scalebox{1.65}{\includegraphics[width=\linewidth]{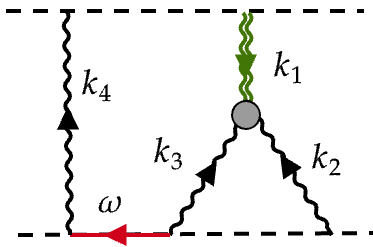}}
 \end{minipage}&
    \end{split}
\end{align}
Proceeding as before, we get the following,
\begin{align}
    \begin{split}
i\chi\Big|_{\mathcal{S}^{0}}&=i\frac{s_1 s_2^3m_1^2 m_2^2}{128 m_p^6}\int e^{-iq\cdot b} \,\hat\delta(q\cdot v_1)\hat\delta(q\cdot v_2)\int_{\ell_{1},\ell_{2}}\left(\frac{1}{(\ell_2\cdot v_2+i\varepsilon)^2}+\frac{1}{(\ell_2\cdot v_2-i\varepsilon)^2}\right)\frac{\hat\delta(\ell_{1}\cdot v_2)\hat\delta(\ell_{2}\cdot v_1)\,}{\ell_{1}^2 \ell_{2}^2 (\ell_{1}+\ell_{2}-q)^2(\ell_{2}-q)^2}\\ &
         \hspace{7 cm}\Bigg((\ell_{1}\cdot v_1)^2\Big[(q-\ell_{1})^2-\ell_2^2-(\ell_1+\ell_2-q)^2\Big]-\frac{1}{d-2}\ell_1^2(\ell_1-q)^2\Bigg)\,,\\ &
        =i\frac{s_1 s_2^3m_1^2 m_2^2}{64 m_p^6}\int e^{-iq\cdot b} \,\hat\delta(q\cdot v_1)\hat\delta(q\cdot v_2)\Bigg[\boldsymbol{\tilde h}_{69}\,\mathbfcal{M}_{0,0;0,0,1,1,1}+\boldsymbol{\tilde h}_{70}\,\,q^2
      \mathbfcal{M}_{0,0;0,0,2,1,1} +\boldsymbol{\tilde h}_{71}\,\,q^2\mathbfcal{M}_{0,0;0,0,1,2,1}\Bigg]\,.
         \end{split}
\end{align}
Then using \eqref{MasterM}. we get, 
\begin{align}
    \begin{split}  \label{eqt9}      \chi\Big|_{\mathcal{S}^{0}}&=\frac{m_1^2m_2^2}{m_p^6}\Delta_{\mathfrak{15}}^{\textrm{poly}}\,,\,\, \Delta_{\mathfrak{15}}^{\textrm{poly}}=\Bigg(\frac{s_{1}s_{2}^3 \left(x^2+1\right)}{4096 \pi ^3 |\boldsymbol{b}|^2 \left(x^2-1\right)}\Bigg)\,.
         \end{split}
\end{align}
Another term that appears in the spinless part from a term proportional to $\ell_{1}^2$ in the numerator. In that case, the integral will be proportional $\mathbfcal{M}_{0,0;0,1,0,1,1}$ which vanishes identically.\par 
Now, the spinning part of the eikonal phase is given by,
\begin{align}
    \begin{split}
       \chi\Big|_{\mathcal{S}^1}=&-\frac{s_1 s_2^3m_1^2 m_2^2}{128 m_p^6}\, S_{1}^{\eta\delta}\int e^{-iq\cdot b} \,\hat\delta(q\cdot v_1)\hat\delta(q\cdot v_2)\int_{\ell_{1},\ell_{2}}\sum_{(\pm)}\frac{\hat\delta(\ell_{1}\cdot v_2)\hat\delta(\ell_{2}\cdot v_1)\,(\ell_{1}\cdot v_1) \,(\ell_2-q)_{[\eta} \, \ell_{1\delta]}}{(\ell_2\cdot v_2\pm i\varepsilon)^2\ell_{1}^2 \,\ell_{2}^2\,(\ell_{1}+\ell_{2}-q)^2}\,\Big[(\ell_{1}-q)^2\\ &
        \hspace{11 cm}-(\ell_1+\ell_2-q)^2-\ell_2^2\Big]\,,\\&
        =-\frac{s_1 s_2^3m_1^2 m_2^2}{64 m_p^6}\int e^{-iq\cdot b} \,\hat\delta(q\cdot v_1)\hat\delta(q\cdot v_2)\Bigg[i \Bigg(\boldsymbol{\tilde h}_{72}\,\mathbfcal{M}_{0,0;0,0,1,1,1}+\boldsymbol{\tilde h}_{73}\,\,q^2
      \mathbfcal{M}_{0,0;0,0,2,1,1}\\& \hspace{6cm}+\boldsymbol{\tilde h}_{74}\,\,q^2\mathbfcal{M}_{0,0;0,0,1,2,1}+\boldsymbol{\tilde h}_{75}\,\,q^2\mathbfcal{M}^{(++)}_{1,1;0,0,1,1,1}
      \Bigg)\big(q\cdot\mathcal{S}_1\cdot v_2\big)\Bigg]\,,\\&=\frac{m_1^2m_2^2}{m_p^6}\Bigg[\Delta_{\mathfrak{16}}^{\textrm{poly}}+\Delta_{\mathfrak{08}}^{\textrm{log}}\log(x)\Bigg],\label{eqt10}      
    \end{split}
\end{align}
where,
\begin{align}
    \begin{split}
    &\Delta_{\mathfrak{16}}^{\textrm{poly}}=\Bigg(\frac{s_1 s_2^3}{131072 \pi ^3 |\boldsymbol{b}|^3 \left(x^2-1\right)^6 \left(x^2+1\right)}\Bigg)\Bigg[x\bigg\{6 \left(65 x^{12}-387 x^8+387 x^4-65\right) \log (|\boldsymbol{b}|)\\ & \hspace{7 cm}+\left(x^4-1\right) \big(x^8 (86-260 \log (2)+65 \log (\pi ))\\& \hspace{7 cm}+1712 x^6+2 x^4 (1466+644 \log (2)-161 \log (\pi ))\\&\hspace{7 cm}+1712 x^2+5 \gamma_E  \left(65 x^8-322 x^4+65\right)+86-260 \log (2)\\& \hspace{7 cm}+65 \log (\pi )\bigg\}\Bigg], \\&\,\Delta_{\mathfrak{08}}^{\textrm{log}}=\Bigg(\frac{s_1 s_2^3}{131072 \pi ^3 |\boldsymbol{b}|^3 \left(x^2-1\right)^6 \left(x^2+1\right)}\Bigg)\Bigg[8 x\Bigg\{20 x^{12}+239 x^{10}+700 x^8+1154 x^6+700 x^4+239 x^2+20\Bigg\}\Bigg]\,.
    \end{split}
\end{align}
\\\\
\textbullet $\,\,$ The next diagrams mentioned below come from the two scalar-graviton interaction vertices (and graviton three vertex) and have $\mathbb{H}$-type topological structure. This diagram will only contribute to the spinless part of the 3PM eikonal phase. For $\mathbb{H}$-type diagram, the worldline propagator will not appear in the amplitude, and we get five independent propagators: $\ell_{1}^2,\,\ell_{2}^2,\,(\ell_{1}+\ell_{2}-q)^2,\,(\ell_{1}-q)^2,\,(q-\ell_{2})^2$. The diagrams are in the following form,
\begin{align}
    \begin{split}
      \hspace{-2.8 cm}   \begin{minipage}[h]{0.12\linewidth}
	\vspace{4pt}
	\scalebox{3.7}{\includegraphics[width=\linewidth]{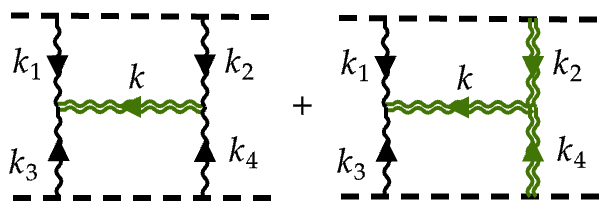}}
 \end{minipage}&
    \end{split}
\end{align}
From the first diagram, the contribution to the eikonal phase is given by,
\begin{align}
    \begin{split}
     i   \chi\Big|_{\mathcal{S}^0}
        =i \frac{s_1^2 s_2^2 m_1^2 m_2^2}{64 m_p^6} \int e^{-iq\cdot b}\hat\delta(q\cdot v_1)\hat\delta(q\cdot v_2)\,\Bigg[&\frac{\epsilon }{4 (\epsilon -1)}\mathbfcal{M}_{0,0;0,0,1,1,1}-\frac{3 q^2}{8 \epsilon }\mathbfcal{M}_{0,0;0,0,2,1,1}\\&+\frac{q^2 (3 \epsilon -4)}{16 (\epsilon -1)}\mathbfcal{M}_{0,0;1,1,0,1,1}+\frac{q^4}{4}\mathbfcal{M}_{0,0;1,1,1,1,1}\Bigg]\,,
    \end{split}
\end{align}
Then using \eqref{MasterM} we get, \\ 
\begin{align}
    \begin{split} \label{eqt11}   
     \chi\Big|_{\mathcal{S}^0}=-\Big(\frac{ m_1^2 m_2^2}{64 m_p^6}\Big)\Delta_{\mathfrak{17}}^{\textrm{poly}}\,, \Delta_{\mathfrak{17}}^{\textrm{poly}}=-\frac{5 x^2s_1^2 s_2^2 \log (x)}{32 \pi ^3 |\boldsymbol{b}|^2 \left(x^2-1\right)^2}\,.
        \end{split}
\end{align}
 The second one with $\mathbb{H}$-type topology may appear at 3PM where we have one scalar-scalar-graviton and one graviton 3-point vertex. This diagram contributes to the spinless as well as the spinning part of the eikonal phase because of the.
The graviton 3-point vertex takes the form,
\begin{align}
    S_{EH}\Big|_{h^3}=\int d^4x \,\,\,\mathcal{U}^{\mu\nu\,\,\alpha\beta\rho\,\,\gamma\delta\sigma}\,h_{\mu\nu}\,\partial_{\rho}h_{\alpha\beta}\partial_{\sigma}h_{\gamma\delta}\,.
\end{align}
In the Fourier space, the vertex factor takes the following form,
\begin{align}
    \begin{split}
        \mathcal{V}(k_1,k_2,k_3)=-\hat\delta^{(4)}(k_1+k_2+k_3)\,\mathcal{U}^{\mu\nu\,\,\alpha\beta\rho\,\,\gamma\delta\sigma}\,k_{2\rho}k_{3\sigma}h_{\mu\nu}(-k_1)\,h_{\alpha\beta}(-k_2)\,h_{\gamma\delta}(-k_3)\,.
    \end{split}
\end{align}
Correspondingly, the contribution to the eikonal phase is given by,
\begin{align}
    \begin{split}
        i\chi=i\frac{m_1^2 m_2^2 s_1 s_2}{64 m_p^6}\int e^{-i q\cdot b}\hat\delta(q\cdot v_1)\hat\delta(q\cdot v_2)\int_{\ell_{1},\ell_{2}}\frac{\hat\delta(\ell_{1}\cdot v_2)\hat\delta(\ell_{2}\cdot v_1)}{\ell_{1}^2 \ell_{2}^2 (\ell_{1}+\ell_{2}-q)^2 (\ell_{1}-q)^2 (\ell_{2}-q)^2}\mathcal{N}\label{4.68j}
    \end{split}
\end{align}
where the numerator $\mathcal{N}$ has the following form,
\begin{align}
    \begin{split}
        \mathcal{N}= \Gamma_{(3)}^{\mu\nu\,\alpha\beta\,\gamma\delta}(k_2,k_4,k)\,k_1^a k_3^b,P_{a b;\mu\nu}P_{\alpha\beta,c d}P_{\gamma\delta;e f}(v_1^{c}v_1^d+i(k_2\cdot \mathcal{S}_1))^{(c}v_1^{d)}\,\,(v_2^{e}v_2^f+i(k_4\cdot \mathcal{S}_2)^{(e}v_2^{f)}),
    \end{split}
\end{align}
where the 3-point vertex function $\Gamma_{(3)}$ takes the form,
\begin{align}
    \begin{split}
    \Gamma^{\mu\nu\,\alpha\beta\,\gamma\delta}=&\textrm{sym}\Big[\frac{1}{2}k_{2}^\mu k_{4}^\nu\eta^{\alpha\beta}\eta^{\gamma\delta}-\frac{1}{4}k_2\cdot k_4\,\eta^{\mu\nu}\eta^{\alpha\beta}\eta^{\gamma\delta}+k_2\cdot k_4 \eta^{\alpha\nu}\eta^{\beta\mu}\eta^{\gamma\delta}-k_2^\delta k_4^\gamma \eta^{\alpha\mu}\eta^{\beta\nu}\\ &
        +\frac{1}{4}(k_2\cdot k_4)\eta^{\mu\nu}\eta^{\alpha\gamma}\eta^{\beta\delta}-k_2^\nu\,k_4^\beta \eta^{\gamma\delta}\eta^{\mu\alpha}-k_{2}^\mu k_4^\nu \eta^{\alpha\delta}\eta^{\beta\gamma}+\frac{1}{2}\,k_2^\alpha\,k_4^\beta\,\eta^{\mu\nu}\eta^{\gamma\delta}\\ &
        -k_2\cdot k_4\,\eta^{\alpha\mu}\eta^{\beta\delta}\eta^{\nu\gamma}-\frac{1}{2}k_2^\gamma\,k_4^\beta\,\eta^{\mu\nu}\eta^{\alpha\delta}+k_2^\gamma k_4^\alpha\,\eta^{\mu\beta}\eta^{\nu\delta}-\frac{1}{2}\,k_2^\mu k_4^\nu \eta^{\alpha\delta}\eta^{\beta \gamma}+2 k_{2}^\nu k_4^\alpha\eta^{\beta\gamma}\eta^{\mu\delta}+\textrm{permutations}\Big]\,.
    \end{split}
\end{align}
Then, proceeding as before, we get, 
\begin{align}
    \begin{split}
        \chi\Big|_{\mathcal{S}^0}= \frac{m_1^2 m_2^2 s_1 s_2}{64 m_p^6}\int e^{-i q\cdot b}\hat\delta(q\cdot v_1)\hat\delta(q\cdot v_2)\Bigg(&\boldsymbol{\tilde h}_{76}\mathbfcal{M}_{0,0;0,0,1,1,1}+\boldsymbol{\tilde h}_{77}\,q^2\mathbfcal{M}_{0,0;0,0,2,1,1}+\boldsymbol{\tilde h}_{78}\,q^2\mathbfcal{M}_{0,0;0,0,1,2,1}\\ &+\boldsymbol{\tilde h}_{79}\,q^2\mathbfcal{M}_{0,0;1,1,0,1,1}+\boldsymbol{\tilde h}_{80}\,q^4\mathbfcal{M}_{0,0;1,1,1,1,1}+\boldsymbol{\tilde h}_{81}\,q^6\mathbfcal{M}_{0,0;1,1,2,1,1}\Bigg)
    \end{split}
\end{align}
and, \begin{align}
    \begin{split}
        \chi\Big|_{\mathcal{S}^1}= \frac{m_1^2 m_2^2 s_1 s_2}{64 m_p^6}\int e^{-i q\cdot b}\hat\delta(q\cdot v_1)\hat\delta(q\cdot v_2)\Bigg[i\,\Bigg(&\boldsymbol{\tilde h}_{82}\mathbfcal{M}_{0,0;0,0,1,1,1}+\boldsymbol{\tilde h}_{83}\,q^2\mathbfcal{M}_{0,0;0,0,2,1,1}\\ &+\boldsymbol{\tilde h}_{84}\,q^2\mathbfcal{M}_{0,0;0,0,1,2,1}+\boldsymbol{\tilde h}_{85}\,q^2\mathbfcal{M}_{0,0;1,1,0,1,1}\\&+\boldsymbol{\tilde h}_{86}\,q^4\mathbfcal{M}_{0,0;1,1,1,1,1}+\boldsymbol{\tilde h}_{87}\,q^6\mathbfcal{M}_{0,0;1,1,2,1,1}\Bigg)\big(q\cdot\mathcal{S}_1\cdot v_2\big)+\\&i\,\Bigg(\boldsymbol{\tilde h}_{88}\mathbfcal{M}_{0,0;0,0,1,1,1}+\boldsymbol{\tilde h}_{89}\,q^2\mathbfcal{M}_{0,0;0,0,2,1,1}\\ &+\boldsymbol{\tilde h}_{90}\,q^2\mathbfcal{M}_{0,0;0,0,1,2,1}+\boldsymbol{\tilde h}_{91}\,q^2\mathbfcal{M}_{0,0;1,1,0,1,1}\\&+\boldsymbol{\tilde h}_{92}\,q^4\mathbfcal{M}_{0,0;1,1,1,1,1}+\boldsymbol{\tilde h}_{93}\,q^6\mathbfcal{M}_{0,0;1,1,2,1,1}\Bigg)\big(q\cdot\mathcal{S}_2\cdot v_1\big)\Bigg]\,.
    \end{split}
\end{align}
Then using \eqref{MasterM} we get, \\ 
\begin{align}
    \begin{split} \label{eqt12}   
        \chi\Big|_{\mathcal{S}^0}=\frac{m_1^2m_2^2}{m_p^6}\Bigg[\Delta_{\mathfrak{18}}^{\textrm{poly}}+\Delta_{\mathfrak{09}}^{\textrm{log}}\log(x)+\Delta_{\mathfrak{03}}^{\textrm{Polylog}}\Big(\log(x)^2+ \textbf{Li}_2\left(1-x^2\right)\Big)\Bigg]\,,
          \end{split}
\end{align}
where, 
\begin{align}
    \begin{split} 
  &  \Delta_{\mathfrak{18}}^{\textrm{poly}}= \frac{s_1s_2 (3-3 x^4)}{16384 \pi ^3 |\boldsymbol{b}|^2 \left(x^2-1\right)^2}\,,   \Delta_{\mathfrak{09}}^{\textrm{log}}=\frac{s_1s_2\bigg\{4 \left(-2 x^2 \log \left(\pi  |\boldsymbol{b}|^6\right)+5 x^4+x^2 (-10 \gamma_E -4+\log (256))+5\right)\bigg\}}{16384 \pi ^3 |\boldsymbol{b}|^2 \left(x^2-1\right)^2}\,,   \\& \Delta_{\mathfrak{03}}^{\textrm{Polylog}}=\frac{4\,s_1s_2\,(x-1) x}{16384 \pi ^3 |\boldsymbol{b}|^2 \left(x^2-1\right)^2}
      \end{split}
\end{align}
and 
\begin{align}
    \begin{split} \label{eqt13}  
        \hspace{-0.7 cm}\chi\Big|_{\mathcal{S}^1}=&\frac{m_1^2m_2^2}{m_p^6}\Bigg\{\Bigg[\Delta_{\mathfrak{19}}^{\textrm{poly}}+\Delta_{\mathfrak{10}}^{\textrm{log}}\log(x)+\Delta_{\mathfrak{04}}^{\textrm{Polylog}}\Big(\log(x)^2+ \textbf{Li}_2\left(1-x^2\right)\Big)\Bigg]\big(\hat{b}\cdot\mathcal{S}_1\cdot v_2\big)\\&\hspace{1cm}+\Bigg[\Delta_{\mathfrak{20}}^{\textrm{poly}}+\Delta_{\mathfrak{11}}^{\textrm{log}}\log(x)+\Delta_{\mathfrak{05}}^{\textrm{Polylog}}\Big(\log(x)^2+ \textbf{Li}_2\left(1-x^2\right)\Big)\Bigg]\big(\hat{b}\cdot\mathcal{S}_2\cdot v_1\big)\Bigg\}
          \end{split}
\end{align}
where, 
\begin{align}
    \begin{split}
   & \Delta_{\mathfrak{19}}^{\textrm{poly}}=\frac{s_{1} s_{2} \left(x^6-10 x^5+3 x^4-49 x^3+3 x^2-10 x+1\right)}{8192 \pi ^3 |\boldsymbol{b}|^3 \left(x^2-1\right)^3},\Delta_{\mathfrak{4}}^{\textrm{Polylog}}=\frac{7 x^2\,s_1s_2\, (x-1) x^2 \left(x^4+4 x^2+1\right)}{8192 \pi ^3 |\boldsymbol{b}|^3 \left(x^2-1\right)^4 \left(x^2+1\right)},\\& \Delta_{\mathfrak{10}}^{\textrm{log}}=-\Bigg(\frac{s_1s_2}{{16384 \pi ^3 |\boldsymbol{b}|^3 \left(x^2-1\right)^4 \left(x^2+1\right)}}\Bigg)\Bigg[x \big(28 \left(x^6+4 x^4+x^2\right) \log \left(\pi  |\boldsymbol{b}|^6\right)-37 x^8+8 x^7+4 x^6 (35 \gamma_E -8-28 \log (2))+24 x^5\\&\hspace{1cm}+x^4 (560 \gamma_E -278-448 \log (2))+24 x^3+4 x^2 (35 \gamma_E -8-28 \log (2))+8 x-37\big)\Bigg]\,, \\&  \Delta_{\mathfrak{11}}^{\textrm{log}}=\Bigg(\frac{s_1s_2}{16384 \pi ^3 |\boldsymbol{b}|^3 \left(x^2-1\right)^6 \left(x^2+1\right)^2}\Bigg)\Bigg[2 \big(48 \left(x^2-1\right)^2 \left(x^6+5 x^4+5 x^2+1\right) x^3 \log ( |\boldsymbol{b}|)-18 x^{16}-13 x^{15}+96 x^{14}\\&\hspace{1cm}+x^{13} (40 \gamma_E -33-32 \log (2)+8 \log (\pi ))+280 x^{12}+x^{11} (120 \gamma_E -47-96 \log (2)+24 \log (\pi ))\\&\hspace{1cm}-992 x^{10}+x^9 (-160 \gamma_E +93+128 \log (2)-32 \log (\pi ))-2604 x^8\\&\hspace{1cm}+x^7 (-160 \gamma_E +93+128 \log (2)-32 \log (\pi ))-992 x^6+x^5 (120 \gamma_E -47-96 \log (2)+24 \log (\pi ))\\&\hspace{1cm}+280 x^4+x^3 (40 \gamma_E -33-32 \log (2)+8 \log (\pi ))+96 x^2-13 x-18\big)\Bigg]\,,\\& \Delta_{\mathfrak{5}}^{\textrm{Polylog}}=-\frac{8 (x-1) x^2 \left(x^6+5 x^4+5 x^2+1\right)\, s_1s_2}{16384 \pi ^3 |\boldsymbol{b}|^3 \left(x^2-1\right)^4 \left(x^2+1\right)^2}\,,\Delta_{\mathfrak{20}}^{\textrm{poly}}=\frac{x  \left(19 x^4-218 x^2+19\right)\,s_1s_2}{16384 \pi ^3 |\boldsymbol{b}|^3 \left(x^2-1\right)^3 }\,.
      \end{split}
\end{align}

\textbullet $\,\,$ At 3PM we face a $\mathbb{N}$-type topology involving scalar and graviton lines. It gives the following three diagrams. 
\begin{center}
    \begin{align}
    \begin{split}
   \hspace{-3.5 cm}\begin{minipage}[h]{0.12\linewidth}
	\vspace{4pt}
	\scalebox{5}{\includegraphics[width=\linewidth]{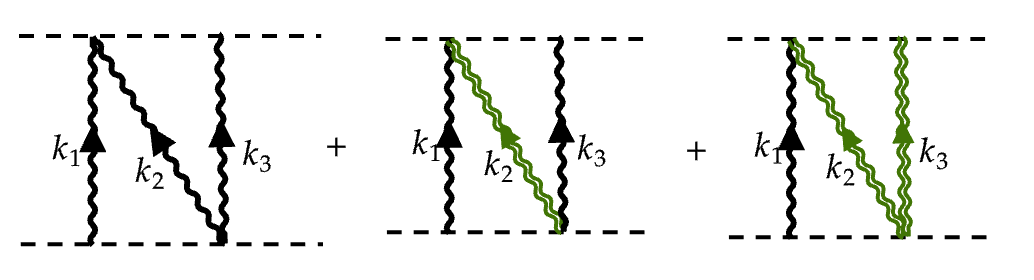}}\end{minipage}&\nonumber
    \end{split}
\end{align}
\end{center}
The spinless contributions come from the first and the second diagrams only. We get, 
\begin{align}
\begin{split} \label{eqt14}  
    \chi\Big|_{\mathcal{S}^0}&=\frac{m_1^2 m_2^2}{16\,m_p^6}\int e^{-i q\cdot b}\hat\delta(q\cdot v_1)\hat\delta(q\cdot v_2)\Bigg(s_1s_2g_1g_2+s_1^2s_2^2\,\boldsymbol{\tilde h}_{94}\Bigg)\mathbfcal{M}_{0,0;0,0,1,1,1}\,,\\&
    =\frac{m_1^2 m_2^2}{m_p^6}\Delta_{\mathfrak{12}}^{\textrm{log}}\,,\,\,\Delta_{\mathfrak{12}}^{\textrm{log}}=\Bigg(\frac{s_{1}s_{2} \log (x) \left(8 g_{1} g_{2} x^2+s_{1} s_{2} \left(x^4+1\right)\right)}{1024 \pi ^3 |\boldsymbol{b}|^2 \left(x^2-1\right)^2}\Bigg)\,.
    \end{split}
\end{align}




The spinning contributions come from the second and third diagrams. We get,
\begin{align}
    \begin{split}
       & \chi\Big|_{\mathcal{S}^{1}}=-\frac{m_1^2 m_2^2 s_1^2  s_2^2}{32 m_p^6}\int e^{-i q\cdot b}\hat\delta(q\cdot v_1)\hat\delta(q\cdot v_2)\Bigg[i\,\Bigg(\boldsymbol{\tilde h}_{95}\mathbfcal{M}_{0,0;0,0,1,1,1}+\boldsymbol{\tilde h}_{96}\,q^2\mathbfcal{M}_{0,0;0,0,1,2,1}\\ &\hspace{7cm}+\boldsymbol{\tilde h}_{97}\,q^2\mathbfcal{M}_{0,0;0,0,2,1,1}\Bigg)\big(q\cdot\mathcal{S}_1\cdot v_2\big)+\\&\hspace{6.5cm}i\,\Bigg\{\Big(\frac{1}{2 s_1 s_2}\boldsymbol{\tilde h}_{98}-\boldsymbol{\tilde h}_{95}\Big)\mathbfcal{M}_{0,0;0,0,1,1,1}+\Big(\frac{1}{2 s_1 s_2}\boldsymbol{\tilde h}_{99}-\boldsymbol{\tilde h}_{97}\Big)\,q^2\mathbfcal{M}_{0,0;0,0,2,1,1}\\ &\hspace{7cm}+\Big(\frac{1}{2 s_1 s_2}\boldsymbol{\tilde h}_{100}-\boldsymbol{\tilde h}_{96}\Big)\,q^2\mathbfcal{M}_{0,0;0,0,1,2,1}\Bigg)\big(q\cdot\mathcal{S}_2\cdot v_1\big)\Bigg\}\Bigg]\,.
    \end{split}
\end{align}
Now using (\ref{MasterM}) we get, 
\begin{align}
    \begin{split} \label{eqt15}     \chi\Big|_{\mathcal{S}^{1}}=\frac{m_1^2m_2^2}{m_p^6}\Bigg\{&\Bigg[\Delta_{\mathfrak{13}}^{\textrm{log}}\log(x)+\Delta_{\mathfrak{06}}^{\textrm{Polylog}}\Big(\log(x)^2+ \textbf{Li}_2\left(1-x^2\right)\Big)\Bigg]\big(\hat{b}\cdot\mathcal{S}_1\cdot v_2\big)+\\&\Bigg[\Delta_{\mathfrak{21}}^{\textrm{poly}}+\Delta_{\mathfrak{14}}^{\textrm{log}}\log(x)+\Delta_{\mathfrak{07}}^{\textrm{Polylog}}\Big(\log(x)^2+ \textbf{Li}_2\left(1-x^2\right)\Big)\Bigg]\big(\hat{b}\cdot\mathcal{S}_2\cdot v_1\big)\Bigg\}\,,
         \end{split}
\end{align}
where, 
\begin{align}
    \begin{split}
 &   \Delta_{\mathfrak{13}}^{\textrm{log}}=-\Bigg(\frac{s_{1}^2 s_{2}^2}{2048 \pi ^3 |\boldsymbol{b}|^3 \left(x^2-1\right)^4}\Bigg)\Bigg[x \left(x^2+1\right) \big(-12 \left(x^4+1\right) \log (|\boldsymbol{b}|)+x^4-10 \gamma_E  \left(x^4+1\right)+8 x^4 \log (2)\\&\hspace{5.5cm}-2 \left(x^4+1\right) \log (\pi )+18 x^2+1+\log (256)\big)\Bigg]\,,\\&
    \Delta_{\mathfrak{06}}^{\textrm{Polylog}}=-\frac{3 s_{1}^2 s_{2}^2 x \left(x^2+1\right) \left(x^4+1\right)}{1024 \pi ^3 |\boldsymbol{b}|^3 \left(x^2-1\right)^4}\,,
     \Delta_{\mathfrak{21}}^{\textrm{poly}}=-\frac{41\,s_{1} s_{2} x \left(x^2+1\right)^2}{2048 \pi ^3 |\boldsymbol{b}|^3 \left(x^2-1\right)^3}\,,\\&
       \Delta_{\mathfrak{14}}^{\textrm{log}}=\Bigg(\frac{s_{1}s_{2}}{4096 \pi ^3 |\boldsymbol{b}|^3 \left(x^2-1\right)^4}\Bigg)\Bigg[-x \left(x^2+1\right) \big(12 \left(x^4+1\right) \log (|\boldsymbol{b}|) (2\,s_{1}s_{2}+1)+10 \gamma_E  \left(x^4+1\right) (2\,s_{1} s_{2}+1)\\&\hspace{5.5cm}+2 \log (\pi ) \left(2\,s_{1} s_{2} \left(x^4+1\right)+x^4\right)-2\,s_{1}s_{2} \left(x^4+8 \left(x^4+1\right) \log (2)+18 x^2+1\right)\\&\hspace{5.5cm}+x^4 (5-8 \log (2))-18 x^2+5-8 \log (2)+2 \log (\pi )\big)\Bigg]\,,\\&
       \Delta_{\mathfrak{07}}^{\textrm{Polylog}}=\frac{3 \,s_{1} s_{2} x \left(x^2+1\right) \left(x^4+1\right) (2 s_{1}s_{2}+1)}{2048 \pi ^3 |\boldsymbol{b}|^3 \left(x^2-1\right)^4 }\,.
      \end{split}
\end{align}
\textbullet $\,\,$ We face the following topology with single 3-point kind of interaction vertex,
\begin{center}
    \begin{align}
    \begin{split}
   \hspace{-3.5 cm}     \begin{minipage}[h]{0.12\linewidth}
	\vspace{4pt}
	\scalebox{5}{\includegraphics[width=\linewidth]{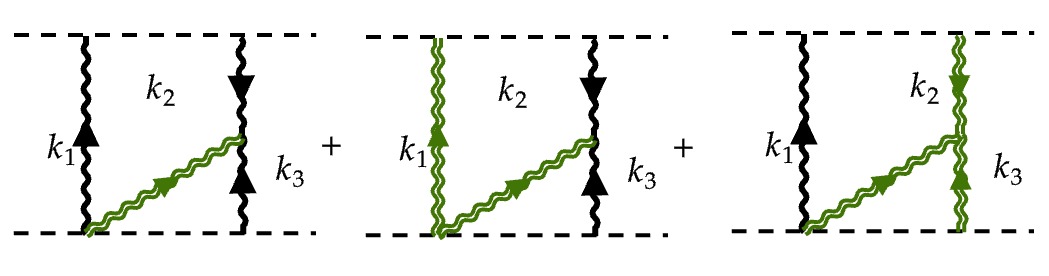}}\end{minipage}&\nonumber
    \end{split}
\end{align}
\end{center}
Before quoting the result for the contribution to the eikonal phase, we note that because of the presence of graviton two-point worldline vertex in the second diagram, this diagram only contributes to the spinning eikonal, and also, there is no term linear in the spin of the first black hole coming from this diagram. Now, we first write down the total spinless contribution from these three diagrams, 
\begin{align}
    \begin{split}  \label{eqt16} 
   \chi\Big|_{\mathcal{S}^0}&=\frac{m_1^2 m_2^2}{m_p^6}\int e^{-i q\cdot b}\hat\delta(q\cdot v_1)\hat\delta(q\cdot v_2)\Bigg[\Bigg(\frac{s_1^2 s_2^2}{128 (1-\epsilon)}-\frac{s_1 s_2}{32} \boldsymbol{\tilde h}_{101}\Bigg)\mathbfcal{M}_{0,0;0,0,1,1,1}-\frac{s_1 s_2}{32}\boldsymbol{\tilde h}_{102}q^2\mathbfcal{M}_{0,0;0,0,1,2,1}\Bigg)\Bigg]\,,\\&
   =\frac{m_1^2 m_2^2}{m_p^6}\Bigg(\Delta_{\mathfrak{22}}^{\textrm{poly}}+\Delta_{\mathfrak{15}}^{\textrm{log}}\log(x)\Bigg)
  \end{split}
\end{align}    
where, 
\begin{align}
    \begin{split}
      \Delta_{\mathfrak{22}}^{\textrm{poly}}= \frac{s_{1} s_{2} \left(-x^8-x^6+x^2+1\right)}{4096 \pi ^3 |\boldsymbol{b}|^2 x^2 \left(x^2-1\right)^2}\,,\Delta_{\mathfrak{15}}^{\textrm{log}}=-\frac{s_{1} s_{2} \left(x^2 (5-s_{1}s_{2})+2 x^4+2\right)}{1024 \pi ^3 |\boldsymbol{b}|^2 \left(x^2-1\right)^2}\,.
    \end{split}
\end{align}
The total contribution to the spinning part is 
, 
\begin{align}
    \begin{split}
     \chi\Big|_{\mathcal{S}^1}=\frac{m_1^2 m_2^2}{m_p^6}\Big(-i\frac{s_1s_2}{32}\Big)\int e^{-i q\cdot b}\hat\delta(q\cdot v_1)\hat\delta(q\cdot v_2)&\Bigg[\Bigg(\boldsymbol{\tilde h}_{103}\mathbfcal{M}_{0,0;0,0,1,1,1}+\boldsymbol{\tilde h}_{104}q^2\mathbfcal{M}_{0,0;0,0,2,1,1}\\&\hspace{1.5cm}+\boldsymbol{\tilde h}_{105}q^2\mathbfcal{M}_{0,0;0,0,1,2,1}\Bigg)\big(q\cdot\mathcal{S}_1\cdot v_2\big)\\&+\Bigg((\boldsymbol{\tilde h}_{106}+\boldsymbol{\tilde h}_{109})\mathbfcal{M}_{0,0;0,0,1,1,1}+(\boldsymbol{\tilde h}_{107}+\boldsymbol{\tilde h}_{110})q^2\mathbfcal{M}_{0,0;0,0,2,1,1}\\&+(\boldsymbol{\tilde h}_{108}+\boldsymbol{\tilde h}_{111})q^2\mathbfcal{M}_{0,0;0,0,1,2,1}\Bigg)\big(q\cdot\mathcal{S}_2\cdot v_1\big)\Bigg]\,,\\&\hspace{-7cm}
     =\frac{m_1^2 m_2^2}{m_p^6}\Bigg\{\Bigg(\Delta_{\mathfrak{23}}^{\textrm{poly}}+\Delta_{\mathfrak{16}}^{\textrm{log}}\log(x)\Bigg)(\hat b\cdot \mathcal{S}_1\cdot v_2)+\Bigg(\Delta_{\mathfrak{24}}^{\textrm{Poly}}+\Delta_{\mathfrak{17}}^{{\textrm{log}}}\log(x)\Bigg)(\hat b\cdot \mathcal{S}_2\cdot v_1)\Bigg\}
      \end{split}
\end{align} 
where, 
\begin{align}
    \begin{split}
   & \Delta_{\mathfrak{23}}^{\textrm{poly}}=-\frac{s_{1}s_{2} \left(x^{10}+9 x^8+32 x^6-32 x^4-9 x^2-1\right)}{2048 \pi ^3 |\boldsymbol{b}|^3 x \left(x^2-1\right)^4 }\,,\Delta_{\mathfrak{16}}^{\textrm{log}}=-\frac{s_{1} s_{2} x \left(x^2+1\right)^3}{128 \pi ^3 |\boldsymbol{b}|^3 \left(x^2-1\right)^4 }\,,
   \\& \Delta_{\mathfrak{24}}^{\textrm{poly}}=\frac{s_{1} s_{2} \left(113 x^6-32 x^5+555 x^4+555 x^2-32 x+113\right)}{32768 \pi ^3 |\boldsymbol{b}|^3 \left(x^2-1\right)^3 }\,,\\&\Delta_{\mathfrak{17}}^{{\textrm{log}}}=\frac{s_{1} s_{2} \left(17 x^8-8 x^7+153 x^6-8 x^5+328 x^4-8 x^3+153 x^2-8 x+17\right)}{8192 \pi ^3 |\boldsymbol{b}|^3 \left(x^2-1\right)^4}\,.  \label{eqt17} 
     \end{split}
\end{align} 
\textbullet $\,\,$ We have the following topologies both involving scalar-graviton and graviton 3-point vertices.
\begin{center}
    \begin{align}
    \begin{split}
   \hspace{-2.2 cm}     \begin{minipage}[h]{0.12\linewidth}
	\vspace{4pt}
	\scalebox{3}{\includegraphics[width=\linewidth]{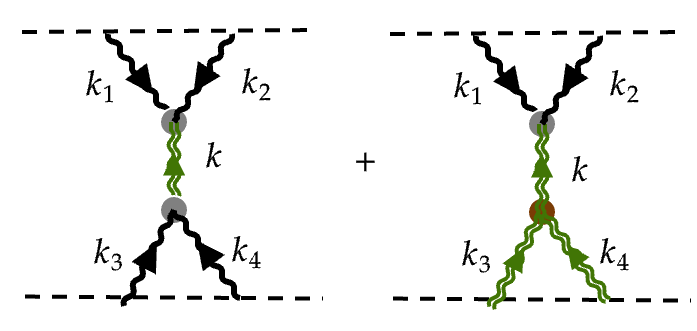}}\end{minipage}&\nonumber
    \end{split}
\end{align}
\end{center}
We can easily check that after evaluating the corresponding integrals for both the spinless and spin part, they vanish identically upto $\mathcal{O}(\epsilon^0)\,.$ Hence, there are no contributions to the eikonal phase from this kind of diagrams for our case. 
\\\\
\textbullet $\,\,$
We now discuss the scalar-graviton diagrams where we have spin contribution coming from the scalar-fermion vertex. The diagrams look like the following. 
\begin{align}
    \begin{split}
     \hspace{-2.5 cm}    
         \begin{minipage}[h]{0.12\linewidth}
	\vspace{4pt}
	\scalebox{3.79}{\includegraphics[width=\linewidth]{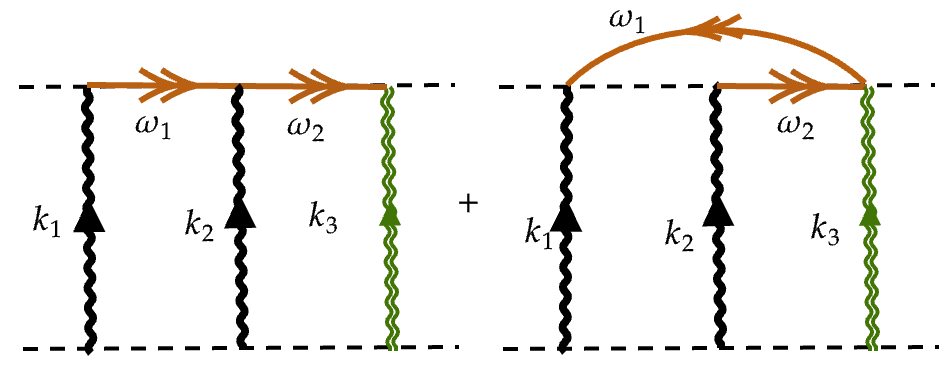}}
 \end{minipage}&\nonumber
    \end{split}
\end{align}
One may notice that, unlike other diagrams, this one consists of $\bar\Psi^\mu\Psi^\nu$. To recover the spin, one needs to take the spinor flow in the opposite direction. So, for this particular diagram, we need to take the antisymmetric combination of $\bar\Psi^\mu\Psi^\nu$ to recover the classical spin.  Moreover, we can easily check that the contribution coming from the first diagram to the eikonal phase gets cancelled by the second diagram. Hence, together, they don't contribute. 
\\\\
\textbullet $\,\,$ Another 3PM spinor diagram that will contribute has the topology of the following form,
\begin{align}
\centering
      \hspace{-3.5 cm}   \begin{minipage}[h]{0.12\linewidth}
	\vspace{4pt}
	\scalebox{4.1}{\includegraphics[width=\linewidth]{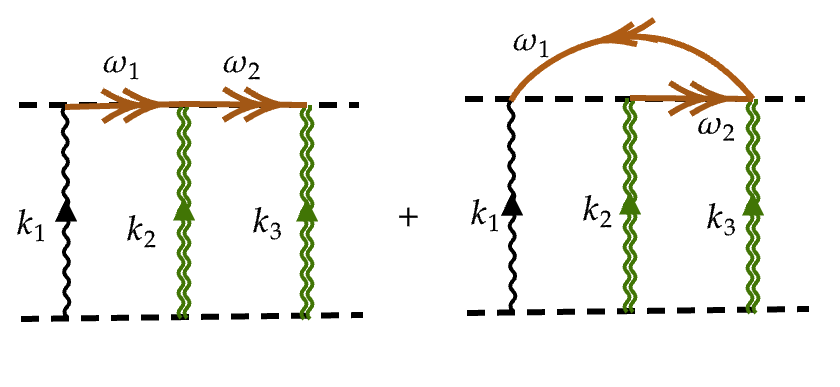}}
 \end{minipage}&\nonumber
\end{align}
Let's start our computations from the first diagram.   The 3-point spin-spin-graviton interaction vertex which contributes to this term comes from the following part of the action,
\begin{align}
    \begin{split}
        S\subset& \,i\frac{m_i}{m_p}\int  d\tau\, \dot x^\mu \omega^{ab}_{\mu}\bar\psi_{a} \psi_{b}\,,\\ &
        = \frac{i m_i}{m_p}v^\mu\int d\tau\, \partial^{[a}h^{b]}_\mu \bar\psi_{[a}\psi_{b]} \xrightarrow[]{F.S.}-\frac{m_a}{m_p}\int_{k,\omega_i}e^{ik\cdot b}\hat\delta(k\cdot v-\omega_1+\omega_2) v^\mu\,k^{[a}h^{b]}_\mu (-k)\bar\psi_{[a}(-\omega_1)\,\psi_{b]}(-\omega_2)  \,.
        \end{split}
\end{align}
Therefore, the vertex factor is given by
\begin{align}
    V: e^{ik\cdot b}\hat\delta(k\cdot v-\omega_1+\omega_2)\,v_i^\mu \,k^a+\textrm{other 3 combinations}\,.
\end{align}
Schematically, the amplitude looks like
\begin{align}
    \begin{split}
       \mathcal{A}\sim\int_{k_i,\omega_i}\Big(\prod\boldsymbol{\delta}^{(6)}\Big)\bar\Psi^\eta\Psi^\sigma\omega_1\langle \psi_\eta\,\bar\psi_{[[a}\rangle_{\omega_1}k_{2[a}\langle  h_{b]\mu}h_{\alpha\beta}\rangle_{k_2} \langle \psi_{b]]}\,\bar\psi_\rho\rangle k_{3[\rho}\delta^{(\chi}_{\sigma]}v_1^{\pi)} \langle  h_{\chi\pi}h_{\kappa\theta}\rangle_{k_3}v_2^\alpha v_2^\beta v_2^\kappa v_2^\theta\,.\nonumber
    \end{split}
\end{align}
Here, $\prod \boldsymbol\delta^6$ stands for the product of all relevant delta functions along with $e^{ik\cdot b}$ coming from the worldline vertices. \textcolor{black}{Note that we take only the antisymmetric combination of $\eta,\sigma$ to relate the combination of background superfield to classical spin, which is eventually important to define the Pauli-Lubanski vector.} 
Then proceeding as before we get, 
\begin{align}
    \begin{split}
i\,\chi\Big|_{\mathcal{S}^1}&=\frac{ m_1 m_2^3 s_1 s_2}{16 m_p^6}\int e^{-iq\cdot b} \hat\delta(q\cdot v_1)\hat\delta(q\cdot v_2)\boldsymbol{\tilde h}_{112} \mathbfcal{L}_{0,0;0,0,1,1,1}\big(q\cdot\mathcal{S}_1\cdot v_2\big)\,.
       \end{split}
\end{align}
Correspondingly, the contribution from the second diagram has the following form,
\begin{align}
    \begin{split}
i\,\chi\Big|_{\mathcal{S}^1}&=\frac{ m_1 m_2^3 s_1 s_2}{8 m_p^6}\int e^{-iq\cdot b} \hat\delta(q\cdot v_1)\hat\delta(q\cdot v_2)\boldsymbol{\tilde h}_{113} \mathbfcal{L}_{0,0;0,0,1,1,1}\big(q\cdot\mathcal{S}_1\cdot v_2\big)\,.
       \end{split}
\end{align}
Then, the total contribution to the spinning part is,
\begin{align}
    \begin{split} \label{eqt18} 
 \chi\Big|_{\mathcal{S}^1}&=\frac{m_1m_2^3}{m_p^6}\Delta_{\mathfrak{25}}^{\textrm{poly}}\,,\quad\,\,\Delta_{\mathfrak{25}}^{\textrm{poly}}=\frac{ s_{1} s_{2}\,9 \left(x^2+1\right) \left(7 x^4+2 x^2+7\right)}{32768 \pi ^3 |\boldsymbol{b}|^3 \left(x^2-1\right)^3}\,.
      \end{split}
\end{align}
Here we have used (\ref{MasterL}).\\\\
\textbullet $\,\,$ Another 3PM diagram having one worldline and one super-field propagator that will contribute has the following topology 
\begin{align}
    \begin{split}
     \hspace{-2.76 cm}  
         \begin{minipage}[h]{0.12\linewidth}
	\vspace{4pt}
	\scalebox{3.9}{\includegraphics[width=\linewidth]{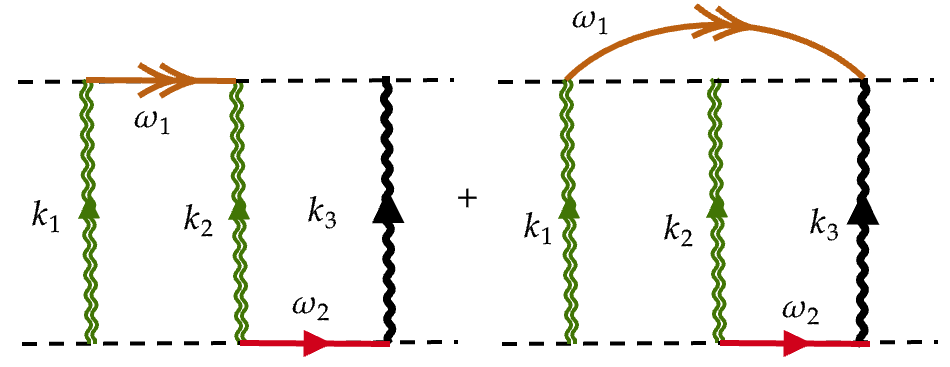}}
 \end{minipage}&\nonumber
 \end{split}
 \end{align}
It is clear from the structure of this diagram that due to the presence of a spinning worldline propagator, we need to take only the spinless terms from the other vertices. Like the previous diagram, this one also has $\bar\Psi^\mu \Psi^\nu$, but for this case, we can add another counterpart, which will have opposite spinor flow. So, in this case, we can also take the antisymmetric combination of $\bar\Psi^\mu \Psi^\nu$ to obtain the classical spin.  Taking all of those things into account, the total contribution to the eikonal phase coming from these two diagrams is given by,
\begin{align}
    \begin{split}
     \chi\Big|_{\mathcal{S}^1}=\frac{m_1^2 m_2^2}{m_p^6}\Big(-i\frac{s_1s_2}{16}\Big)\int e^{-i q\cdot b}\hat\delta(q\cdot v_1)\hat\delta(q\cdot v_2)&\Bigg[\Bigg(\boldsymbol{\tilde h}_{114}\mathbfcal{M}_{0,0;0,0,1,1,1}+\boldsymbol{\tilde h}_{115}q^2\mathbfcal{M}_{0,0;0,0,2,1,1}\\&\hspace{0.25cm}+\boldsymbol{\tilde h}_{116}q^2\mathbfcal{M}_{0,0;0,0,1,2,1}-\frac{1}{2}\boldsymbol{\tilde h}_{117}q^2\mathbfcal{M}^{(++)}_{1,1;0,0,1,1,1}\Bigg)\big(q\cdot\mathcal{S}_1\cdot v_2\big)\,.
     \end{split}
\end{align}
Finally, we get, 
\begin{align}
    \begin{split}  \label{eqt19} 
     \chi\Big|_{\mathcal{S}^1}=\frac{m_1^2 m_2^2}{m_p^6}\Bigg(\Delta_{\mathfrak{26}}^{\textrm{poly}}+\Delta_{\mathfrak{18}}^{\textrm{log}}\log(x)\Bigg)\big(\hat b\cdot\mathcal{S}_1\cdot v_2\big)\,,
      \end{split}
\end{align} 
where, 
\begin{align}
    \begin{split}
  &\Delta_{\mathfrak{26}}^{\textrm{poly}}=\frac{s_1s_2}{32768 \pi ^3 |\boldsymbol{b}|^3 \left(x^2-1\right)^5}\Bigg[-192 \left(x^4-1\right)^2 x \log (|\boldsymbol{b}|)-48 x^{10}+x^9 (-160 \gamma_E +407+128 \log (2)\\&\hspace{5cm}-32 \log (\pi ))-144 x^8+1520 x^7+192 x^6+x^5 (320 \gamma_E +2226-256 \log (2)\\&\hspace{5cm}+64 \log (\pi ))+192 x^4+1520 x^3-144 x^2+x (-160 \gamma_E +407+128 \log (2)\\&\hspace{5cm}-32 \log (\pi ))-48\Bigg]\,,\\&
\Delta_{\mathfrak{18}}^{\textrm{log}}=\frac{s_{1} s_{2} x}{4096 \pi ^3 |\boldsymbol{b}|^3 \left(x^2-1\right)^6 \left(x^2+1\right)}\Bigg(20 x^{12}-32 x^{11}+239 x^{10}-32 x^9+700 x^8+64 x^7+1154 x^6\\&\hspace{6.5cm}+64 x^5+700 x^4-32 x^3+239 x^2-32 x+20\Bigg)\,.
      \end{split}
\end{align}
\textbullet $\,\,$ Two 3PM diagrams coming from graviton-scalar and scalar-scalar worldline interaction vertex have the following topology,
\begin{align}
    \begin{split}
      \hspace{-2.8 cm}   \begin{minipage}[h]{0.12\linewidth}
	\vspace{4pt}
	\scalebox{4}{\includegraphics[width=\linewidth]{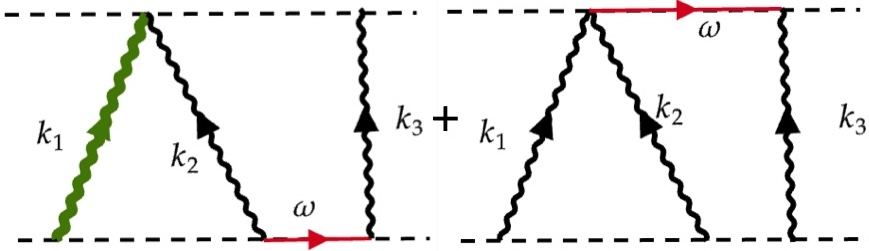}} \label{NewDiag}
 \end{minipage}&\nonumber
 \end{split}
 \end{align}
For the first diagram, we get the following contribution for the spinless part of the eikonal phase,\\
 \begin{align}
     \begin{split}
         i\chi\Big|_{\mathcal{S}^{0}}=& -i\frac{s_1^2 s_2^2 m_1^2 m_2^2}{32 m_p^6} \frac{\gamma^2(d-2)-1}{2(d-2)}\int e^{-iq\cdot b}\hat\delta(q\cdot v_1)\hat\delta(q\cdot v_2)\Bigg(\frac{\epsilon  (2 \epsilon +1)}{2 \epsilon -1}\mathbfcal{M}_{0,0;0,0,1,1,1}+\frac{q^2 (2 \epsilon +1)}{2-4 \epsilon }\mathbfcal{M}_{0,0;0,0,2,1,1}\\ &\hspace{8.5cm}+\frac{q^2 (2 \epsilon +1)}{2 \left(\gamma ^2-1\right) \epsilon }\mathbfcal{M}_{0,0;0,0,1,2,1}\Bigg)\,.
     \end{split}
 \end{align}
 Using (\ref{MasterM}) we get, 
 \begin{align}
     \begin{split}  \label{eqt20} 
     \chi\Big|_{\mathcal{S}^{0}}=-\frac{m_1^2m_2^2}{m_p^6}\Delta_{\mathfrak{27}}^{\textrm{poly}},\,\,\,\Delta_{\mathfrak{27}}^{\textrm{poly}}=\Bigg(\frac{s_{1}^2 s_{2}^2 \left(x^2+1\right) \left(x^4+1\right)}{2048 \pi ^3 |\boldsymbol{b}|^2 \left(x^2-1\right)^3}\Bigg).&
      \end{split}
 \end{align}
The contribution to the spinning part is given by,
 \begin{align}
     \begin{split}
         i\chi\Big|_{\mathcal{S}^{1}}= -\frac{\gamma s_1^2 s_2^2 m_1^2 m_2^2}{64 m_p^6}\int e^{-i q\cdot b}\hat\delta(q\cdot v_1)\hat\delta(q\cdot v_2)\,&\Bigg(\boldsymbol{\tilde h}_{118}\mathbfcal{M}_{0,0;0,0,1,1,1}+q^2 \boldsymbol{\tilde h}_{119}\mathbfcal{M}_{0,0;0,0,2,1,1} +q^2 \boldsymbol{\tilde h}_{120}\mathbfcal{M}_{0,0;0,0,1,2,1}\Bigg)\\&\Bigg(q\cdot\mathcal{S}_1\cdot v_2- \frac{1-2\epsilon}{4\,\gamma\,(1-\epsilon)}q\cdot\mathcal{S}\cdot v_1\Bigg)\,.
     \end{split}
 \end{align}
 Then using \eqref{MasterM} we get, \\
 \hfsetfillcolor{gray!10}
\hfsetbordercolor{black!150}
\begin{align}
     \begin{split}  \label{eqt21} 
 \chi\Big|_{\mathcal{S}^{1}}= \frac{m_1^2 m_2^2}{m_p^6}\Bigg(\Delta_{\mathfrak{28}}^{\textrm{poly}}(\hat{b}\cdot\mathcal{S}_1\cdot v_2)+\Delta_{\mathfrak{29}}^{\textrm{poly}} (\hat{b}\cdot\mathcal{S}_2\cdot v_1)\Bigg),\,
     \end{split}
 \end{align}
 where,
 \begin{align}
     \Delta_{\mathfrak{28}}^{\textrm{poly}}=-\Bigg(\frac{s_1^2s_2^2\, x(x^2+1)^2}{1024 \pi ^3 |\boldsymbol{b}|^3 \left(x^2-1\right)^3}\Bigg), \quad \Delta_{\mathfrak{29}}^{\textrm{poly}}= -\frac{x}{2(x^2+1)}\Delta_{\mathfrak{28}}^{\textrm{poly}}.
 \end{align}
 Proceeding as before, for the second diagram, we have only  $\mathcal{O}(\mathcal{S}^0)\,.$  contribution.
 \begin{align}
     \begin{split}  \label{eqt22} 
    \chi \Big|_{\mathcal{S}^0}=\frac{m_1m_2^3}{m_p^6}\Delta_{\mathfrak{30}}^{\textrm{poly}},\quad \Delta_{\mathfrak{30}}^{\textrm{poly}}=-\Bigg(\frac{g_{1} s_{1} s_{2}^3 x^2}{256 \pi ^3 |\boldsymbol{b}|^2 \left(x^2-1\right)^2}\Bigg)\,.
    \end{split}
 \end{align}
 Again, one also needs to add the contributions from 3PM diagrams obtained by interchanging the worldlines one and two in all the diagrams mentioned above. We will not show them explicitly as the contributions from those diagrams will be similar to those we have already computed and can be obtained easily by exchanging labels one and two. 
\\\\
\textbf{3PM eikonal phase from Chern-Simons interaction vertices:}\\
\\ Now let us focus on the dCS contribution to the eikonal vertex. The corresponding vertex term is given by,
 \begin{align}
    \begin{split}
        \mathcal{V}_{\textrm{dCS}}(k_1,k_2,q)=\epsilon^{\chi\varepsilon\mu\nu}\Bigg(k_{1\mu}k_{1\beta}k_{2\chi}k_{2\sigma}h^{\sigma}_{\nu}(-k_1)h_{\delta}^{\beta}(-k_2)-k_{1\mu}(k_1\cdot k_2)\,k_{2\chi}h^{\sigma}_{\nu}(-k_1)h_{\delta\sigma}(-k_2)
        \Bigg)\varphi(-q)\,\hat\delta^{(4)}(k_1+k_2+q)
    \end{split}
\end{align}
With this vertex, we will now proceed with the computation of the contribution of this dCS term to the eikonal phase. Below, we list all possible diagrams for the same. 
 \\\\
\textbullet $\,\,$ The following two diagrams will contribute to the eikonal phase at 3 PM involving Chern-Simons interaction  :
\begin{align}
    \begin{split}
    \hspace{-2 cm}
         \begin{minipage}[h]{0.12\linewidth}
	\vspace{4pt}
	\scalebox{2.7}{\includegraphics[width=\linewidth]{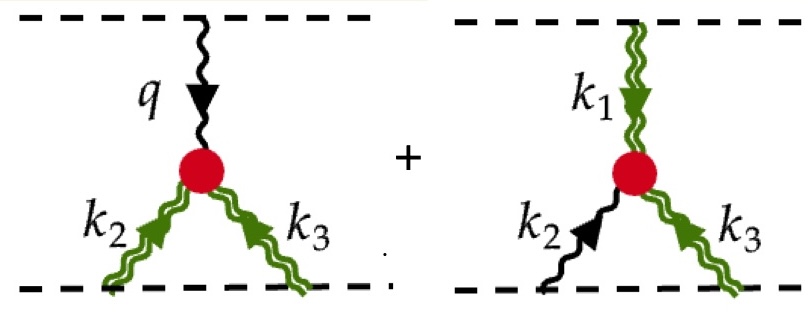}}
 \end{minipage}&\nonumber
 \end{split}
 \end{align}
Contribution to the eikonal phase from the first diagram is given by,
\begin{align}
    \begin{split}
      i  \chi=-l_{dCS}^2\frac{s_1 m_1 m_2^2}{8 m_p^6} \int e^{iq\cdot b}\frac{\hat\delta(q\cdot v_1)\hat\delta(q\cdot v_2)}{q^2}\int_{k_2}\hat\delta(k_2\cdot v_2)\frac{\mathcal{N}}{k_2^2 (k_2+q)^2}
    \end{split}
\end{align}
where the numerator takes the form,
\begin{align}
\begin{split}
    \mathcal{N}&=-\frac{i}{2}\epsilon^{\gamma\delta\tau\sigma}(q+k_2)\cdot k_2(q+k_2)_{\tau}k_{2\gamma}\,\Big(v_{2\sigma}(k_2\cdot \mathcal{S}_2)_{\delta}-[(q+k_2)\cdot \mathcal{S}_2 ]_{\sigma}\,\,v_{2\delta}\Big)\,,\\ &
  \to  \frac{i}{2}\epsilon^{\gamma\delta\tau\sigma}q^2\, q_\tau \,k_{2\gamma} v_{2\sigma}(k_2\cdot \mathcal{S}_2)_{\delta}\,\,(\textrm{ignoring the massless tadpoles})\,.
  \end{split}
\end{align}
Now, doing the $k_2$  and $q$ integral and using the anti-symmetric property of epsilon, it is easy to show that the contribution to the eikonal phase from this diagram is given by,\\
\begin{align}
\begin{split} \label{eqt23} 
 \chi\Big|_{\mathcal{S}^1}&=l_{dCS}^2\frac{s_1 m_1 m_2^2}{8 m_p^6}\frac{ 3 x\,}{64\pi (1-x^2)\,|\boldsymbol{b}|^4}\,\epsilon^{\gamma\delta\tau\sigma}v_{2\sigma}\mathcal{S}_{2\gamma\delta}\,\hat{b}_\tau\,,\\ &=\Delta_{\mathfrak{1}}^{\textrm{dCS}}\left(\mathcal{S}_2\wedge \hat b\right).
\end{split}
\end{align}
where,
\begin{align}
\Delta_{\mathfrak{1}}^{\textrm{dCS}}=l_{dCS}^2\frac{s_1 }{8 }\frac{ 3 x\,}{64\pi (1-x^2)\,|\boldsymbol{b}|^4}
\end{align}
Correspondingly, the second diagram gives,
\begin{align}
    \begin{split}
      i  \chi=-i\,l_{dCS} ^2\frac{s_2m_1 m_2^2}{m_p^6}\int_{k_1}e^{ik_1\cdot b}\frac{\hat\delta(k_1\cdot v_1)\hat\delta(k_1\cdot v_2)}{k_1^2}\int_{k_2}\hat\delta(k_2\cdot v_2)\frac{\mathcal{N}}{k_2^2(k_1+k_2)^2}
    \end{split}
\end{align}
where the numerator is given by,
\begin{align}
    \begin{split}
        \mathcal{N}=&\epsilon^{\gamma\delta\tau\sigma}\Bigg(v_1^\mu v_1^\nu+i(k_1\cdot \mathcal{S}_1^{(\mu})\,v_1^{\nu)}-\frac{1}{2}(k_1\cdot \mathcal{S}_1)^\mu (k_1\cdot \mathcal{S}_1)^{\nu}\Bigg)\,P_{\mu\nu,\rho\sigma}\Bigg(k_{1\tau}k_{2\gamma}k_1^{\chi}k_2^\rho-\eta^{\rho\chi}(k_1\cdot k_2)\Bigg)\\ &
        \times P_{\chi\varepsilon,\alpha\beta}\Bigg(v_2^\alpha v_2^\beta+i(k_2\cdot \mathcal{S}_2^{(\alpha})\,v_2^{\beta)}-\frac{1}{2}(k_2\cdot \mathcal{S}_2)^\alpha (k_2\cdot \mathcal{S}_2)^{\beta}\Bigg)\,.
    \end{split}
\end{align}
The numerator can be decomposed in different orders of spin and are given by,
\begin{align}
    \begin{split}
&\mathcal{N}\Big|_{\mathcal{S}^0}=\gamma  \left(  k_1 \cdot k_2\right)  \epsilon_{\alpha \beta\rho\sigma} k_1^\alpha k_2^\beta v_1^\rho v_2^\sigma\,, \\ &
\mathcal{N}\Big|_{\mathcal{S}^1}=\frac{1}{4}i\Big(2\gamma(k_1\cdot k_2)(k_1\cdot \mathcal{S}_1)^\mu\epsilon_{\mu \beta\rho\sigma}k_1^\beta k_2^\rho v_2^\sigma+2(k_1\cdot k_2)\epsilon_{\alpha \beta\rho\sigma} k_1^\alpha k_2^\beta v_1^\rho v_2^\sigma (k_1\cdot \mathcal{S}_1\cdot v_2)\,\\&
\hspace{1.2 cm}-2\gamma(k_1\cdot k_2) (k_2\cdot{\mathcal{S}_2})^\alpha\epsilon_{\alpha\beta\rho\sigma}{k_1}^\beta {k_2}^\rho {v_1}^\sigma+2 (k_1\cdot \mathcal{S}_2\cdot k_2)(k_2\cdot v_1)\epsilon_{\alpha\beta\rho\sigma} k_1^\alpha k_2^\beta v_1^\rho v_2^\sigma\\&\hspace{1.0 cm}+2(k_1\cdot k_2)(k_2\cdot \mathcal{S}_2\cdot v_1) \epsilon_{\alpha \beta\rho\sigma} k_1^\alpha k_2^\beta v_1^\rho v_2^\sigma \Big)\,.
\end{split}
\end{align}
Therefore, the contribution to the eikonal phase takes the form, 

\begin{align}
\begin{split}
i\chi\Big|_{\mathcal{S}^0}&
    \to\frac{\gamma}{2}\epsilon_{i j\rho\sigma}\delta_{mn}\,v_1^\rho v_2^\sigma\int_{k_1}e^{ik_1\cdot b}{\hat\delta(k_1\cdot v_1)\hat\delta(k_1\cdot v_2)k_1^i\,k_1^{m}}\underbrace{\int_{\vec k_2}\frac{k_2^j \,k_2^n}{\vec k_2^2(\vec k_2+\vec k_1)^2}}_{\propto ()\delta^{jn}+()k_1^j k_1^n}=0\,.\label{3.79}
    \end{split}
\end{align}
%
Now, let us concentrate on the spin-orbit contribution to the eikonal phase coming from the Chern-Simons interaction vertices. 
\begin{align}
    \begin{split}
        \chi\Big|_{\mathcal{S}^1}=\sum_{i=1}^{3}\chi_{\mathcal{S}^1}^{(i)}
    \end{split}
\end{align}
where,
\begin{align}
    \begin{split}
  \\
      \bullet \,\,\, \chi_{\mathcal{S}^1}^{(1)}&\sim -\frac{i}{4}\epsilon_{\alpha\beta\rho\sigma}(\mathcal{S}_2)^{\eta\tau}\,v_{1\tau}v_1^\rho v_2^\sigma \int_{k_1}e^{ik_1\cdot b}\hat\delta(k_1\cdot v_1)\hat\delta(k_1\cdot v_2)k_1^\alpha \int_{k_2}\frac{\hat\delta(k_2\cdot v_2)k_2^\beta k_{2\eta}}{k_2^2(k_2+k_1)^2}\,,\\ &
      \xrightarrow[]{\epsilon\to 0}\frac{3}{128\pi |\boldsymbol{b}|^4} \Bigg(\frac{2x}{1-x^2}\Bigg)(\overrightarrow{\mathcal{S}_2\cdot v_1})\cdot (\vec v_1\times \hat b)\,.
      \\ &
  \\
 \bullet \,\,\, \chi_{\mathcal{S}^1}^{(2)}&\sim \frac{i}{2}\epsilon_{\alpha\beta \rho\sigma}v_1^\rho v_2^\sigma v_{1\chi}(\mathcal{S}_2)_{\eta\delta}\int e^{i k_1\cdot b}\frac{\hat\delta(k_1\cdot v_1)\hat\delta(k_1\cdot v_2)\,k_1^\eta k_1^\alpha}{k_1^2}\int \frac{\hat\delta(k_2\cdot v_2)k_2^\delta\, k_2^\beta k_2^\chi}{k_2^2\,(k_2+k_1)^2}\,,\\ &
 =-\frac{3}{128\pi |\boldsymbol{b}|^4}\Bigg(\frac{2x}{1-x^2}\Bigg)\Bigg(\frac{1}{2}\vec v_1\cdot \hat b\times (\overrightarrow {v_1\cdot \mathcal{S}_2})\Bigg)=\frac{3}{256\pi |\boldsymbol{b}|^4}(\overrightarrow{\mathcal{S}_2\cdot v_1})\cdot (\vec v_1\times \hat b  )\,.\nonumber
\end{split}
\end{align}
\begin{align}
\begin{split}
\hspace{0.7 cm} \bullet \,\,\, \chi_{\mathcal{S}^1}^{(3)}&\sim -\frac{i \gamma}{2}\epsilon_{\alpha\beta \rho\sigma}v_1^\sigma \mathcal{S}_2^{\eta\alpha}\int e^{ik_1\cdot b}\frac{\hat\delta(k_1\cdot v_1)\hat\delta(k_1\cdot v_2)k_1^\beta}{k_1^2}\int \frac{\hat\delta(k_2\cdot v_2)\,k_1\cdot k_2}{k_2^2(k_2+k_1)^2}k_2^\eta k_{2\rho}\,,\\ &
 \\ &=-\frac{3}{512 \pi|\boldsymbol{b}|^4\sqrt{\gamma^2-1}}\Big(\hat b\cdot (\vec{\mathcal{S}}_2\times \vec v_1)+\gamma \hat{\mathcal{S}}_2\wedge \hat b\Big),\,\,\textrm{with},\,\,\hat{\mathcal{S}}_2\wedge \hat b=\hat b^j\,\epsilon_{jik}\,\mathcal{S}_2^{ik}
      \label{3.81}
    \end{split}
\end{align}
  Therefore, the total contribution coming from the dCS term at $\mathcal{O}(\mathcal{S}^1)$ after inserting all the vertex factors is given by,
\begin{align}
    \begin{split}
 \hspace{-0.5 cm}i\chi\Big|_{\mathcal{S}^1}&=-il_{dCS} ^2\frac{s_2m_1 m_2^2}{m_p^6}\Bigg(\frac{2x}{1-x^2}\Bigg)\Bigg[\frac{9}{256 \pi |\boldsymbol{b}|^4} (\overrightarrow{\mathcal{S}_2\cdot v_1})\cdot (\vec v_1\times \hat b)-\frac{3}{512\,\pi |\boldsymbol{b}|^4}\Big(\hat b\cdot (\vec{\mathcal{S}}_2\times \vec v_1)+\frac{1+x^2}{2x} \hat{\mathcal{S}}_2\wedge \hat b\Big)\Bigg]\\ &
 =
 \frac{m_1 m_2^2}{m_p^6}\left[\Delta_{\mathfrak{2}}^{\textrm{dCS}}(\overrightarrow{\mathcal{S}_2\cdot v_1})\cdot (\vec v_1\times \hat b)+\Delta_{\mathfrak{3}}^{\textrm{dCS}}\left(\hat b\cdot \left(\vec{\mathcal{S}}_2\times \vec v_1\right)\right)+\Delta_{\mathfrak{4}}^{\textrm{dCS}}\left(\hat{\mathcal{S}}_2\wedge \hat b\right)\right]
    \end{split} \label{eqt24} 
\end{align}\\
where, $(\overrightarrow{\mathcal{S}_2\cdot v_1})\equiv (\mathcal{S}_2\cdot v_1)^{i},\,\,\vec {\mathcal{S}}_{1,2}\equiv \mathcal{S}_{1,2}^{0i},\,\textrm{and,} \,\,\mathcal{S}_2\wedge \hat b=\epsilon^{\tau\gamma\delta\sigma}v_{2\sigma}\mathcal{S}_{\gamma\delta}\hat b_{\tau}$ 
and,
\begin{align}
    \begin{split}
\Delta_{\mathfrak{2}}^{\textrm{dCS}}=-l_{\textrm{dCS}}^2 \frac{18 xs_2}{256\pi |\boldsymbol{b}|^4(1-x^2)},\,\Delta_{\mathfrak{3}}^{\textrm{dCS}}=l_{\textrm{dCS}}^2 \frac{6 s_2 x}{512 \pi |\boldsymbol{b}|^4(1-x^2)},\,\textrm{and,}\,\Delta_{\mathfrak{4}}^{\textrm{dCS}}=l_{\textrm{dCS}}^2\frac{3s_2(1+x^2)}{512 \pi|\boldsymbol{b}|^4 (1-x^2)}.
    \end{split}
\end{align}
Let us make some comments on the computation of \eqref{3.81}:
\begin{itemize}
\item Due to the presence of anti-symmetric $\epsilon$ tensor, there are no contribution from the dCS term to the eikonal phase at $\mathcal{O}(\mathcal{S}^0)\,.$ It gives contribution only when there is a spin. 
    \item We ignore the contribution of a two-loop integral, which is proportional to  $k_1^i k_1^j$  because of the antisymmetric properties of the epsilon tensor. 
    \item Also it is evident from \eqref{eqt24}, that if we take (anti-)aligned spin, the contribution will again give zero due to the presence of binormal like term. In the case of aligned spins, we define
    \begin{align}
\mathcal{S}_1^{\mu\nu}=\frac{2\,\tilde{s}_1}{|\boldsymbol{b}|\sqrt{\gamma^2-1} }b^{[\mu}(\gamma v_1-v_2)^{\nu]} \,,\hspace{0.4 cm }\mathcal{S}_2^{\mu\nu}=\frac{2\,\tilde{s}_2}{|\boldsymbol{b}|\sqrt{\gamma^2-1} }b^{[\mu}( v_1-\gamma v_2)^{\nu]}, \,\,\, \textrm{with},\,\, {\textrm{tr\,}\,(\mathcal{S}_i\cdot\mathcal{S}_i)=-2 \,\tilde{s}_i^2}.
    \end{align}
      \end{itemize}
      \vspace{-0.89 cm}
   After using this parametrization, one can see that the \textit{3PM contribution from the Chern-Simons type term (arising at the bulk vertices) gives a vanishing contribution. }So if two black holes being scattered have the aligned spin in the dCS theory, there will be no contribution to the eikonal phase and eventually to the scattering angle.
Now, adding all spinless contributions at 3PM, we have,\\
\begin{tcolorbox}[enhanced,  height= 5 cm, width=16.8 cm, colback=brown!2!white, colframe=black!2000!brown, title=\textit{Spinless Eikonal at 3PM}, breakable]
\begin{align}  \label{dCSne3waaa}
 \chi\Big|_{\mathcal{S}^0}=&\frac{m_1m_2^3}{m_p^6}\left[\Delta_{\mathfrak{01}}^{\textrm{poly}}+\Delta_{\mathfrak{02}}^{\textrm{poly}}+\Delta_{\mathfrak{05}}^{\textrm{poly}}\right]
+\\&\frac{m_1^2 m_2^2}{m_p^6}\Big[\Delta_{\mathfrak{08}}^{\textrm{poly}}+\Delta_{\mathfrak{09}}^{\textrm{poly}}+\Delta_{\mathfrak{12}}^{\textrm{poly}}+\Delta_{\mathfrak{15}}^{\textrm{poly}}+\Delta_{\mathfrak{17}}^{\textrm{poly}}+\Delta_{\mathfrak{17}}^{\textrm{poly}}+\Delta_{\mathfrak{18}}^{\textrm{poly}}+\Delta_{\mathfrak{22}}^{\textrm{poly}}+\Delta_{\mathfrak{27}}^{\textrm{poly}}+\Delta_{\mathfrak{30}}^{\textrm{poly}}\nonumber+\\ &\hspace{1.2 cm}\log(x)\Big(\Delta_{\mathfrak{01}}^{\textrm{log}}+\Delta_{\mathfrak{02}}^{\textrm{log}}+\Delta_{\mathfrak{05}}^{\textrm{log}}+\Delta_{\mathfrak{09}}^{\textrm{log}}+\Delta_{\mathfrak{12}}^{\textrm{log}}+\Delta_{\mathfrak{15}}^{\textrm{log}}\Big)+(\log^2(x)+\textbf{Li}_2(1-x^2))\left(\Delta_{\mathfrak{03}}^{\textrm{Polylog}}\right)\Big]\,.\nonumber
    \end{align}
\end{tcolorbox}
Finally, adding all the spinning contributions to the eikonal phase at 3PM we will get,
\begin{tcolorbox}[enhanced, height= 8.9 cm, width=16.8 cm, colback=brown!2!white, colframe=black!2000!brown, title=\textit{Spinning Eikonal at 3PM }, breakable]
\begin{align}\label{dC7Snew}
 \chi\Big|_{\mathcal{S}^1}=&\frac{m_1 m_2^3}{m_p^6}\Big[(\hat b\cdot\mathcal{S}_1\cdot v_2)\Big\{ \Delta_{\mathfrak{03}}^{\textrm{poly}}+\Delta_{\mathfrak{06}}^{\textrm{poly}} \Big\}+(\hat b\cdot\mathcal{S}_2\cdot v_1)\Big\{ \Delta_{\mathfrak{04}}^{\textrm{poly}}+\Delta_{\mathfrak{07}}^{\textrm{poly}} \Big\}+\Delta_{\mathfrak{25}}^{\textrm{poly}}\Big]+ \\ &
\frac{m_1^2 m_2^2}{m_p^6}\Big[(\hat b\cdot \mathcal{S}_1\cdot v_2)\Big\{\Delta_{\mathfrak{10}}^{\textrm{poly}}+\Delta_{\mathfrak{13}}^{\textrm{poly}}+\Delta_{\mathfrak{16}}^{\textrm{poly}}+\Delta_{\mathfrak{19}}^{\textrm{poly}}+\Delta_{\mathfrak{23}}^{\textrm{poly}}+\Delta_{\mathfrak{26}}^{\textrm{poly}}+\Delta_{\mathfrak{28}}^{\textrm{poly}}\nonumber+\\ &\hspace{3cm}\log(x)\Big(\Delta_{\mathfrak{3}}^{\textrm{log}}+\Delta_{\mathfrak{06}}^{\textrm{log}}+\Delta_{\mathfrak{08}}^{\textrm{log}}+\Delta_{\mathfrak{10}}^{\textrm{log}}+\Delta_{\mathfrak{13}}^{\textrm{log}}+\Delta_{\mathfrak{16}}^{\textrm{log}}+\Delta_{\mathfrak{18}}^{\textrm{log}}\Big)\nonumber+\\ &\hspace{3cm}
(\log^2 x+\textbf{Li}_2(1-x^2))\Big(\Delta_{\mathfrak{1}}^{\textrm{Polylog}}+\Delta_{\mathfrak{4}}^{\textrm{Polylog}}+\Delta_{\mathfrak{6}}^{\textrm{Polylog}}\Big)+\boldsymbol{\mathfrak{U}_1}^{\textrm{LI}_3}\,\boldsymbol{\mathfrak{(UT)}}_3\Big\}\nonumber+\\ & \hspace{1.2cm}(\hat b\cdot \mathcal{S}_2\cdot v_1)\Big\{\Delta_{\mathfrak{11}}^{\textrm{poly}}+\Delta_{\mathfrak{14}}^{\textrm{poly}}+\Delta_{\mathfrak{20}}^{\textrm{poly}}+\Delta_{\mathfrak{21}}^{\textrm{poly}}+\Delta_{\mathfrak{24}}^{\textrm{poly}}+\Delta_{\mathfrak{29}}^{\textrm{poly}}\nonumber+\\&\hspace{3cm}\log(x)\Big(\Delta_{\mathfrak{4}}^{\textrm{log}}+\Delta_{\mathfrak{7}}^{\textrm{log}}+\Delta_{\mathfrak{11}}^{\textrm{log}}+\Delta_{\mathfrak{14}}^{\textrm{log}}+\Delta_{\mathfrak{17}}^{\textrm{log}}\Big)\nonumber+\\ &\hspace{3cm}(\log^2 x+\textbf{Li}_2(1-x^2))\Big(\Delta_{\mathfrak{2}}^{\textrm{Polylog}}+\Delta_{\mathfrak{5}}^{\textrm{Polylog}}+\Delta_{\mathfrak{7}}^{\textrm{polylog}}\Big)+\boldsymbol{\mathfrak{U}_2}^{\textrm{LI}_3}\,\boldsymbol{\mathfrak{(UT)}}_3\Big\}\Big]\nonumber+\\ &
\frac{m_1 m_2^2 }{m_p^6}\left[\Delta_{\mathfrak{1}}^{\textrm{dCS}}(\mathcal{S}_2\wedge \hat b )+\Delta_{\mathfrak{2}}^{\textrm{dCS}}(\overrightarrow{\mathcal{S}_2\cdot v_1})\cdot (\vec v_1\times \hat b)+\Delta_{\mathfrak{3}}^{\textrm{dCS}}\left(\hat b\cdot \left(\vec{\mathcal{S}}_2\times \vec v_1\right)\right)+\Delta_{\mathfrak{4}}^{\textrm{dCS}}\left(\hat{\mathcal{S}}_2\wedge \hat b\right)\right]
\,\nonumber
    \end{align} 
\end{tcolorbox}

\noindent

where, $\boldsymbol{\mathfrak{(UT)}}_3$ a polynomial consisting of a particular combination of functions with transcendental weight three (UT-3) and defined as,
\begin{align}
    \begin{split}
\boldsymbol{\mathfrak{(UT)}}_3&=54\,\textbf{Li}_3\left(x^2\right)-216\, \textbf{Li}_3\left(\frac{x+1}{1-x}\right)
\\ &+432\, \textbf{Li}_3(1-x)+216\, \textbf{Li}_3\left(\frac{x+1}{x-1}\right)+432\, \textbf{Li}_3(x+1)+216 \Big(-\textbf{Li}_2\left(\frac{x+1}{1-x}\right)-\textbf{Li}_2(1-x)\\ &+\textbf{Li}_2\left(\frac{x+1}{x-1}\right)+\textbf{Li}_2(x+1)\Big) \log \left(\frac{2}{x+1}-1\right)-54 \left(\pi ^2-2 \log ^2(x)\right) \log \left(1-x^2\right)-20 \log ^3(x)\\ &+108 i \pi  \log ^2(1-x)+108 i \pi  \log ^2(x+1)-11 \pi ^2 \log (x)-432 \zeta (3)\label{4.133a}\,,
    \end{split}
\end{align}
and, also note that one needs to add the terms ($1\leftrightarrow 2$) with \eqref{dCSne3waaa} and \eqref{dC7Snew} to get the complete answer.
On the first note,  it may seem that the polynomial $\boldsymbol{\mathfrak{(UT)}_3}$ in \eqref{4.133a} is complex, in general. However, it can be easily checked (numerically) that it is actually real in the domain $x\in(0,1)$. \textit{To the best of our knowledge, this is the first example for which, unlike GR, we get UT-3 polynomials in the Eikonal phase at 3PM.} 
Again to remind the readers, $\gamma=\frac{1+x^2}{2\,x}\,.$ As we notice that in the expression of eikonal phase, we encounter $\log(b)$ type terms, and one should put an IR cutoff, $\mu_{IR}$, as $\log(\mu_{IR}b)$ to make the argument dimensionless. However, for computing observables like impulses, one needs to take a derivative with respect to $|b|$, and hence, it is independent of the IR cutoff. Also note that, the $\Delta's$ mentioned above are defined in \cref{eqt1,eqt2,eqt3,eqt4,eqt5,eqt6,eqt7,eqt8,eqt9,eqt10,eqt11,eqt12,eqt13,eqt14,eqt15,eqt16,eqt17,eqt18,eqt19,eqt20,eqt21,eqt22,eqt23}.
  
\section{Conclusion and outlook}\label{con}
In \cite{Jakobsen:2021zvh}, the authors introduced spin degrees of freedom in WQFT by implementing  $\mathcal{N}=2$ supersymmetry as an alternative approach to describe a compact spinning object, considering terms up to quadratic-in-spin. This work motivates us to study the scattering event of two spinning black holes up to $3$PM in a gravity theory modified by a dynamical Chern-Simons (dCS) term using the WQFT approach. 
The introduction of a parity-violating Chern-Simons term in Einstein gravity, which couples to gravity via a scalar field, opens up new avenues in the study of astrophysical events, especially gravitational waves. While the Chern-Simons term does not affect the non-rotating black hole solutions, it induces a correction to the rotating metric. Therefore, it becomes important for the study of spinning black holes. 

\textbullet$\,\,$ We start with the Einstein-Hilbert action modified by the dCS term and formulate the worldline theory for a spinning particle by introducing the anti-commuting worldline vectors $\psi^{a}$. Upon modification, $\mathcal{N}=1$  supersymmetry is broken (approximate), but without loss of generality, we define the classical spin of the body $(\mathcal{S}^{ab}=-2i\psi^{[a}\,\psi^{b]})$. Although, as we have mentioned, the SSC is also approximate due to the broken (approximate) supersymmetry, it holds for small coupling (sensitivity parameters) and leading order in a spin throughout the trajectory, but the asymptotic SUSY is still there. It provides the framework to study the scattering of two spinning black holes using the WQFT formalism. 

\textbullet $\,\,$ Our computation deals with two-loop Feynman integrals. We discuss a mathematical framework \cite{Dlapa:2023hsl,Henn:2014qga} to solve multi-loop integral via the methodology of IBP reduction and differential equations and hence analyze the two-loop integrals required for our computation. \textit{We illustrate the aforementioned techniques to solve \textit{four} master integrals in higher order of $\epsilon$}.

\textbullet$\,\,$Next, we explicitly found the eikonal phase for the spinning black hole system up to 3PM order. We have shown the correction to spinning eikonal due to the presence of a scalar field upto 3PM. \textit{To the best of our knowledge, this is the first result for spinning eikonal for massless scalar field upto 3PM order. Also, we find out from the eikonal phase that the Chern-Simons interaction does not contribute to the non-spinning black holes, nor does it contribute if one considers (anti-)aligned spins.}\\\\
\textbf{\textit{$i\varepsilon$ prescription and the IR problem}:}\\\\
In the eikonal phase computation, one is expected to use the Feynman $i\varepsilon$ prescription to get sensible results. However,  a naive choice of Feynman $i\varepsilon$ prescription could result in unwanted IR divergences ($\propto \frac{1}{\epsilon}$) in the eikonal phase, which we have also encountered in our computations, specifically for those diagrams with the worldline propagator. More precisely, let's say, after IBP reduction, we face the following master integral (in some diagram): $\mathbfcal{M}_{1,1;0,0,1,1,1}$. Now, using the Feynman prescription amounts to the following substitution:
\begin{align}
    \mathbfcal{M}_{1,1;0,0,1,1,1}\to \frac{1}{2}\left( \mathbfcal{M}_{1,1;0,0,1,1,1}^{(++)}-\mathbfcal{M}_{1,1;0,0,1,1,1}^{(+-)}\right)\,. \label{6.1k}
\end{align}
Then, for some (or most) of the cases, this gives rise to IR divergences. As a possible resolution, the authors of \cite{Kim:2024svw} have proposed the notion of Magnus expansion which fixes the choice of the $i\varepsilon$ prescription for cancelling IR divergences. In general, one needs to consider the following prescription for precise IR cancellation,
\begin{align}
    \mathbfcal{M}_{1,1;0,0,1,1,1}\to \left( w_{(++)}\mathbfcal{M}_{1,1;0,0,1,1,1}^{(++)}+w_{(+-)}\mathbfcal{M}_{1,1;0,0,1,1,1}^{(+-)}\right)\,, \label{6.2k}
\end{align}
where the weight factors ($w$ 's) are determined by the Magnus expansion. However, these weight factors may depend on the underlying input, or it might happen that the prescription also may not cancel the IR divergences (i.e., the theory itself has the IR problem). So, it needs to be properly investigated, especially for the theory considered in this paper. For the moment, we have only considered the finite part of the regulated (dimensionally regularized) integral, giving rise to the eikonal phase and leaving aside a more systematic approach to cancel IR divergences for future studies.\\\\
\textbf{\textit{Comments on computation of observables from eikonal:}}\\\\
In this paper, we mainly computed the eikonal phase using WQFT. Now, the question is, how does one extract the observables, such as impulse (scattering angle), from the eikonal? A possible answer to this question has been discussed recently in \cite{Kim:2024grz,Kim:2024svw}. The eikonal phase can be understood as a generator of the canonical transformations that can transform the states from in-state to out-state in some scattering event. Any instantaneous change of observable in the classical limit can be written as,
    \begin{align}
        \begin{split}
            \Delta\mathcal{O}=\lim_{\hbar\to0}\left[\langle\psi|S^\dagger\mathcal{O}S|\psi\rangle-\langle\psi|O|\psi\rangle\right]\,.
        \end{split}
    \end{align}
 We know that in the classical limit, the S-matrix enjoys eikonal exponentiation, $S=e^{i\chi/\hbar}$. Therefore, 
 \begin{align}
     \begin{split}
         \mathcal{O}_{out}&=e^{i\chi/\hbar}\mathcal{O}_{in}e^{i\chi/\hbar}\\ &=\mathcal{O}_{in}+\frac{1}{i\hbar}[\chi,\mathcal{O}_{in}]+\frac{1}{2(i\hbar)^2}[\chi,[\chi,\mathcal{O}_{in}]]+\cdots\\ &
         \xrightarrow[]{\hbar\to 0}\mathcal{O}_{in}+\{\chi,\mathcal{O}_{in}\}_{P.B.}+\frac{1}{2}\{\chi,\{\chi,\mathcal{O}_{in}\}\}_{P.B.}+\cdots
     \end{split}
 \end{align}
 Now, in WQFT formalism, the eikonal phase can be identified as the free energy of the theory, $\chi=-i\log\,Z_{WQFT}$. Therefore, the observable, e.g impulse, can be computed as,
 \begin{align}
     \Delta p_i^\mu=\{\chi,p_i^\mu\}_{P.B.}+\frac{1}{2}\{\chi,\{\chi,p_i^\mu\}\}_{P.B.}+\cdots
 \end{align}
 In Post-Minkowskian formalism, one can expand the eikonal perturbatively in $G_N$,
 \begin{align}
     \begin{split}
         \chi=\chi_{1PM}+\chi_{2PM}+\chi_{3PM}+\cdots
     \end{split}
 \end{align}
 Therefore, the impulse can be computed as,
 \begin{align}
     \begin{split}
         &\Delta_{(1PM)}\,p_i^\mu=\{\chi_{(1PM)},p_i^\mu\},\\ &
         \Delta_{(2PM)}p_i^\mu=\{\chi_{(2PM)},p_i^\mu\}+\frac{1}{2}\{\chi_{(1PM)},\{\chi_{(1PM)},p_i^\mu\},\\ &
         \Delta_{(3PM)}p_i^\mu=\{\chi_{(3PM)},p_i^\mu\}+\frac{1}{2}\{\chi_{(2PM)},\{\chi_{(1PM)},p_i^\mu\}+\frac{1}{2}\{\chi_{(1PM)},\{\chi_{(2PM)},p_i^\mu\}+\frac{1}{3!}\{\chi_{1PM},\{\chi_{1PM},\{\chi_{1PM},p_i^\mu\}\}\},\\ &
         \vdots\label{5.5a}
     \end{split}
 \end{align}
 To compute the brackets in \eqref{5.5a}, one needs to use the fundamental Poisson bracket relations mentioned in \cite{Kim:2024grz}. We hope to use this methodology to compute the impulse for our case and report it in the near future. \\\\

Our work shows the path of several follow-up opportunities. The immediate one is to find the Eikonal phase in $\mathcal{N}=2$ supersymmetric worldline theory, which deals with the interaction of spin-1 particles or quadratic order in spin, including the scalar dipole term in the action. One can also apply the spinning WQFT formalism to find out observables for the dCS theory for higher PM order. Our work can also be naturally extended to find out the observables, e.g., Impulse $(\Delta P^{\mu}),$ the spin kick $(\Delta S^{\mu\nu})$ from the Eikonal Phase and the waveform up to 3PM. More importantly, one needs to be careful about the $i\varepsilon$ prescription and find a way to handle the IR divergences along the way of \cite{Kim:2024svw}.  We hope to report on this soon.

\section*{Acknowledgements}
The authors would like to thank Tianheng Wang for useful email correspondence. S.G (PMRF ID: 1702711) and S.P (PMRF ID: 1703278)
are supported by the Prime Minister’s Research Fellowship of the Government of India. S.G would like to thank Prof. Jan Plefka at Humboldt University, Prof. Niels Emil Bjerrum-Bohr at Niels Bohr Institute and Prof. Rafael Porto at DESY for their kind hospitality during the course of the work and their insightful comments on the previous version of the manuscript. S.G would also like to thank Gustav Jacobsen, Gregor Kälin and Christoph Dlapa for discussions and comments on related topics. A.B is supported by the Core Research Grant (CRG/2023/005112) and Mathematical Research Impact Centric Support Grant (MTR/2021/ 000490) by the Department of Science and Technology Science and Engineering Research Board (India). D.G is supported by the National Postdoctoral Fellowship (N-PDF) (PDF/2023/001163) from the Science and Engineering Research Board (SERB), Dept. of Science and Technology, Govt. of India. S.G. and S.P. would like to thank the “Strings 2025” organizing committee for
giving the opportunity to present posters and NYUAD (New York University Abu Dhabi)
for their kind hospitality during the course of the work. We also thank the participants of the (virtual) workshop “Testing Aspects of General Relativity-II”,
“Testing Aspects of General Relativity-III” and “New insights into particle physics from quantum information and gravitational waves” at Lethbridge University, Canada, funded by McDonald
Research Partnership-Building Workshop grant by McDonald Institute for valuable discussions. A.B. acknowledge the associateship program of the Indian Academy of Science, Bengaluru.
\appendix 
\section{Useful Feynman integrals} \label{D}
In the main text, we extensively use the following integrals. 
 We take the integrals from \cite{Levi:2011eq}.
\begin{eqnarray}
\label{tensorfourierindentity}
\hspace{-2 cm}I&\equiv& \int \frac{d^d \boldsymbol{k}}{(2\pi)^d}\frac{e^{i\bf{k}\cdot\bf{r}}}{({\bf{k}}^2)^\alpha}=\frac{1}{(4\pi)^{d/2}}\frac{\Gamma(d/2-\alpha)}{\Gamma(\alpha)}\Bigg(\frac{\bf{r}^2}{4}\Bigg)^{\alpha-d/2}, \label{int1}\\ 
I^i\equiv\rmint\frac{d^d\bf{k}}{(2\pi)^d}\frac{k^ie^{i\bf{k}\cdot\bf{r}}}{({\bf{k}}^2)^\alpha}&=&\frac{i}{(4\pi)^{d/2}}\frac{\Gamma(d/2-\alpha+1)}{\Gamma(\alpha)}\left(\frac{{\bf{r}}^2}{4}\right)^{\alpha-d/2-1/2}n^i, \label{int}\\
I^{ij}\equiv\rmint\frac{d^d\bf{k}}{(2\pi)^d}\frac{k^ik^je^{i\bf{k}\cdot\bf{r}}}{({\bf{k}}^2)^\alpha}&=&\frac{1}{(4\pi)^{d/2}}\frac{\Gamma(d/2-\alpha+1)}{\Gamma(\alpha)}\left(\frac{{\bf{r}}^2}{4}\right)^{\alpha-d/2-1}\left(\frac{1}{2}\delta^{ij}+(\alpha-1-d/2)n^in^j\right),\\
I^{ijl}\equiv\rmint\frac{d^d\bf{k}}{(2\pi)^d}\frac{k^ik^jk^le^{i\bf{k}\cdot\bf{r}}}{({\bf{k}}^2)^\alpha}&=&\frac{i}{(4\pi)^{d/2}}\frac{\Gamma(d/2-\alpha+2)}{\Gamma(\alpha)}\left(\frac{{\bf{r}}^2}{4}\right)^{\alpha-d/2-3/2}\nonumber\\
&&\,\,\,\,\,\,\,\,\,\,\,\,\,\,\,\,\,\,\,\,\,\,\,\,\,\,\,\,\,\,\,\
\times\left(\frac{1}{2}\left(\delta^{ij}n^l+\delta^{il}n^j+\delta^{jl}n^i\right)+(\alpha-d/2-2)n^in^jn^l\right),\\
I^{ijlm}\textstyle{\equiv\rmint\frac{d^d\bf{k}}{(2\pi)^d}\frac{k^ik^jk^lk^me^{i\bf{k}\cdot\bf{r}}}{({\bf{k}}^2)^\alpha}}&=&\textstyle{\frac{1}{(4\pi)^{d/2}}\frac{\Gamma(d/2-\alpha+2)}{\Gamma(\alpha)}\left(\frac{{\bf{r}}^2}{4}\right)^{\alpha-d/2-2}\left(\frac{1}{4}\left(\delta^{ij}\delta^{lm}+\delta^{il}\delta^{jm}+\delta^{im}\delta^{jl}\right)\right.} \\
&&
\textstyle{+\frac{\alpha-d/2-2}{2}\left(\delta^{ij}n^ln^m+\delta^{il}n^jn^m+\delta^{im}n^jn^l+\delta^{jl}n^in^m+\delta^{jm}n^in^l+\delta^{lm}n^in^j\right) }\nonumber\\
&&\left.\,\,\,\,\,\,\,\,\,\,\,\,\,\,\,\,\,\,\,\,\,\,\,\,\,\,\,\,\,\,\,\,\,\,\,\,\,\,\,\,\,\,\,\,\,\,\,\,\,\,\,\,\,\,\,\,\,\,
+(\alpha-d/2-2)(\alpha-d/2-3)n^in^jn^ln^m\right).\nonumber
\end{eqnarray}
We use the d-dimensional master formula for one-loop scalar integrals given by ,
\begin{align}
\begin{split}
   & J\equiv\rmint \frac{d^d\bf{k}}{(2\pi)^d}\frac{1}{\left[{\bf{k}}^2\right]^\alpha\left[({\bf{k}-\bf{q}})^2\right]^\beta}=  \frac{1}{(4\pi)^{d/2}}\frac{\Gamma(\alpha+\beta-d/2)}{\Gamma(\alpha)\Gamma(\beta)}\frac{\Gamma(d/2-\alpha)\Gamma(d/2-\beta)}{\Gamma(d-\alpha-\beta)}\left(q^2\right)^{d/2-\alpha-\beta}.\label{eq:1loop}
    \end{split}
\end{align}
The d-dimensional master formula for one-loop tensor integrals is taken from \cite{Levi:2011eq}. 
Similarly, one can also derive the following d-dimensional formulas for the one-loop tensor integrals: 
\begin{align}
\begin{split}
J^i\equiv\rmint \frac{d^d\bf{k}}{(2\pi)^d}\frac{k^i}{\left[{\bf{k}}^2\right]^\alpha\left[({\bf{k}-\bf{q}})^2\right]^\beta}=\frac{1}{(4\pi)^{d/2}}\frac{\Gamma(\alpha+\beta-d/2)}{\Gamma(\alpha)\Gamma(\beta)}\frac{\Gamma(d/2-\alpha+1)\Gamma(d/2-\beta)}{\Gamma(d-\alpha-\beta+1)}\\ \left(q^2\right)^{d/2-\alpha-\beta}q^i\,,\end{split}
\end{align}
\begin{align}
\begin{split}
J^{ij}\equiv\rmint \frac{d^d\bf{k}}{(2\pi)^d}\frac{k^ik^j}{\left[{\bf{k}}^2\right]^\alpha\left[({\bf{k}-\bf{q}})^2\right]^\beta}=\frac{1}{(4\pi)^{d/2}}\frac{\Gamma(\alpha+\beta-d/2-1)}{\Gamma(\alpha)\Gamma(\beta)}\frac{\Gamma(d/2-\alpha+1)\Gamma(d/2-\beta)}{\Gamma(d-\alpha-\beta+2)}\left(q^2\right)^{d/2-\alpha-\beta}\\
\times\left(\frac{d/2-\beta}{2}q^2\delta^{ij}+(\alpha+\beta-d/2-1)(d/2-\alpha+1)q^iq^j\right),\end{split}
\end{align}
\begin{align}
\begin{split}
J^{ijl}\equiv\rmint \frac{d^d\bf{k}}{(2\pi)^d}\frac{k^ik^jk^l}{\left[{\bf{k}}^2\right]^\alpha\left[({\bf{k}-\bf{q}})^2\right]^\beta}=\frac{1}{(4\pi)^{d/2}}\frac{\Gamma(\alpha+\beta-d/2-1)}{\Gamma(\alpha)\Gamma(\beta)}\frac{\Gamma(d/2-\alpha+2)\Gamma(d/2-\beta)}{\Gamma(d-\alpha-\beta+3)}\left(q^2\right)^{d/2-\alpha-\beta} \\ 
\times\left(\frac{d/2-\beta}{2}q^2\left(\delta^{ij}q^l+\delta^{il}q^j+\delta^{jl}q^i\right)\right.  \left. 
+(\alpha+\beta-d/2-1)(d/2-\alpha+2)q^iq^jq^l\right),
\end{split}
\end{align}

\begin{align}
\begin{split}
&J^{ijlm}\equiv\rmint \frac{d^d\bf{k}}{(2\pi)^d}\frac{k^ik^jk^lk^m}{\left[{\bf{k}}^2\right]^\alpha\left[({\bf{k}-\bf{q}})^2\right]^\beta}=\frac{1}{(4\pi)^{d/2}}\frac{\Gamma(\alpha+\beta-d/2-2)}{\Gamma(\alpha)\Gamma(\beta)}\\& \frac{\Gamma(d/2-\alpha+2)\Gamma(d/2-\beta)}{\Gamma(d-\alpha-\beta+4)}\left(q^2\right)^{d/2-\alpha-\beta}\,\,\,\,
\times\left(\frac{(d/2-\beta)(d/2-\beta+1)}{4}q^4\left(\delta^{ij}\delta^{lm}+\delta^{il}\delta^{jm}+\delta^{jl}\delta^{im}\right)\right.\\
& \textstyle{
+(\alpha+\beta-d/2-2)(d/2-\alpha+2)\frac{d/2-\beta}{2}q^2 
\times\left(\delta^{ij}q^lq^m+\delta^{il}q^jq^m+\delta^{im}q^jq^l+\delta^{jl}q^iq^m+\delta^{jm}q^iq^l+\delta^{lm}q^iq^j\right)}\\
&\left.\,\,\,\,\,\,\,\,\,
+(\alpha+\beta-d/2-2)(\alpha+\beta-d/2-1)(d/2-\alpha+2)(d/2-\alpha+3)\right.
\left.
\times q^iq^jq^lq^m\right).
\end{split}
\end{align}
More details of  Feynman integrals can be found in \cite{smirnov}.
Another important two loop integral that we use to compute in the potential region is given by,
\begin{align}
    \begin{split}
        \mathbfcal{M}^{(++)}=\int_{\ell_{1},\ell_{2}}\frac{1}{(\ell_{1}\cdot n+i\varepsilon)(\ell_{2}\cdot n+i\varepsilon)}\frac{1}{\ell_{1}^2\, \ell_{2}^2\, (\ell_{1}+\ell_{2}-q)^2}\,.\label{intextra}
    \end{split}
\end{align}
Now, introducing an auxiliary delta functio,n we can write the above integral as \cite{Saotome:2012vy, Parra-Martinez:2020dzs},
\begin{align}
    \begin{split}
        \mathbfcal{M}^{(++)}=\frac{1}{3!}\int_{\ell_{1},\ell_{2},l_3}\Bigg(\frac{1}{(\ell_{1}\cdot n+i\varepsilon)(\ell_{2}\cdot n+i\varepsilon)}+\textrm{perms.}\Bigg)\frac{\delta^{(d)}(l_{123}-q)}{\ell_{1}^2\, \ell_{2}^2\, l_3^2}\,.
    \end{split}
\end{align}
Now using the following $\delta$-function identities,
\begin{align}
    \begin{split}
        &\delta(z_1+z_2+z_3)\Bigg(\frac{1}{z_1+i\varepsilon}\frac{1}{z_{12}+i\varepsilon}+\textrm{perms.}\Bigg)=(2\pi i)^2 \delta(z_1)\delta(z_2)\delta(z_3)\,,\\ &
\delta(z_1+z_2+z_3)\Bigg(\frac{1}{z_1+i\varepsilon}\frac{1}{z_{2}+i\varepsilon}+\textrm{perms.}\Bigg)=2(2\pi i)^2 \delta(z_1)\delta(z_2)\delta(z_3),
    \end{split}
\end{align}
one can rewrite the integral as,
\begin{align}
    \begin{split}
        \mathbfcal{M}^{(++)}=\frac{(2\pi i)^2}{6}\int_{\boldsymbol{\ell_1},\boldsymbol{\ell_2}}\frac{1}{\boldsymbol{\ell_1}^2\,\boldsymbol{\ell_2}^2\,(\boldsymbol{\ell_1}+\boldsymbol{\ell_2}-\boldsymbol{q})^2}\sim -\frac{\Gamma^3(-\epsilon)\Gamma(1+2\epsilon)}{\Gamma(-3\epsilon)}\,.
    \end{split}
\end{align}
In principle, the integral \eqref{intextra}, can be computed directly using Schwinger parametrization, but the above-mentioned approach is useful when we go beyond two loops.
\section{Boundary integrals in potential mode relevant for solving the differential equation}\label{AppB}
\textbf{Computation of boundary integrals in potential region:}\\ Now we compute the boundary integrals relevant to the four master integrals as mentioned previously for our subsequent computations.
We start with the master with two linear propagators:
$\,\,$\begin{align}
    \begin{split}
        \mathbfcal{M}_{1,1;0,0,1,1,1}^{(-+)}\Big|_{v_{\infty}\to 0}^{\textrm{pot.}}&=-v_{\infty}^{-2}\int_{\boldsymbol{\ell_1},\boldsymbol{\ell_2}}\frac{1}{(-\boldsymbol{\ell_1}\cdot \boldsymbol{n}+i\varepsilon)(\boldsymbol{\ell_2}\cdot \boldsymbol{n}+i\varepsilon)(\boldsymbol{\ell_1}+\boldsymbol{\ell_2}-\boldsymbol{q})^2(\boldsymbol{q}-\boldsymbol{\ell_1})^2(\boldsymbol{q}-\boldsymbol{\ell_2})^2}\,.\label{2.30}
    \end{split}
\end{align}
In \eqref{2.30}, the $i\varepsilon$  prescription is very important, and one needs to carefully take care of it. Our goal is to make relation with $\mathcal{M}^{(++)}$ with $\mathcal{M}^{+-}$. Now doing the following change of variable ($\boldsymbol{q}-\boldsymbol{\ell_i}\to \boldsymbol{\ell_i}$) we left with 
\begin{align}
    \begin{split}
        \mathbfcal{M}_{1,1;0,0,1,1,1}^{(-+)}\Big|_{v_{\infty}\to 0}^{\textrm{pot.}}&=-v_{\infty}^{-2}\int_{\boldsymbol{\ell_1},\boldsymbol{\ell_2}}\frac{1}{(\boldsymbol{\ell_1}\cdot \boldsymbol{n}+i\varepsilon)(-\boldsymbol{\ell_2}\cdot \boldsymbol{n}+i\varepsilon)(\boldsymbol{\ell_1}+\boldsymbol{\ell_2}-\boldsymbol{q})^2\,\boldsymbol{\ell_1}^2\,\boldsymbol{\ell_2}^2}\,.
    \end{split}
\end{align}
Now again we do the following relabelling: $\boldsymbol{\ell_1}\to \boldsymbol{\ell_1}+\boldsymbol{\ell_2}-\boldsymbol{q}$, $\boldsymbol{\ell_2}\to -\boldsymbol{\ell_2}$ and, ${\boldsymbol{q}\to -\boldsymbol{q}}\,.$ The we get,
\begin{align}
    \begin{split}
        \mathbfcal{M}_{1,1;0,0,1,1,1}^{(-+)}\Big|_{v_{\infty}\to 0}^{\textrm{pot.}}=-v_\infty^{-2}\int_{\boldsymbol{\ell_1},\boldsymbol{\ell_2}}\frac{1}{(\boldsymbol{\ell_1}\cdot \boldsymbol{n}+\boldsymbol{\ell_2}\cdot \boldsymbol{n}+i\varepsilon)(\boldsymbol{\ell_2}\cdot \boldsymbol{n}+i\varepsilon)(\boldsymbol{\ell_1}+\boldsymbol{\ell_2}-\boldsymbol{q})^2\,\boldsymbol{\ell_1}^2\,\boldsymbol{\ell_2}^2}\,.
    \end{split}
\end{align}
Now interchanging $\boldsymbol{\ell_{1}}\leftrightarrow \boldsymbol{\ell}_2$ and adding we will get,
\begin{align}
    \begin{split}
        & 2\mathbfcal{M}_{1,1;0,0,1,1,1}^{(-+)}\Big|_{v_{\infty}\to 0}^{\textrm{pot.}}=-v_\infty^{-2}\int_{\boldsymbol{\ell_{1},\ell_{2}}}\frac{1}{(\boldsymbol{\ell_1}\cdot \boldsymbol{n}+i\varepsilon)(\boldsymbol{\ell_2}\cdot \boldsymbol{n}+i\varepsilon)(\boldsymbol{\ell_1}+\boldsymbol{\ell_2}-\boldsymbol{q})^2\,\boldsymbol{\ell_1}^2\,\boldsymbol{\ell_2}^2}\equiv\mathbfcal{M}_{1,1;0,0,1,1,1}^{(++)}\Big|_{v_{\infty}\to 0}^{\textrm{pot.}}\,
    \end{split}
\end{align}
and the integral is given by,
\begin{align}
    \mathbfcal{M}^{(++)}_{1,1;0,0,1,1,1}=\frac{1}{v_\infty^2}\frac{1}{(4\pi)^{2-2\epsilon}}\frac{\Gamma^3(-\epsilon)\Gamma(1+2\epsilon)}{3\Gamma(-3\epsilon)}\,.
\end{align}
In the main text, we mainly face the following form of two-loop boundary integral.
\begin{align}
    &\mathbfcal{M}^{(\pm\pm)\textrm{pot.}}_{\alpha_1,\alpha_2;\beta_1,\cdots \beta_5}=\int_{\boldsymbol{\ell_2},\boldsymbol{\ell_2}}\frac{1}{(\pm \boldsymbol{\ell_2}\cdot \boldsymbol{n}+i\varepsilon)^{\alpha_1}(\pm \boldsymbol{\ell_2}\cdot \boldsymbol{n}+i\varepsilon)^{\alpha_2}\,\boldsymbol{D}_1^{\beta_1}\cdots \boldsymbol{D}_5^{\beta_5}}
\end{align}
To compute the integral, one can further use the IBP reduction procedure with the following 7 basis functions
\begin{align}
    \{D\}: \{\boldsymbol{\ell_1}\cdot \boldsymbol{n},\,\boldsymbol{\ell_2}\cdot \boldsymbol{n},\,\boldsymbol{\ell_2}^2,\,\boldsymbol{\ell_2}^2,\,(\boldsymbol{\ell_2}+\boldsymbol{\ell_2}-\boldsymbol{q})^2,\,(\boldsymbol{\ell_2}-\boldsymbol{q})^2,\,(\boldsymbol{\ell_2}-\boldsymbol{q})^2\}\,.
\end{align}
After doing the IBP reduction one will get 15 unique sectors with 22 master integrals.
Particularly, we need the following two loop boundary integrals,
\begin{align}
    \begin{split}
        &\mathbfcal{M}^{(\pm\pm)\textrm{pot.}}_{0,0;1,1,1,1,1}=-\frac{2 \left(1-36 \epsilon ^2\right)}{\left(\boldsymbol q^2\right)^2 (2 \epsilon +1)^2}\mathbfcal{M}^{(\pm\pm)\textrm{pot.}}_{0,0;0,0,1,1,1}-\frac{4 \epsilon }{\boldsymbol q^2 (2 \epsilon +1)}\mathbfcal{M}^{(\pm\pm)\textrm{pot.}}_{0,0;1,1,0,1,1}\,,\\ &
        \mathbfcal{M}^{(\pm\pm)\textrm{pot.}}_{0,0;1,1,2,1,1}=\frac{12 (\epsilon +1) (2 \epsilon -1) (36 \epsilon^2 -1) }{\left(\boldsymbol q^2\right)^3 (2 \epsilon +1) (2 \epsilon +3)^2}\mathbfcal{M}^{(\pm\pm)\textrm{pot.}}_{0,0;0,0,1,1,1}+\frac{8 \epsilon }{\left(\boldsymbol q^2\right)^2 (2 \epsilon +3)}\mathbfcal{M}^{(\pm\pm)\textrm{pot.}}_{0,0;1,1,0,1,1}\,,\\ &
  \mathbfcal{M}^{(\pm\pm)\textrm{pot.}}_{0,0;0,0,2,1,1}=      \frac{2 \epsilon  (6 \epsilon -1)}{\boldsymbol q^2 (2 \epsilon +1)}\mathbfcal{M}^{(\pm\pm)\textrm{pot.}}_{0,0;0,0,1,1,1}=\mathbfcal{M}^{(\pm\pm)\textrm{pot.}}_{0,0;0,0,1,2,1}
    \end{split}
\end{align}
where,
\begin{align}
    \begin{split}
        &\mathbfcal{M}^{(\pm\pm)\textrm{pot.}}_{0,0;0,0,1,1,1}= \frac{1}{(4\pi)^{3-2\epsilon}}\frac{\Gamma(1/2-\epsilon)^3\Gamma(2\epsilon)}{\Gamma(3/2-3\epsilon)}\frac{1}{\boldsymbol{q}^{4\epsilon}}\,,\\ &
        \mathbfcal{M}^{(\pm\pm)\textrm{pot.}}_{0,0;1,1,0,1,1}=\frac{1}{(4\pi)^{3-2\epsilon}}\frac{\Gamma^4(1/2-\epsilon)\Gamma^2(1/2+2\epsilon)}{\Gamma^2(1-2\epsilon)}\,.
    \end{split}
\end{align}
Apart from that we will face the following integrals,
\begin{align}
    \begin{split}
        &
        \mathbfcal{M}^{(\pm\pm)\textrm{pot.}}_{0,0;0,1,1,0,1}=0\,,\\ &
\mathbfcal{M}^{(\pm\pm)\textrm{pot.}}_{0,1;0,0,1,1,1}=-i \frac{\sqrt{\pi}}{(4\pi)^3}\frac{\Gamma(1/2-2\epsilon)\Gamma^2(1/2-\epsilon)\Gamma(-\epsilon)\Gamma(1/2+2\epsilon)}{\Gamma(1/2-3\epsilon)\Gamma(1-2\epsilon)}\,,\\ &    \mathbfcal{M}^{(\pm\pm)\textrm{pot.}}_{0,2;1,0,1,1,0}=-\frac{1}{(4\pi)^3}\frac{4\epsilon\Gamma(2\epsilon)\Gamma^2(-2\epsilon)\Gamma(1/2-\epsilon)\Gamma(1/2+\epsilon)}{\Gamma(-4\epsilon)}\,,\\ &
\mathbfcal{M}^{(\pm\pm)\textrm{pot.}}_{1,1;1,1,0,1,1}=-\frac{\pi}{(4\pi)^3}\frac{\Gamma^4(-\epsilon)\Gamma^2(\epsilon+1)}{\Gamma^2(-2\epsilon)}\,,\\ &
\mathbfcal{M}^{(\pm\pm)\textrm{pot.}}_{0,1;1,0,1,1,0}=-i\frac{\pi}{(4\pi)^3}\frac{\Gamma(\epsilon)\Gamma(1/2-2\epsilon)\Gamma(1/2+2\epsilon)}{\Gamma(1-\epsilon)}\,,\\ &
\mathbfcal{M}^{(++)\textrm{pot.}}_{1,1;0,0,1,1,1}=2\mathbfcal{M}^{(+-)\textrm{pot.}}_{1,1;0,0,1,1,1}=\frac{1}{(4\pi)^{2-2\epsilon}}\frac{\Gamma^3(-\epsilon)\Gamma(1+2\epsilon)}{3\Gamma(-3\epsilon)}\,.
    \end{split}
\end{align}
\section{$ A(x,\epsilon)$ in $\epsilon$ factorized form} \label{Ca}
\vspace{-0.5 cm}

As mentioned in the main text, the $A(x,\epsilon)$ matrix constructed from the coupled differential equations required for solving the master integrals reduces to an  $\epsilon$ factorized form ($\epsilon\,\mathbb{S}(x)$) in UT basis. Here, we explicitly  write down the expression for $\mathbb{S}(x)$, which we have used to solve the master integrals. 

\begin{equation}
    \hspace{-1.8 cm}   \mathbb{S}(x)=\left(
\begin{array}
{cccccccccccccccc}
 -\frac{18 x^2+9}{2 x-2 x^3} & -\frac{2 x^2+3}{4 x-4 x^3} & -\frac{6 x}{x^2-1} & 0 & 0 & 0 & 0 & 0 & 0 & 0 & 0 & 0 & 0 & 0 & 0 & 0 \\
 \frac{6 x^2+15}{x-x^3} & \frac{6 x^2+5}{2 x-2 x^3} & \frac{12 x}{x^2-1} & 0 & 0 & 0 & 0 & 0 & 0 & 0 & 0 & 0 & 0 & 0 & 0 & 0 \\
 \frac{6}{x} & -\frac{1}{3 x} & -\frac{2}{x} & 0 & 0 & 0 & 0 & 0 & 0 & 0 & 0 & 0 & 0 & 0 & 0 & 0 \\
 0 & 0 & 0 & \frac{2 (x-4) x+2}{x \left(x^2-1\right)} & 0 & 0 & 0 & 0 & 0 & 0 & 0 & 0 & 0 & 0 & 0 & 0 \\
 0 & 0 & 0 & 0 & \frac{2 \left(x^2+1\right)}{x-x^3} & 0 & 0 & 0 & 0 & 0 & 0 & 0 & 0 & 0 & 0 & 0 \\
 0 & 0 & 0 & 0 & 0 & \frac{2 \left(x^2+1\right)}{x-x^3} & 0 & 0 & 0 & 0 & 0 & 0 & 0 & 0 & 0 & 0 \\
 0 & 0 & 0 & 0 & 0 & 0 & 0 & 0 & 0 & 0 & 0 & 0 & 0 & 0 & 0 & 0 \\
 0 & 0 & 0 & -\frac{1}{x} & 0 & 0 & 0 & 0 & 0 & 0 & 0 & 0 & 0 & 0 & 0 & 0 \\
 0 & 0 & 0 & 0 & 0 & -\frac{1}{x} & 0 & 0 & 0 & 0 & 0 & 0 & 0 & 0 & 0 & 0 \\
 0 & 0 & 0 & 0 & \frac{336}{x^3-x} & 0 & 0 & 0 & 0 & 0 & 0 & 0 & 0 & 0 & 0 & 0 \\
 \frac{211-286 x^2}{15 x-15 x^3} & \frac{117-242 x^2}{30 x-30 x^3} & -\frac{97-122 x^2}{5 x-5 x^3} & 0 & \frac{84}{x-x^3} & 0 & 0 & 0 & 0 & 0 & -\frac{1-5 x^2}{x-x^3} & -\frac{2}{3 x} & 0 & 0 & 0 & 0 \\
 -\frac{3983-3481 x^2}{20 x-20 x^3} & \frac{2687 x^2+231}{40 x-40 x^3} & \frac{1821-4371 x^2}{20 x-20 x^3} & 0 & \frac{126}{x-x^3} & 0 & \frac{24}{x^2-1} & 0 & 0 & 0 & \frac{3-75 x^2}{2 x-2 x^3} & \frac{1-5 x^2}{x-x^3} & 0 & 0 & 0 & 0 \\
 0 & 0 & 0 & -\frac{16 (x (7 x+6)+7)}{5 x \left(x^2-1\right)} & 0 & \frac{64}{x^2-1} & 0 & 0 & 0 & 0 & 0 & 0 & -\frac{4}{x^2-1} & 0 & 0 & 0 \\
 0 & 0 & 0 & 0 & 0 & 0 & 0 & 0 & 0 & 0 & 0 & 0 & 0 & 0 & 0 & 0 \\
 \frac{5}{x} & \frac{25}{6 x} & -\frac{5}{x} & 0 & 0 & 0 & 0 & 0 & 0 & 0 & 0 & 0 & 0 & 0 & 0 & 0 \\
 0 & 0 & 0 & 0 & 0 & 0 & 0 & 0 & 0 & 0 & 0 & 0 & 0 & 0 & 0 & 0 
\end{array}\right)\,.
\end{equation}
\vspace{7 cm}
\section{Iterative UT integrals}\label{D1}
\begin{align}
    \begin{split}
     &   \boldsymbol{\mathscr{J}}(\{-1,-1,0\},x)\equiv  -\textbf{Li}_3\left(\frac{1}{x+1}\right)+\frac{1}{6} \log ^3(x+1)-\frac{1}{12} \pi ^2 \log (x+1)+\frac{7 \zeta (3)}{8},\\ &
     \boldsymbol{\mathscr{J}}(\{-1,0,0\},x)\equiv -\textbf{Li}_3(-x)+\textbf{Li}_2(-x) \log (x)+\frac{1}{2} \log (x+1) \log ^2(x)-\frac{3 \zeta (3)}{4},\\ &
    \boldsymbol{\mathscr{J}}(\{-1,1,0\},x)\equiv  \textbf{Li}_3\left(\frac{x+1}{2}\right)+\textbf{Li}_3\left(\frac{1+x}{x}\right)+\textbf{Li}_3(x)-\textbf{Li}_3\left(\frac{x+1}{2 x}\right)+\log (2) \textbf{Li}_2\left(\frac{x+1}{2}\right)\\ &-\log (2) \textbf{Li}_2(x)-\log (2) \textbf{Li}_2\left(\frac{x+1}{2 x}\right)-\textbf{Li}_2\left(\frac{x+1}{2}\right) \log (x+1)+\textbf{Li}_2\left(\frac{1+x}{x}\right) \log \left(\frac{2 x}{x+1}\right)-\textbf{Li}_2(1-x) \log (x+1)\\ &-\textbf{Li}_2(x) \log (x)-\textbf{Li}_2\left(\frac{x+1}{2 x}\right) \log (x)+\textbf{Li}_2\left(\frac{x+1}{2 x}\right) \log (x+1)-\log ^2(2) \log (x)-\frac{1}{2} \log (2) \log ^2(x)\\ &+\log (2) \log (x) \log (x+1)-\log (1-x) \log (x) \log (x+1)-\frac{15 \zeta (3)}{8}+\frac{1}{2} i \pi  \log ^2(2)-\frac{1}{12} \pi ^2 \log (2),\\ &
      \boldsymbol{\mathscr{J}}(\{0,-1,0\},x)\equiv 2 \textbf{Li}_3(-x)-\textbf{Li}_2(-x) \log (x)+\frac{1}{12} \pi ^2 \log (x)+\frac{3 \zeta (3)}{2},\\ &
       \boldsymbol{\mathscr{J}}(\{0,0,0\},x)\equiv \frac{\log ^3(x)}{6},\\ &
       \boldsymbol{\mathscr{J}}(\{0,1,0\},x)\equiv 2 \textbf{Li}_3(x)-\textbf{Li}_2(x) \log (x)-\frac{1}{6} \pi ^2 \log (x)-2 \zeta (3),\\ &
   \boldsymbol{\mathscr{J}}(\{1,0,0\},x)\equiv     -\textbf{Li}_3(x)+\textbf{Li}_2(x) \log (x)+\frac{1}{2} \log (1-x) \log ^2(x)+\zeta (3),\\ &
   \boldsymbol{\mathscr{J}}(\{1,1,0\},x)\equiv-\textbf{Li}_3(1-x),\\ &
   \boldsymbol{\mathscr{J}}(\{1,-1,0\},x)\equiv-i \pi  \textbf{Li}_2\left(\frac{x}{x-1}\right)+i \pi  \textbf{Li}_2\left(\frac{2 x}{x-1}\right)+i \pi  \textbf{Li}_2\left(\frac{1}{x+1}\right)+\textbf{Li}_3\left(\frac{1-x}{2}\right)+\textbf{Li}_3\left(\frac{x}{x-1}\right)-\textbf{Li}_3\left(\frac{2 x}{x-1}\right)\\ &+\textbf{Li}_3(-x)-\log (2) \textbf{Li}_2\left(\frac{x}{x-1}\right)+\log (2) \textbf{Li}_2\left(\frac{2 x}{x-1}\right)-\textbf{Li}_2\left(\frac{2 x}{x-1}\right) \log (1-x)-\textbf{Li}_2\left(\frac{x}{x-1}\right) \log (x)\\ &+\textbf{Li}_2\left(\frac{1-x}{2}\right) \left(\log \left(-\frac{2}{x-1}\right)+i \pi \right)+\textbf{Li}_2\left(\frac{x}{x-1}\right) \log (1-x)+\textbf{Li}_2\left(\frac{2 x}{x-1}\right) \log (x)+\textbf{Li}_2\left(\frac{1}{x+1}\right) \log (1-x)\\ &+\textbf{Li}_2(-x) (-\log (2 x)-2 i \pi )-\log ^2(2) \log (1-x)+\frac{1}{2} \log (2) \log ^2(1-x)+\frac{1}{2} i \pi  \log ^2(x+1)+\frac{1}{2} \log ^2(x+1) \log (1-x)\\ &-i \pi  \log (2) \log (1-x)-i \pi  \log (x) \log (x+1)-\frac{1}{12} \pi ^2 \log (1-x)-\log (x) \log (x+1) \log (1-x)+\frac{3 \zeta (3)}{4}-\frac{i \pi ^3}{4}\\ &+\frac{\log ^3(2)}{3}+\frac{1}{2} i \pi  \log ^2(2)-\frac{1}{4} \pi ^2 \log (2).
    \end{split}
\end{align}
\newpage
\section{$\boldsymbol{\tilde{h}}$ coefficients} \label{AppE}
Note that, $\gamma=\frac{x^2+1}{2 x}\,.$
\begin{align}
    \begin{split}
   &  \boldsymbol{\tilde h_1}=4,\,\,
      \boldsymbol{\boldsymbol{\tilde h_2}}   =0,
     \boldsymbol{\tilde h}_3=\frac{72 \epsilon ^2-2}{12 \left(\gamma ^2-1\right)^2},\,\,
        \boldsymbol{\tilde h}_4=\frac{6 \gamma ^2+2 \epsilon -3}{12 \left(\gamma ^2-1\right)},\\& \boldsymbol{\tilde h}_5=\frac{1}{24 \left(\gamma ^2-1\right)^2 (\epsilon -1) \epsilon }\Bigg[6 \gamma ^4+72 \gamma ^2 \epsilon ^4-10 \gamma ^2-36 \left(\gamma ^4+\gamma ^2-1\right) \epsilon ^3\\&+\left(-36 \gamma ^6+150 \gamma ^4-116 \gamma ^2\right) \epsilon ^2+\left(36 \gamma ^6-120 \gamma ^4+110 \gamma ^2-25\right) \epsilon +4\Bigg],
       \boldsymbol{\tilde h}_6=-\frac{4 \gamma ^2 \epsilon ^2+2 \left(\gamma ^2+1\right) \epsilon -4 \gamma ^2+1}{24 \left(\gamma ^2-1\right)^2 (\epsilon -1)}\,,\\ &
       \boldsymbol{\tilde h}_7=\frac{-6 \gamma ^3 (\epsilon  (4 \epsilon -3)+1)-6 \gamma ^2 \epsilon +\gamma  \left(2 \epsilon  \left(\epsilon  \left(-24 \epsilon ^2+22 \epsilon +21\right)-14\right)+7\right)+3 \epsilon  (6 \epsilon +1)}{24 \left(\gamma ^2-1\right)^2 (\epsilon -1) \epsilon }\,,\\ &
   \boldsymbol{\tilde h}_8    =\frac{\gamma  \left(8 \epsilon ^3-13 \epsilon +4\right)}{48 \left(\gamma ^2-1\right) (\epsilon -1) (3 \epsilon -1)}\,,
    \boldsymbol{\tilde h}_9 =\frac{\gamma  \left(-9 \gamma ^2-2 (\epsilon -5) \epsilon +8\right)}{6 \left(\gamma ^2-1\right)^2}\,,
    \\&
    \boldsymbol{\tilde h}_{10}  =-\frac{\gamma}{24 \left(\gamma ^2-1\right)^2 (3 \epsilon -1)}\Bigg[8 \gamma ^4 \left(18 \epsilon ^2-21 \epsilon +5\right) \epsilon +3 \gamma ^2 \left(48 \epsilon ^4-208 \epsilon ^3+216 \epsilon ^2-63 \epsilon +4\right)\\&-288 \epsilon ^5+216 \epsilon ^4+464 \epsilon ^3-558 \epsilon ^2+173 \epsilon -14\Bigg]\,,
    \\ &\boldsymbol{\tilde h}_{11}=\frac{-2 \gamma ^4+72 \gamma ^2 \epsilon ^4+2 \gamma ^2-36 \left(\gamma ^4+\gamma ^2-1\right) \epsilon ^3+\left(12 \gamma ^6+6 \gamma ^4-20 \gamma ^2\right) \epsilon ^2+\left(-12 \gamma ^6+32 \gamma ^4-18 \gamma ^2-1\right) \epsilon }{24 \left(\gamma ^2-1\right)^2 (\epsilon -1) \epsilon}\,,\\ &
   \boldsymbol{\tilde h}_{12}=\frac{-8 \gamma ^4+4 \gamma ^2 \epsilon ^2+8 \gamma ^2+\left(8 \gamma ^4-6 \gamma ^2+2\right) \epsilon -3}{24 \left(\gamma ^2-1\right)^2 (\epsilon -1)}\,,\\&
  \boldsymbol{\tilde h}_{13}  =\frac{\gamma  \left(4 \gamma ^3 (\epsilon -1) (4 \epsilon -1)-12 \gamma ^2 \epsilon +2 \gamma  (\epsilon -1) (4 \epsilon -1) (4 \epsilon  (3 \epsilon +1)-3)+\epsilon  (42 \epsilon +5)\right)}{48 \left(\gamma ^2-1\right)^2 (\epsilon -1) \epsilon }\,,\\ &
 \boldsymbol{\tilde h}_{14}=\frac{\gamma  \left(8 \gamma  (\epsilon -1) \left(\gamma ^2 (8 \epsilon -2)-2 \epsilon  (\epsilon +5)+3\right)-\epsilon \right)}{96 \left(\gamma ^2-1\right) (\epsilon -1) (3 \epsilon -1)}\,,
  \\ &
  \boldsymbol{\tilde h}_{15} =\frac{\gamma  \left(4 \gamma ^4 \epsilon ^2+\gamma ^2 (2 \epsilon  (\epsilon  (12 \epsilon -11)-3)+2)+\epsilon  (16 \epsilon  (\epsilon  (2-3 \epsilon )+1)-1)-1\right)}{12 \left(\gamma ^2-1\right)^2 (\epsilon -1) \epsilon }\,,\\ &
  \boldsymbol{\tilde h}_{16}=\frac{\gamma  \left(\gamma ^4 (4-20 \epsilon )+4 \gamma ^2 (2 \epsilon  (\epsilon +6)-3)-\epsilon  (8 \epsilon +27)+8\right)}{24 \left(\gamma ^2-1\right)^2 (3 \epsilon -1)}\,,\\&\boldsymbol{\tilde h}_{17}=\frac{1}{8 \left(\gamma ^2-1\right)^2 (\epsilon -1)^2 (\epsilon +1) (2 \epsilon +1)}\Bigg[-4 \gamma ^6 \left(108 \epsilon ^5-132 \epsilon ^4-109 \epsilon ^3+142 \epsilon ^2+\epsilon -10\right)\\&+2 \gamma ^4 \left(432 \epsilon ^6-432 \epsilon ^5-792 \epsilon ^4+308 \epsilon ^3+539 \epsilon ^2-16 \epsilon -39\right)\\&+2 \gamma ^2 \left(576 \epsilon ^5+156 \epsilon ^4-596 \epsilon ^3-317 \epsilon ^2+20 \epsilon +21\right)+216 \epsilon ^4+324 \epsilon ^3+138 \epsilon ^2-9 \epsilon -4\Bigg]\,,\nonumber
       \end{split}
\end{align}
\begin{align}
    \begin{split}
&\boldsymbol{\tilde h}_{18}=-\frac{\left(2 \gamma ^2 (\epsilon -1)+1\right) \left(4 \gamma ^4 \left(\epsilon ^2-1\right)-2 \gamma ^2 \left(\epsilon  \left(6 \epsilon ^2+\epsilon -3\right)-2\right)-3 \epsilon  (2 \epsilon +1)+1\right)}{16 \left(\gamma ^2-1\right) (\epsilon -1)^2 (\epsilon +1)}\,,\\&\boldsymbol{\tilde h}_{19}=-\frac{1}{96 (\gamma -1)^2 (\gamma +1)^2 (\epsilon -1)^2 \epsilon  (\epsilon +1) (2 \epsilon +1)}\Bigg[2 \gamma ^5 \epsilon  \big(1152 \epsilon ^7+4312 \epsilon ^6-4644 \epsilon ^5-7706 \epsilon ^4+3361 \epsilon ^3\\&+3347 \epsilon ^2+107 \epsilon +23\big)+2 \gamma ^3 \big(1728 \epsilon ^9-16416 \epsilon ^8+1912 \epsilon ^7+26492 \epsilon ^6+3490 \epsilon ^5-12697 \epsilon ^4-5305 \epsilon ^3\\&+282 \epsilon ^2+35 \epsilon -1\big)+192 \gamma ^2 (\epsilon -1)^2 \epsilon  \left(72 \epsilon ^5+108 \epsilon ^4-70 \epsilon ^3-99 \epsilon ^2+13 \epsilon +6\right)+\gamma  \big(-10368 \epsilon ^9+11520 \epsilon ^8\\&-1888 \epsilon ^7-20272 \epsilon ^6+8000 \epsilon ^5+14004 \epsilon ^4+3092 \epsilon ^3-630 \epsilon ^2-93 \epsilon +1\big)+96 (\epsilon -1)^2 \epsilon  \big(72 \epsilon ^4+172 \epsilon ^3+82 \epsilon ^2\\&-27 \epsilon -9\big)\Bigg]\,,\\&\boldsymbol{\tilde h}_{20}=\frac{1}{96 \left(\gamma ^2-1\right)^2 (\epsilon -1)^2 \epsilon  (\epsilon +1)}\Bigg[\gamma  \big(12 \gamma ^8 \epsilon ^3 \left(\epsilon ^2-1\right)-36 \gamma ^6 \epsilon ^2 \left(2 \epsilon ^4-4 \epsilon ^3-9 \epsilon ^2+4 \epsilon +7\right)\\&+4 \gamma ^4 \epsilon  \left(90 \epsilon ^5-126 \epsilon ^4-276 \epsilon ^3+114 \epsilon ^2+187 \epsilon +11\right)-2 \gamma ^2 \left(252 \epsilon ^6-192 \epsilon ^5-726 \epsilon ^4+204 \epsilon ^3+439 \epsilon ^2+22 \epsilon +1\right)\\&+216 \epsilon ^6-180 \epsilon ^5-576 \epsilon ^4+192 \epsilon ^3+360 \epsilon ^2-11 \epsilon +1\big)\Bigg]\,,\\&\boldsymbol{\tilde h}_{21}=\frac{1}{48 (\gamma -1)^2 (\gamma +1)^2 (\epsilon -1)^2 \epsilon  (\epsilon +1) (2 \epsilon +1)}\Bigg[4 \gamma ^5 \big(1296 \epsilon ^7-360 \epsilon ^6-2348 \epsilon ^5+304 \epsilon ^4+1049 \epsilon ^3+48 \epsilon ^2-6 \epsilon\\& -1\big)+360 \gamma ^4 (\epsilon -1)^2 \epsilon  \left(2 \epsilon ^2+3 \epsilon +1\right)-4 \gamma ^3 \big(2592 \epsilon ^8+864 \epsilon ^7-7644 \epsilon ^6-1168 \epsilon ^5+5281 \epsilon ^4-24 \epsilon ^3-36 \epsilon ^2\\&+100 \epsilon -1\big)-2 \gamma ^2 \left(360 \epsilon ^6-60 \epsilon ^5-694 \epsilon ^4-193 \epsilon ^3+333 \epsilon ^2+253 \epsilon +1\right)-2 \gamma  \epsilon  \big(3456 \epsilon ^6+5136 \epsilon ^5-4484 \epsilon ^4\\&-5268 \epsilon ^3+1339 \epsilon ^2+45 \epsilon -188\big)-408 \epsilon ^5-324 \epsilon ^4+250 \epsilon ^3+323 \epsilon ^2+158 \epsilon +1\Bigg]\,,\\&\boldsymbol{\tilde h}_{22}=\frac{1}{48 \left(\gamma ^2-1\right)^2 (\epsilon -1)^2 \epsilon  (\epsilon +1) (2 \epsilon +1)}\Bigg[8 \gamma ^5 \left(6 \epsilon ^5-3 \epsilon ^4-9 \epsilon ^3+\epsilon ^2+3 \epsilon +2\right)-4 \gamma ^3 \big(36 \epsilon ^6-12 \epsilon ^5-45 \epsilon ^4\\&-21 \epsilon ^3+17 \epsilon ^2+18 \epsilon +7\big)+2 \gamma ^2 (\epsilon -1)^2 \left(2 \epsilon ^2+3 \epsilon +1\right)-6 \gamma  (2 \epsilon +1)^2 \left(3 \epsilon ^3-2 \epsilon ^2-\epsilon -1\right)+2 \epsilon ^3+\epsilon ^2-2 \epsilon -1\Bigg]\,,\\&
        \boldsymbol{\tilde h}_{23}=\frac{4 \gamma ^4 \left(2 \epsilon ^3-3 \epsilon ^2+5 \epsilon -4\right)+2 \gamma ^2 \left(4 \epsilon ^2-8 \epsilon +7\right)+8 \gamma ^6 (\epsilon -1)^2+2 \epsilon -3}{48 \left(\gamma ^2-1\right) (\epsilon -1)^2}\,,\\&\boldsymbol{\tilde h}_{24}-\frac{\left(2 \gamma ^2 (\epsilon -1)+1\right) \left(4 \gamma ^4 \left(\epsilon ^2-1\right)-2 \gamma ^2 \left(\epsilon  \left(6 \epsilon ^2+\epsilon -3\right)-2\right)-3 \epsilon  (2 \epsilon +1)+1\right)}{16 \left(\gamma ^2-1\right) (\epsilon -1)^2 (\epsilon +1)}\,,\\&  \boldsymbol{\tilde h}_{25}= \frac{1}{96 \left(\gamma ^2-1\right)^2 (\epsilon -1)^2 \epsilon  (3 \epsilon -1)}\Bigg[\gamma  \big(-2 \gamma ^2 \left(288 \epsilon ^5-552 \epsilon ^4+534 \epsilon ^3-515 \epsilon ^2+256 \epsilon -27\right) \epsilon\\& +\gamma ^4 \left(372 \epsilon ^4-658 \epsilon ^3+308 \epsilon ^2-28 \epsilon +2\right)-288 \epsilon ^5+288 \epsilon ^4-288 \epsilon ^3+208 \epsilon ^2-33 \epsilon -1\big)\Bigg]\,,\\&  \boldsymbol{\tilde h}_{26}=\frac{\gamma  \left(-4 \gamma ^2 \left(\epsilon  \left(4 \epsilon ^4-3 \epsilon ^3+\epsilon -4\right)+2\right)-2 \gamma ^4 (\epsilon -1)^2 (\epsilon  (6 \epsilon +5)-4)-\epsilon  \left(8 (\epsilon -1) \epsilon ^2+\epsilon +7\right)+4\right)}{96 \left(\gamma ^2-1\right) (\epsilon -1)^2 \epsilon  (3 \epsilon -1)}\,,\\& \boldsymbol{\tilde h}_{27}=\frac{\gamma}{24 \left(\gamma ^2-1\right)^2 (\epsilon -1)^2 \epsilon}\Bigg[-4 \gamma ^4 \left(24 \epsilon ^4-57 \epsilon ^3+41 \epsilon ^2-5 \epsilon -2\right)+\gamma ^2 \left(96 \epsilon ^5-88 \epsilon ^4-224 \epsilon ^3+290 \epsilon ^2-64 \epsilon -8\right)\\&+48 \epsilon ^4+12 \epsilon ^3-90 \epsilon ^2+31 \epsilon +1\Bigg]\,,\nonumber\end{split}\end{align}
        \begin{align}\begin{split}
    & \boldsymbol{\tilde h}_{28}= \frac{\gamma ^5 \left(4 \epsilon ^2-6 \epsilon +2\right)+\gamma ^3 \left(4 \epsilon  \left(4 \epsilon ^2-6 \epsilon +3\right)+1\right)+4 \gamma  \epsilon  (2 \epsilon -3)+\gamma }{48 \left(\gamma ^2-1\right) (\epsilon -1) (3 \epsilon -1)}\,,\\&
   \boldsymbol{\tilde h}_{29}=\frac{4 \epsilon  (2 \epsilon +1)}{\gamma ^2-1}\,, \boldsymbol{\tilde h}_{30}=-\frac{2 (6 \epsilon +5)}{3 \left(\gamma ^2-1\right)}\,, \boldsymbol{\tilde h}_{31}=\frac{2 \left(12 \epsilon ^2+8 \epsilon +1\right)}{3 \left(\gamma ^2-1\right)^2 \epsilon }\,, \boldsymbol{\tilde h}_{32}=-\frac{2 \gamma  (2 \epsilon +3)}{3 \left(\gamma ^2-1\right)}\,,\\& \boldsymbol{\tilde h}_{33}=\frac{18 \epsilon ^3+19 \epsilon ^2+3 \epsilon -1}{4 \left(\gamma ^2-1\right) (\epsilon +1)}\,, \boldsymbol{\tilde h}_{34}=-\frac{54 \epsilon ^3+75 \epsilon ^2+25 \epsilon +1}{24 \left(\gamma ^2-1\right) \epsilon  (\epsilon +1)}\,, \boldsymbol{\tilde h}_{35}=\frac{108 \epsilon ^4+132 \epsilon ^3+37 \epsilon ^2-3 \epsilon -1}{24 \left(\gamma ^2-1\right)^2 \epsilon ^2 (\epsilon +1)}\,,\\& \boldsymbol{\tilde h}_{36}=-\frac{\gamma  \left(18 \epsilon ^2+19 \epsilon +7\right)}{24 \left(\gamma ^2-1\right) (\epsilon +1)}\,,\boldsymbol{\tilde h}_{37}=\frac{\epsilon  (2 \epsilon +1)}{2 \epsilon -1},\,\,\boldsymbol{\tilde h}_{38}=\frac{ (2 \epsilon +1)}{2(1-2 \epsilon) },\,\,\boldsymbol{\tilde h}_{39}=\frac{ (2 \epsilon +1)}{2 \left(\gamma ^2-1\right) \epsilon }\,,\\&\boldsymbol{\tilde h}_{40}=\frac{2 \epsilon +1}{2 (1-3 \epsilon )}\,,\boldsymbol{\tilde h}_{41}=\frac{\epsilon  (2 \epsilon +1) \left(\gamma ^2 (\epsilon -1) (2 \epsilon -1)-\epsilon  (2 \epsilon +5)+1\right)}{\left(\gamma ^2-1\right) (\epsilon -1) (2 \epsilon -1) (3 \epsilon -1)}\,,\boldsymbol{\tilde h}_{42}=\frac{(2 \epsilon +1) (\epsilon  (4 (\epsilon -1) \epsilon +5)-1)}{2 \left(\gamma ^2-1\right) \epsilon  (\epsilon -1) (2 \epsilon -1) (3 \epsilon -1)}\,,\\&\boldsymbol{\tilde h}_{43}=\frac{\gamma ^4 \left(7 \epsilon ^3+4 \epsilon ^2-7 \epsilon -4\right)+\gamma ^2 \left(50 \epsilon ^4-3 \epsilon ^3-49 \epsilon ^2+2\right)+31 \epsilon ^3+31 \epsilon ^2+5 \epsilon +2}{\left(\gamma ^2-1\right) \left(\epsilon ^2-1\right)}\,,\\& \boldsymbol{\tilde h}_{44}=\frac{-21 \gamma ^4 \epsilon  \left(\epsilon ^2-1\right)+\gamma ^2 \left(-150 \epsilon ^4-41 \epsilon ^3+149 \epsilon ^2+44 \epsilon -2\right)-93 \epsilon ^3-135 \epsilon ^2-43 \epsilon +2}{6 \left(\gamma ^2-1\right) \epsilon  \left(\epsilon ^2-1\right)}\,,\\&\boldsymbol{\tilde h}_{45}=\frac{1}{6 \left(\gamma ^2-1\right)^2 \epsilon ^2 \left(\epsilon ^2-1\right)}\Bigg[3 \gamma ^4 \epsilon  \left(14 \epsilon ^3-3 \epsilon ^2-14 \epsilon +3\right)+\gamma ^2 \left(300 \epsilon ^5+32 \epsilon ^4-265 \epsilon ^3-45 \epsilon ^2-20 \epsilon -2\right)\\&+186 \epsilon ^4+225 \epsilon ^3+61 \epsilon ^2+9 \epsilon +2\Bigg]\,,\\& \boldsymbol{\tilde h}_{46}=-\frac{\gamma  \left(\gamma ^2 \left(50 \epsilon ^3+19 \epsilon ^2-44 \epsilon -25\right)+36 \epsilon ^2+43 \epsilon +13\right)}{6 \left(\gamma ^2-1\right) \left(\epsilon ^2-1\right)}\\&\boldsymbol{\tilde h}_{47}=\frac{1}{48 \gamma  \left(\gamma ^2-1\right)^2 (\epsilon -1) \epsilon ^2 (2 \epsilon -1) (3 \epsilon -1)}\Bigg[32 \gamma ^6 \epsilon ^2 \left(5 \epsilon ^2-7 \epsilon +2\right)+2 \gamma ^5 \left(60 \epsilon ^4-116 \epsilon ^3+41 \epsilon ^2+9 \epsilon -4\right)\\&+\gamma ^4 \left(2248 \epsilon ^5-4548 \epsilon ^4+3722 \epsilon ^3-2080 \epsilon ^2+700 \epsilon -96\right)+\gamma ^3 \left(60 \epsilon ^4+42 \epsilon ^3-40 \epsilon ^2-6 \epsilon +4\right)\\&+\gamma ^2 \left(-2248 \epsilon ^5+4388 \epsilon ^4-3434 \epsilon ^3+1884 \epsilon ^2-668 \epsilon +96\right)+\gamma  \left(-180 \epsilon ^4+150 \epsilon ^3+4 \epsilon ^2-31 \epsilon +7\right)\\&-16 \epsilon  \left(7 \epsilon ^2-9 \epsilon +2\right)\Bigg]\,,\\& \boldsymbol{\tilde h}_{48}=\frac{1}{48 (\gamma -1)^2 \gamma  (\gamma +1)^2 (\epsilon -1) \epsilon ^2 (2 \epsilon -1) (3 \epsilon -1)}\Bigg[24 \gamma ^6 \epsilon ^3 \left(30 \epsilon ^3-45 \epsilon ^2+16 \epsilon -1\right)\\&-2 \gamma ^5 \epsilon  \left(180 \epsilon ^4-504 \epsilon ^3+353 \epsilon ^2-81 \epsilon +4\right)+8 \gamma ^4 \epsilon  \left(120 \epsilon ^5-164 \epsilon ^4-9 \epsilon ^3+80 \epsilon ^2-30 \epsilon +3\right)\\&+\gamma ^3 \epsilon  \left(360 \epsilon ^4-648 \epsilon ^3+356 \epsilon ^2-63 \epsilon +1\right)-\gamma ^2 \left(1680 \epsilon ^6-2392 \epsilon ^5+440 \epsilon ^4+520 \epsilon ^3-252 \epsilon ^2+27 \epsilon +1\right)\\&+\gamma  \epsilon  \left(-360 \epsilon ^3+350 \epsilon ^2-99 \epsilon +7\right)+8 \epsilon ^2 \left(7 \epsilon ^2-9 \epsilon +2\right)\Bigg]\,,\\& \boldsymbol{\tilde h}_{49}=\frac{1}{48 (\gamma -1)^2 \gamma  (\gamma +1)^2 (\epsilon -1) \epsilon ^2 (2 \epsilon -1) (3 \epsilon -1)}\Bigg[-60 \gamma ^6 \epsilon ^2 \left(6 \epsilon ^3-11 \epsilon ^2+6 \epsilon -1\right)+30 \gamma ^5 \epsilon ^2 \left(6 \epsilon ^2-5 \epsilon +1\right)\\&+\gamma ^4 \epsilon  \left(240 \epsilon ^4+12 \epsilon ^3-306 \epsilon ^2+77 \epsilon -1\right)+\gamma ^3 \left(-720 \epsilon ^5+1056 \epsilon ^4-448 \epsilon ^3+94 \epsilon ^2-38 \epsilon +8\right)\\&+\gamma ^2 \left(-4800 \epsilon ^6+7864 \epsilon ^5-3904 \epsilon ^4+1058 \epsilon ^3-217 \epsilon ^2-47 \epsilon +24\right)+\gamma  \left(180 \epsilon ^4-82 \epsilon ^3+2 \epsilon ^2-17 \epsilon +7\right)\\&+8 \epsilon  \left(-14 \epsilon ^3+11 \epsilon ^2+5 \epsilon -2\right)\Bigg]\,,\nonumber
 \end{split}
\end{align}
\begin{align}
\begin{split}
&\boldsymbol{\tilde h}_{50}=\frac{-40 \gamma ^3 \left(3 \epsilon ^2-4 \epsilon +1\right)+\gamma ^2 \left(-400 \epsilon ^3+160 \epsilon ^2+372 \epsilon -125\right)+4 \gamma  (25 \epsilon -8)-8 (\epsilon -1) \epsilon }{48 (\gamma -1) (\gamma +1) (\epsilon -1) (3 \epsilon -1)}\,,\\&\boldsymbol{\tilde h}_{51}=\frac{1}{96 \gamma  \left(\gamma ^2-1\right)^2 (\epsilon -1)^2 \epsilon ^2 \left(6 \epsilon ^2-5 \epsilon +1\right)}\Bigg[-16 \gamma ^6 (\epsilon -1)^2 \epsilon ^2 \left(90 \epsilon ^3-75 \epsilon ^2-5 \epsilon +8\right)\\&+\gamma ^4 \left(2880 \epsilon ^7-4960 \epsilon ^6-1448 \epsilon ^5+8824 \epsilon ^4-9099 \epsilon ^3+5167 \epsilon ^2-1568 \epsilon +192\right)\\&-8 \gamma ^3 \epsilon  \left(12 \epsilon ^4-52 \epsilon ^3+67 \epsilon ^2-32 \epsilon +5\right)+\gamma ^2 \big(-1440 \epsilon ^7+880 \epsilon ^6+5080 \epsilon ^5-9562 \epsilon ^4+8769 \epsilon ^3-5245 \epsilon ^2\\&+1720 \epsilon -220\big)-8 \gamma  \left(6 \epsilon ^4-29 \epsilon ^3+39 \epsilon ^2-19 \epsilon +3\right)+2 \epsilon  \left(-103 \epsilon ^3+241 \epsilon ^2-173 \epsilon +32\right)\Bigg]\,,\\&\boldsymbol{\tilde h}_{52}=\frac{1}{96 (\gamma -1)^2 \gamma  (\gamma +1)^2 (\epsilon -1)^2 \epsilon ^2 \left(6 \epsilon ^2-5 \epsilon +1\right)}\Bigg[-4800 \gamma ^2 \left(\gamma ^2-1\right) \epsilon ^7-32 \gamma ^2 \big(15 \gamma ^4-9 \gamma ^3-364 \gamma ^2\\&+9 \gamma +349\big)\epsilon ^6+16 \left(72 \gamma ^6-78 \gamma ^5-384 \gamma ^4+63 \gamma ^3+319 \gamma ^2+15 \gamma -7\right) \epsilon ^5-8 \big(108 \gamma ^6-201 \gamma ^5+500 \gamma ^4\\&+128 \gamma ^3-567 \gamma ^2+73 \gamma -32\big) \epsilon ^4+2 \left(96 \gamma ^6-384 \gamma ^5+2172 \gamma ^4+132 \gamma ^3-2165 \gamma ^2+252 \gamma -91\right) \epsilon ^3\\&+\left(120 \gamma ^5-1120 \gamma ^4+64 \gamma ^3+1109 \gamma ^2-184 \gamma +39\right) \epsilon ^2+3 \left(24 \gamma ^4-8 \gamma ^3-26 \gamma ^2+8 \gamma +1\right) \epsilon -1\Bigg]\,,\\&\boldsymbol{\tilde h}_{53}=\frac{1}{96 (\gamma -1)^2 \gamma  (\gamma +1)^2 (\epsilon -1)^2 \epsilon ^2 \left(6 \epsilon ^2-5 \epsilon +1\right)}\Bigg[-60 \gamma ^8 \epsilon ^3 \left(6 \epsilon ^3-11 \epsilon ^2+6 \epsilon -1\right)\\&+60 \gamma ^6 \epsilon ^2 \left(30 \epsilon ^4-67 \epsilon ^3+52 \epsilon ^2-17 \epsilon +2\right)-3 \gamma ^4 \epsilon ^2 \left(520 \epsilon ^4-836 \epsilon ^3+172 \epsilon ^2+185 \epsilon -48\right)\\&-24 \gamma ^3 \epsilon  \left(24 \epsilon ^5-116 \epsilon ^4+162 \epsilon ^3-87 \epsilon ^2+18 \epsilon -1\right)+\gamma ^2 \big(9600 \epsilon ^7-26696 \epsilon ^6+26436 \epsilon ^5-12388 \epsilon ^4+3812 \epsilon ^3\\&-727 \epsilon ^2-117 \epsilon +60\big)+8 \gamma  \left(12 \epsilon ^5-16 \epsilon ^4+19 \epsilon ^3-29 \epsilon ^2+17 \epsilon -3\right)+\epsilon  \left(224 \epsilon ^4-400 \epsilon ^3+99 \epsilon ^2+107 \epsilon -31\right)\Bigg]\,,\\&\boldsymbol{\tilde h}_{54}=\frac{32 \gamma ^2 (\epsilon -1)^2 \left(25 \epsilon ^2+20 \epsilon -9\right)-40 \gamma  \left(6 \epsilon ^3-17 \epsilon ^2+14 \epsilon -3\right)+16 \epsilon ^3-32 \epsilon ^2+28 \epsilon -5}{96 (\gamma -1) (\gamma +1) (\epsilon -1)^2 (3 \epsilon -1)}\,,\\& \boldsymbol{\tilde h}_{55}=\frac{-2 \gamma ^4 \left(6 \epsilon ^4-7 \epsilon ^3-2 \epsilon ^2+7 \epsilon -4\right)+2 \gamma ^2 \epsilon  \left(104 \epsilon ^4-42 \epsilon ^3-99 \epsilon ^2+33 \epsilon +4\right)+104 \epsilon ^4+40 \epsilon ^3-36 \epsilon ^2+11 \epsilon -8}{2 \left(\gamma ^2-1\right) (\epsilon -1) (\epsilon +1) (2 \epsilon -1)}\,,\\&
       \boldsymbol{\tilde h}_{56}=\frac{1}{12 \left(\gamma ^2-1\right) (\epsilon -1) \epsilon  (\epsilon +1) (2 \epsilon -1)}\Bigg[6 \gamma ^4 \epsilon  \left(6 \epsilon ^3+\epsilon ^2-6 \epsilon -1\right)+\gamma ^2 \big(-624 \epsilon ^5+44 \epsilon ^4+738 \epsilon ^3-34 \epsilon ^2-132 \epsilon\\& +8\big)-312 \epsilon ^4-288 \epsilon ^3+100 \epsilon ^2+77 \epsilon -8\Bigg]\,,\\&\boldsymbol{\tilde h}_{57}=\frac{\gamma ^4 \left(-4 \epsilon ^3+2 \epsilon ^2+4 \epsilon -2\right)+\gamma ^2 \left(-192 \epsilon ^4-28 \epsilon ^3+178 \epsilon ^2+38 \epsilon +4\right)-96 \epsilon ^3-112 \epsilon ^2-22 \epsilon -1}{4 \left(\gamma ^2-1\right)^2 \epsilon  \left(\epsilon ^2-1\right)}\,,\\&\boldsymbol{\tilde h}_{58}=-\frac{\gamma  \left(2 \gamma ^2 \left(8 \epsilon ^3+5 \epsilon ^2-7 \epsilon -6\right)+4 \epsilon ^2+11 \epsilon +8\right)}{4 \left(\gamma ^2-1\right) (\epsilon -1) (\epsilon +1)}\,,\\& \boldsymbol{\tilde h}_{59}=\frac{1}{4 \gamma  \left(\gamma ^2-1\right) (\epsilon -1) (2 \epsilon -1) (3 \epsilon -1)}\Bigg[2 \gamma ^4 \epsilon  \left(4 \epsilon ^3-2 \epsilon ^2-3 \epsilon +1\right)-2 \gamma ^3 \epsilon  \left(4 \epsilon ^3-3 \epsilon ^2-2 \epsilon +1\right)\\&+\gamma ^2 \big(-16 \epsilon ^5+20 \epsilon ^4+48 \epsilon ^3-26 \epsilon ^2+7 \epsilon -1\big)-2 \gamma  \epsilon  \left(5 \epsilon ^2-6 \epsilon +1\right)+\epsilon  \left(-28 \epsilon ^2+31 \epsilon -7\right)\Bigg]\,,\nonumber
    \end{split}
\end{align}
\begin{align}
    \begin{split}
       & 
       \boldsymbol{\tilde h}_{60}=\frac{1}{24 \gamma  \left(\gamma ^2-1\right) (\epsilon -1) \epsilon  (2 \epsilon -1) (3 \epsilon -1)}\Bigg[-6 \gamma ^4 \epsilon  \left(4 \epsilon ^3-2 \epsilon ^2-3 \epsilon +1\right)+6 \gamma ^3 \epsilon  \left(4 \epsilon ^3-3 \epsilon ^2-2 \epsilon +1\right)\\&+\gamma ^2 \left(48 \epsilon ^5-44 \epsilon ^4+48 \epsilon ^3-62 \epsilon ^2+27 \epsilon -5\right)-2 \gamma  \epsilon  \left(5 \epsilon ^2-6 \epsilon +1\right)+\epsilon  \left(-28 \epsilon ^2+31 \epsilon -7\right)\Bigg]\,,
       \\&\boldsymbol{\tilde h}_{61}=-\frac{1}{\left.24 \gamma  \left(\gamma ^2-1\right)^2 (\epsilon -1) \epsilon ^2 (2 \epsilon -1) (3 \epsilon -1)\right)}\Bigg[(2 \epsilon +1) \big(-6 \gamma ^4 \epsilon  \left(4 \epsilon ^3-14 \epsilon ^2+7 \epsilon -1\right)\\&+6 \gamma ^3 (\epsilon -1)^2 \epsilon  (4 \epsilon -1)+\gamma ^2 \left(48 \epsilon ^5-76 \epsilon ^4-60 \epsilon ^3+98 \epsilon ^2-39 \epsilon +5\right)+2 \gamma  \epsilon  \left(5 \epsilon ^2-6 \epsilon +1\right)\\&+\epsilon  \left(28 \epsilon ^2-31 \epsilon +7\right)\big)\Bigg]\,,\\&\boldsymbol{\tilde h}_{62}=\frac{\gamma ^2 \left(4 \epsilon ^3-6 \epsilon +2\right)+2 \gamma  (\epsilon -1) \epsilon +\epsilon  (2 \epsilon -1)}{12 \left(\gamma ^2-1\right) (\epsilon -1) (3 \epsilon -1)}\,,\boldsymbol{\tilde h}_{63}=\frac{\gamma ^2 (4 \epsilon -1)}{3 \left(\gamma ^2-1\right) (3 \epsilon -1)}\,,\\&\boldsymbol{\tilde h}_{64}=\frac{1}{8 \gamma  \left(\gamma ^2-1\right)^2 (\epsilon -1) (\epsilon +1) (2 \epsilon -1) (3 \epsilon -1)}\Bigg[-4 \gamma ^6 \epsilon  \left(4 \epsilon ^4+2 \epsilon ^3-5 \epsilon ^2-2 \epsilon +1\right)\\&+\gamma ^4 \left(256 \epsilon ^6-184 \epsilon ^5-260 \epsilon ^4+27 \epsilon ^3+83 \epsilon ^2-45 \epsilon +7\right)-4 \gamma ^3 \epsilon ^2 \left(2 \epsilon ^2+\epsilon -1\right)\\&-2 \gamma ^2 \left(128 \epsilon ^6-100 \epsilon ^5-138 \epsilon ^4+27 \epsilon ^3+47 \epsilon ^2-29 \epsilon +5\right)+4 \gamma  \epsilon ^2 \left(2 \epsilon ^2+\epsilon -1\right)\\&-8 \epsilon ^4+7 \epsilon ^3+19 \epsilon ^2+3 \epsilon -1\Bigg]\,,\\&\boldsymbol{\tilde h}_{65}=\frac{1}{48 (\gamma -1) \gamma  (\gamma +1) (\epsilon -1) \epsilon  (\epsilon +1) (2 \epsilon -1) (3 \epsilon -1)}\Bigg[12 \gamma ^4 \epsilon  \left(4 \epsilon ^4+2 \epsilon ^3-5 \epsilon ^2-2 \epsilon +1\right)\\&+\gamma ^2 \left(-768 \epsilon ^6+344 \epsilon ^5+676 \epsilon ^4-281 \epsilon ^3-99 \epsilon ^2+39 \epsilon +5\right)+4 \gamma  \epsilon  \left(6 \epsilon ^3+5 \epsilon ^2-2 \epsilon -1\right)\\&+8 \epsilon ^4-7 \epsilon ^3-19 \epsilon ^2-3 \epsilon +1\Bigg]\,,\\& \boldsymbol{\tilde h}_{66}=\frac{1}{48 (\gamma -1)^2 \gamma  (\gamma +1)^2 (\epsilon -1) \epsilon ^2 (\epsilon +1) (2 \epsilon -1) (3 \epsilon -1)}\Bigg[2 \gamma ^4 \big(2400 \epsilon ^6-1180 \epsilon ^5-2616 \epsilon ^4\\&+1385 \epsilon ^3+345 \epsilon ^2-205 \epsilon +15\big) \epsilon +\gamma ^2 \big(6336 \epsilon ^7-4808 \epsilon ^6-4708 \epsilon ^5+3302 \epsilon ^4-401 \epsilon ^3+133 \epsilon ^2-21 \epsilon \\&-13\big)-4 \gamma  \left(12 \epsilon ^4+8 \epsilon ^3-7 \epsilon ^2-2 \epsilon +1\right) \epsilon -44 \epsilon ^5+40 \epsilon ^4+19 \epsilon ^3-77 \epsilon ^2-5 \epsilon +7\Bigg]\,,\\
    &\boldsymbol{\tilde h}_{67}=\frac{\gamma ^2 \left(-8 \epsilon ^3+13 \epsilon -5\right)+2 \gamma  \epsilon -\epsilon +1}{24 (\gamma -1) (\gamma +1) (\epsilon -1) (3 \epsilon -1)}\,,\\&\boldsymbol{\tilde h}_{68}=\frac{-8 \gamma ^2 \left(107 \epsilon ^4+27 \epsilon ^3-114 \epsilon ^2-45 \epsilon +25\right)+100 \epsilon ^3-74 \epsilon ^2-127 \epsilon +47}{24 (\gamma -1) (\gamma +1) (\epsilon -1) (\epsilon +1) (3 \epsilon -1)}\,,\\& \boldsymbol{\tilde h}_{69}=-\frac{\epsilon  \left(\gamma ^2 \left(2 \epsilon ^2-3 \epsilon +1\right)-10 \epsilon ^2-9 \epsilon -5\right)}{2 (\epsilon -1) (2 \epsilon -3) (2 \epsilon -1)}\,,\boldsymbol{\tilde h}_{70}=\frac{ \left(\gamma ^2 (\epsilon -1)-\epsilon -1\right)}{4 (\epsilon -1) (2 \epsilon -3)}\,,\boldsymbol{\tilde h}_{71}=-\frac{ \left(4 \epsilon ^3+5 \epsilon +3\right)}{4 (\epsilon -1) \epsilon  (2 \epsilon -3) (2 \epsilon -1)}\,,\\& \boldsymbol{\tilde h}_{72}=\frac{1}{256 \left(\gamma ^2-1\right)^2 (\epsilon -1)^2 (\gamma  \epsilon  (6 \epsilon -5)+\gamma )}\Bigg[-64 \gamma ^8 (\epsilon -1)^2 \epsilon ^2+\gamma ^6 \big(160 \epsilon ^5+1872 \epsilon ^4-5282 \epsilon ^3+4581 \epsilon ^2\\&-1490 \epsilon +159\big)-2 \gamma ^4 \left(80 \epsilon ^5-3816 \epsilon ^4+8202 \epsilon ^3-5819 \epsilon ^2+1453 \epsilon -120\right)+\gamma ^2 \big(800 \epsilon ^4-2506 \epsilon ^3+1597 \epsilon ^2\\&-184 \epsilon -27\big)-4 \left(22 \epsilon ^2+7 \epsilon -3\right)\Bigg]\,,\nonumber 
 \end{split}
\end{align}
\begin{align}
    \begin{split}
  &\boldsymbol{\tilde h}_{73}=\frac{1}{1536 \gamma  \left(\gamma ^2-1\right) (\epsilon -1)^2 \epsilon  (\epsilon  (6 \epsilon -5)+1)}\Bigg[192 \gamma ^6 (\epsilon -1)^2 \epsilon ^2+\gamma ^4 \big(-480 \epsilon ^5-944 \epsilon ^4+4078 \epsilon ^3-3551 \epsilon ^2\\&+896 \epsilon +1\big)+\gamma ^2 \left(-1440 \epsilon ^4+2762 \epsilon ^3-1733 \epsilon ^2+292 \epsilon -1\right)+4 \left(22 \epsilon ^2+7 \epsilon -3\right)\Bigg]\,,\\&\boldsymbol{\tilde h}_{74}=\frac{1}{1536 \left(\gamma ^2-1\right)^2 (\epsilon -1)^2 \epsilon ^2 (\gamma  \epsilon  (6 \epsilon -5)+\gamma )}\Bigg[(2 \epsilon +1) \big(-192 \gamma ^6 (\epsilon -1)^2 \epsilon ^2+\gamma ^4 \big(480 \epsilon ^5-5360 \epsilon ^4\\&+9646 \epsilon ^3-5663 \epsilon ^2+896 \epsilon +1\big)+\gamma ^2 \left(-480 \epsilon ^4+1850 \epsilon ^3-1493 \epsilon ^2+292 \epsilon -1\right)+4 \left(22 \epsilon ^2+7 \epsilon -3\right)\big)\Bigg]\,,\\&\boldsymbol{\tilde h}_{75}=\frac{4 (\epsilon -3)-\gamma ^2 (\epsilon -1) \left(\gamma ^2 \left(320 \epsilon ^2-418 \epsilon +97\right)-336 \epsilon ^2+466 \epsilon -97\right)}{768 \left(\gamma ^2-1\right) (\epsilon -1)^2 (3 \epsilon -1)}\,,\\&\boldsymbol{\tilde h}_{76}=\frac{\gamma ^2 \left(-200 \epsilon ^6+264 \epsilon ^5+75 \epsilon ^4-184 \epsilon ^3+34 \epsilon ^2+16 \epsilon -5\right)-72 \epsilon ^6-328 \epsilon ^5+117 \epsilon ^4+124 \epsilon ^3-57 \epsilon ^2+3 \epsilon +3}{8 (1-2 \epsilon )^2 (\epsilon -1)^2 (\epsilon +1) (2 \epsilon +1)}\,,\\&\boldsymbol{\tilde h}_{77}=\frac{\gamma ^2 (\epsilon -1)^2 \left(232 \epsilon ^4+90 \epsilon ^3-123 \epsilon ^2-10 \epsilon +19\right)-8 \epsilon ^6+110 \epsilon ^5-49 \epsilon ^4-132 \epsilon ^3+60 \epsilon ^2+16 \epsilon -9}{16 (1-2 \epsilon )^2 (\epsilon -1)^2 \epsilon  (\epsilon +1) (2 \epsilon +1)}\,,\\&\boldsymbol{\tilde h}_{78}=-\frac{64 \epsilon ^7-280 \epsilon ^6+22 \epsilon ^5+115 \epsilon ^4-60 \epsilon ^3+5 \epsilon ^2+7 \epsilon +1}{16 (1-2 \epsilon )^2 (\epsilon -1)^2 \epsilon ^2 (\epsilon +1) (2 \epsilon +1)}\,,\\&\boldsymbol{\tilde h}_{79}=\frac{-2 \gamma ^2 (1-2 \epsilon )^2 \left(10 \epsilon ^3-19 \epsilon ^2+4 \epsilon +8\right)+32 \epsilon ^5-168 \epsilon ^4+228 \epsilon ^3-106 \epsilon ^2+7 \epsilon +7}{32 (1-2 \epsilon )^2 (\epsilon -1)^2 (2 \epsilon +1)}\,,\\&\boldsymbol{\tilde h}_{80}=\frac{\left(4-36 \gamma ^2\right) \epsilon ^4+\left(74 \gamma ^2-26\right) \epsilon ^3+\left(21-48 \gamma ^2\right) \epsilon ^2+2 \left(5 \gamma ^2-2\right) \epsilon -1}{16 (1-2 \epsilon )^2 (\epsilon -1) \epsilon }\,,\boldsymbol{\tilde h}_{81}=-\frac{\left(\gamma ^2-1\right) \left(4 \epsilon ^3+\epsilon ^2-2 \epsilon +1\right)}{16 (1-2 \epsilon )^2 \epsilon  (2 \epsilon +1)}\,,\\&\boldsymbol{\tilde h}_{82}=\frac{1}{128 (\gamma^2 -1) \gamma \left(2 \epsilon ^2-3 \epsilon +1\right)^2 (\epsilon ^2 (6 \epsilon +7)-1)}\Bigg[2 \gamma ^4 \big(3624 \epsilon ^7-5990 \epsilon ^6+1394 \epsilon ^5\\&+2233 \epsilon ^4-1628 \epsilon ^3+362 \epsilon ^2+18 \epsilon -13\big)+16 \gamma ^3 (\epsilon -1)^2 \left(72 \epsilon ^5+36 \epsilon ^4-50 \epsilon ^3-5 \epsilon ^2+8 \epsilon -1\right)\\&+\gamma ^2 \left(-21216 \epsilon ^7+22288 \epsilon ^6+17516 \epsilon ^5-23897 \epsilon ^4+3172 \epsilon ^3+3674 \epsilon ^2-1392 \epsilon +143\right)\\&+3 \left(-3408 \epsilon ^7+4148 \epsilon ^6+1264 \epsilon ^5-2739 \epsilon ^4+652 \epsilon ^3+180 \epsilon ^2-80 \epsilon +7\right)\Bigg]\,,\\&\boldsymbol{\tilde h}_{83}=-\frac{1}{256 (\gamma^2 -1) \gamma \left(2 \epsilon ^2-3 \epsilon +1\right)^2 \epsilon  (\epsilon ^2 (6 \epsilon +7)-1)}\Bigg[2 \gamma ^4 \big(4380 \epsilon ^7-7886 \epsilon ^6+1546 \epsilon ^5\\&+4549 \epsilon ^4-2920 \epsilon ^3+10 \epsilon ^2+402 \epsilon -81\big)+\gamma ^2 \big(-6192 \epsilon ^7+5784 \epsilon ^6+8610 \epsilon ^5-12461 \epsilon ^4+3572 \epsilon ^3\\&+1740 \epsilon ^2-1226 \epsilon +197\big)+888 \epsilon ^7+388 \epsilon ^6-2142 \epsilon ^5+107 \epsilon ^4+1206 \epsilon ^3-514 \epsilon ^2+36 \epsilon +7\Bigg]\,,\\&\boldsymbol{\tilde h}_{84}=-\frac{1}{256 (\gamma^2 -1) \gamma \left(2 \epsilon ^2-3 \epsilon +1\right)^2 \epsilon ^2 (\epsilon ^2 (6 \epsilon +7)-1)}\Bigg[16 \gamma ^3 \left(-2 \epsilon ^2+\epsilon +1\right)^2 \left(6 \epsilon ^3+\epsilon ^2-4 \epsilon +1\right) \epsilon\\& +2 \gamma ^2 \left(-6096 \epsilon ^8+7564 \epsilon ^7+1670 \epsilon ^6-4134 \epsilon ^5+1011 \epsilon ^4+52 \epsilon ^3+36 \epsilon ^2-38 \epsilon +7\right)-9696 \epsilon ^8+10792 \epsilon ^7\\&+4068 \epsilon ^6-6210 \epsilon ^5+801 \epsilon ^4+272 \epsilon ^3+104 \epsilon ^2-66 \epsilon +7\Bigg]\,,\nonumber
    \end{split}
\end{align}
\begin{align}
    \begin{split}
    &\boldsymbol{\tilde h}_{85}=\frac{1}{256 \gamma  \left(\gamma ^2-1\right) (2 \epsilon +1) \left(2 \epsilon ^2-3 \epsilon +1\right)^2}\Bigg[\gamma ^4 (1-2 \epsilon )^2 \left(56 \epsilon ^3-144 \epsilon ^2+32 \epsilon +59\right)\\&+\gamma ^2 \left(-2048 \epsilon ^5+4672 \epsilon ^4-3376 \epsilon ^3+448 \epsilon ^2+444 \epsilon -143\right)-8 \epsilon  (1-2 \epsilon )^2 \left(15 \epsilon ^2-22 \epsilon +7\right)\Bigg]\,,\\&\boldsymbol{\tilde h}_{86}=\frac{1}{256 \gamma  \left(\gamma ^2-1\right) (1-2 \epsilon )^2 (\epsilon -1) \epsilon}\Bigg[4 \gamma ^4 \epsilon  \left(66 \epsilon ^3-133 \epsilon ^2+84 \epsilon -17\right)+\gamma ^2 \big(-720 \epsilon ^4+1104 \epsilon ^3-566 \epsilon ^2\\&+72 \epsilon +14\big)-(1-2 \epsilon )^2 \left(30 \epsilon ^2+\epsilon -7\right)\Bigg]\,,\\&\boldsymbol{\tilde h}_{87}=\frac{(\epsilon +1) \left(2 \gamma ^2 \left(42 \epsilon ^3-60 \epsilon ^2+33 \epsilon -7\right)+(15 \epsilon -7) (1-2 \epsilon )^2\right)}{256 \gamma  (1-2 \epsilon )^2 (\epsilon -1) \epsilon  (2 \epsilon +1)}\,,\\&\boldsymbol{\tilde h}_{88}=-\frac{1}{64 \gamma ^2 \left(\gamma ^2-1\right)^2 (1-2 \epsilon )^4 (\epsilon -1)^3 \left(\epsilon ^2 (6 \epsilon +7)-1\right)^2}\Bigg[\gamma ^8 \big(-156 \epsilon ^6+404 \epsilon ^5-332 \epsilon ^4+43 \epsilon ^3+92 \epsilon ^2\\&-63 \epsilon +12\big)^2+4 \gamma ^7 (\epsilon -1)^2 (2 \epsilon -1)^3 \big(1800 \epsilon ^8+396 \epsilon ^7-3464 \epsilon ^6-1260 \epsilon ^5+1299 \epsilon ^4+345 \epsilon ^3\\&-183 \epsilon ^2-21 \epsilon +8\big)+2 \gamma ^6 \big(112320 \epsilon ^{13}-537984 \epsilon ^{12}+889584 \epsilon ^{11}-486508 \epsilon ^{10}-237128 \epsilon ^9\\&+414152 \epsilon ^8-159781 \epsilon ^7-11177 \epsilon ^6+15554 \epsilon ^5+602 \epsilon ^4+4123 \epsilon ^3-5533 \epsilon ^2+2016 \epsilon -240\big)\\&-4 \gamma ^5 (\epsilon -1)^2 (2 \epsilon -1)^3 \left(7704 \epsilon ^8+8172 \epsilon ^7-5770 \epsilon ^6-5225 \epsilon ^5+2756 \epsilon ^4+1375 \epsilon ^3-511 \epsilon ^2-110 \epsilon +33\right)\\&+\gamma ^4 \big(518400 \epsilon ^{14}-2505600 \epsilon ^{13}+3769056 \epsilon ^{12}-1383360 \epsilon ^{11}-1057208 \epsilon ^{10}+1569336 \epsilon ^9-1608359 \epsilon ^8\\&+817324 \epsilon ^7+308191 \epsilon ^6-483302 \epsilon ^5+121399 \epsilon ^4+31648 \epsilon ^3-15071 \epsilon ^2+162 \epsilon +328\big)\\&+4 \gamma ^3 (\epsilon -1)^2 (2 \epsilon -1)^3 \left(4824 \epsilon ^8+7092 \epsilon ^7+508 \epsilon ^6-1106 \epsilon ^5+1265 \epsilon ^4+316 \epsilon ^3-358 \epsilon ^2-38 \epsilon +25\right)\\&-6 \gamma ^2 \big(172800 \epsilon ^{14}-826560 \epsilon ^{13}+1067520 \epsilon ^{12}+363056 \epsilon ^{11}-1442532 \epsilon ^{10}+407040 \epsilon ^9+477315 \epsilon ^8\\&-177793 \epsilon ^7-64099 \epsilon ^6-4775 \epsilon ^5+24081 \epsilon ^4-135 \epsilon ^3-3513 \epsilon ^2+703 \epsilon -20\big)+12 \gamma  \epsilon  (2 \epsilon -1)^3 (10 \epsilon -17)\\& \left(6 \epsilon ^4+\epsilon ^3-7 \epsilon ^2-\epsilon +1\right)^2+9 \left(-240 \epsilon ^7+620 \epsilon ^6-96 \epsilon ^5-565 \epsilon ^4+212 \epsilon ^3+76 \epsilon ^2-32 \epsilon +1\right)^2\Bigg]\,,\\& \boldsymbol{\tilde h}_{89}=\frac{1}{64 \gamma  \left(\gamma ^2-1\right) (1-2 \epsilon )^2 (\epsilon -1)^2 \epsilon  \left(\epsilon ^2 (6 \epsilon +7)-1\right)}\Bigg[\gamma ^4 \big(1416 \epsilon ^7-2900 \epsilon ^6+596 \epsilon ^5+1956 \epsilon ^4\\&-1131 \epsilon ^3-82 \epsilon ^2+175 \epsilon -30\big)+\gamma ^2 \left(-1344 \epsilon ^7+2432 \epsilon ^6+938 \epsilon ^5-3411 \epsilon ^4+1117 \epsilon ^3+500 \epsilon ^2-291 \epsilon +35\right)\\&+312 \epsilon ^7-716 \epsilon ^6-246 \epsilon ^5+975 \epsilon ^4-116 \epsilon ^3-248 \epsilon ^2+62 \epsilon +1\Bigg]\,,\\&\boldsymbol{\tilde h}_{90}=\frac{1}{64 \gamma  \left(\gamma ^2-1\right) (1-2 \epsilon )^2 (\epsilon -1)^2 \epsilon ^2 \left(\epsilon ^2 (6 \epsilon +7)-1\right)}\Bigg[\gamma ^2 \big(-1536 \epsilon ^8+1528 \epsilon ^7-244 \epsilon ^6-240 \epsilon ^5\\&+696 \epsilon ^4-323 \epsilon ^3-46 \epsilon ^2+19 \epsilon +2\big)-1152 \epsilon ^8+2232 \epsilon ^7+908 \epsilon ^6-2398 \epsilon ^5-303 \epsilon ^4+626 \epsilon ^3+78 \epsilon ^2-64 \epsilon +1\Bigg]\,,\\&\boldsymbol{\tilde h}_{91}=\frac{1}{64 \gamma  \left(\gamma ^2-1\right) (2 \epsilon +1) \left(2 \epsilon ^2-3 \epsilon +1\right)^2}\Bigg[\gamma ^4 \left(-(1-2 \epsilon )^2\right) \left(20 \epsilon ^3-48 \epsilon ^2+7 \epsilon +18\right)+\gamma ^2 \big(304 \epsilon ^5-816 \epsilon ^4\\&+648 \epsilon ^3-72 \epsilon ^2-97 \epsilon +30\big)+8 \epsilon  \left(2 \epsilon ^2-3 \epsilon +1\right)^2\Bigg]\,,\nonumber
    \end{split}
\end{align}
\begin{align}
    \begin{split}
  &\boldsymbol{\tilde h}_{92}=\frac{-4 \gamma ^4 \epsilon  \left(18 \epsilon ^2-19 \epsilon +5\right)+2 \gamma ^2 \left(64 \epsilon ^3-52 \epsilon ^2+11 \epsilon +1\right)+(2 \epsilon +1) (1-2 \epsilon )^2}{64 \gamma  \left(\gamma ^2-1\right) (1-2 \epsilon )^2 \epsilon }\,,\\
   & \boldsymbol{\tilde h}_{93}=-\frac{(\epsilon +1) \left(2 \gamma ^2 \left(6 \epsilon ^2-4 \epsilon +1\right)+(1-2 \epsilon )^2\right)}{64 \gamma  (1-2 \epsilon )^2 \epsilon  (2 \epsilon +1)}\,, \boldsymbol{\tilde h}_{94}=\gamma ^2+\frac{1}{2 (\epsilon -1)}\,,\\&\boldsymbol{\tilde h}_{95}=-\frac{\gamma  (2 \epsilon +1) \left(\gamma ^2 (2 \epsilon -1) (7 \epsilon -2)+10 \epsilon ^2+\epsilon -1\right)}{8 \left(\gamma ^2-1\right) (\epsilon -1) \epsilon  (2 \epsilon -1) (3 \epsilon -1)}\,,\\&\boldsymbol{\tilde h}_{96}=\frac{\gamma  (2 \epsilon +1)^2 (4 \epsilon -1)}{8 \left(\gamma ^2-1\right) (\epsilon -1) \epsilon  (2 \epsilon -1) (3 \epsilon -1)}\,,\boldsymbol{\tilde h}_{97}=\frac{2 \gamma  \epsilon +\gamma }{24 \epsilon ^2-32 \epsilon +8}\,,\\&\boldsymbol{\tilde h}_{98}=-\frac{\gamma^2  (2 \epsilon +1)^2 \left(\gamma ^2 (2 \epsilon -1) (7 \epsilon -2)+10 \epsilon ^2+\epsilon -1\right)}{8 \left(\gamma ^2-1\right) (\epsilon -1)^2 \epsilon  (\epsilon  (6 \epsilon -5)+1)}\,,\boldsymbol{\tilde h}_{99}=\frac{\gamma^2  (2 \epsilon +1)^2}{8 (\epsilon -1)^2 (3 \epsilon -1)}\,,\\&\boldsymbol{\tilde h}_{100}=\frac{\gamma^2  (2 \epsilon +1)^3 (4 \epsilon -1)}{8 \left(\gamma ^2-1\right) (\epsilon -1)^2 \epsilon  (\epsilon  (6 \epsilon -5)+1)}\,,\\&
    \boldsymbol{\tilde h}_{101}=\frac{8 \gamma ^2 (\epsilon -1)^2 \left(6 \epsilon ^2-6 \epsilon +1\right)+36 \epsilon ^4-60 \epsilon ^3+23 \epsilon ^2-\epsilon +1}{4 (\epsilon -1)^3 (2 \epsilon -1)}\,,\boldsymbol{\tilde h}_{102}=\frac{-4 \gamma ^2 (\epsilon -1)^2 (2 \epsilon +1)-12 \epsilon ^3+5 \epsilon +1}{8 (\epsilon -1)^2 \epsilon  (2 \epsilon -1)}\,,\\&\boldsymbol{\tilde h}_{103}=\frac{1}{8 \gamma  \left(\gamma ^2-1\right) (\epsilon -1)^2 (\epsilon  (6 \epsilon -5)+1)}\Bigg[-2 \gamma ^4 \left(228 \epsilon ^4-464 \epsilon ^3+313 \epsilon ^2-85 \epsilon +8\right)\\&+\gamma ^2 \epsilon  \left(-930 \epsilon ^3+1265 \epsilon ^2-454 \epsilon +47\right)-9 \epsilon  \left(18 \epsilon ^3-27 \epsilon ^2+13 \epsilon -2\right)\Bigg]\,,\\&\boldsymbol{\tilde h}_{104}=\frac{2 \gamma ^2 \left(18 \epsilon ^2-23 \epsilon +5\right)+3 \left(9 \epsilon ^2-9 \epsilon +2\right)}{16 \gamma  (\epsilon -1)^2 (3 \epsilon -1)}\,,\\&\boldsymbol{\tilde h}_{105}=\frac{(2 \epsilon +1) \left(4 \gamma ^4 \left(6 \epsilon ^3-11 \epsilon ^2+6 \epsilon -1\right)+2 \gamma ^2 \left(90 \epsilon ^3-119 \epsilon ^2+38 \epsilon -3\right)+54 \epsilon ^3-81 \epsilon ^2+39 \epsilon -6\right)}{16 (\gamma -1) \gamma  (\gamma +1) (\epsilon -1)^2 \epsilon  (2 \epsilon -1) (3 \epsilon -1)}\,,\\&\boldsymbol{\tilde h}_{106}=\frac{1}{32 \left(\gamma ^2-1\right) (\epsilon -1)^3 (\epsilon  (6 \epsilon -5)+1)}\Bigg[2 \gamma ^4 (\epsilon -1)^2 \left(228 \epsilon ^3-314 \epsilon ^2+131 \epsilon -16\right)\\&-4 \gamma ^3 (\epsilon -1)^2 \left(12 \epsilon ^2-8 \epsilon +1\right) \epsilon +\gamma ^2 \left(2640 \epsilon ^5-7108 \epsilon ^4+7052 \epsilon ^3-3179 \epsilon ^2+621 \epsilon -42\right)\\&-4 \gamma  (\epsilon -1)^2 \left(132 \epsilon ^3-76 \epsilon ^2+15 \epsilon -1\right)+2088 \epsilon ^5-4072 \epsilon ^4+2922 \epsilon ^3-1037 \epsilon ^2+162 \epsilon -7\Bigg]\,,\\&\boldsymbol{\tilde h}_{107}=\frac{-2 \gamma ^2 (\epsilon -1)^2 \left(84 \epsilon ^2-98 \epsilon +27\right)+4 \gamma  (\epsilon -1)^2 \left(12 \epsilon ^2-8 \epsilon +1\right)-264 \epsilon ^4+600 \epsilon ^3-482 \epsilon ^2+181 \epsilon -27}{64 (\epsilon -1)^3 \left(6 \epsilon ^2-5 \epsilon +1\right)}\,,\\&\boldsymbol{\tilde h}_{108}=-\frac{1}{\left.64 \left(\gamma ^2-1\right) (\epsilon -1)^3 \epsilon  (\epsilon  (6 \epsilon -5)+1)\right)}\Bigg[(2 \epsilon +1) \big(6 \gamma ^2 \left(64 \epsilon ^4-174 \epsilon ^3+169 \epsilon ^2-68 \epsilon +9\right)\\&-4 \gamma  (\epsilon -1)^2 \left(24 \epsilon ^2-10 \epsilon +1\right)+480 \epsilon ^4-932 \epsilon ^3+648 \epsilon ^2-215 \epsilon +27\big)\Bigg]\,,\\&\boldsymbol{\tilde h}_{109}=\frac{1}{64 \gamma  \left(\gamma ^2-1\right) (\epsilon -1) (2 \epsilon -1) (3 \epsilon -1)}\Bigg[\gamma ^5 \left(36 \epsilon ^3-62 \epsilon ^2+29 \epsilon -4\right)+8 \gamma ^4 \left(14 \epsilon ^2-11 \epsilon +2\right)\\&+\gamma ^3 \left(108 \epsilon ^3-106 \epsilon ^2+25 \epsilon -1\right)-8 \gamma ^2 \left(36 \epsilon ^3-19 \epsilon ^2-5 \epsilon +2\right)+24 \epsilon  (1-3 \epsilon )\Bigg]\,,\\&\boldsymbol{\tilde h}_{110}=\frac{\gamma ^3 \left(-12 \epsilon ^2+18 \epsilon -5\right)+\gamma ^2 (8-16 \epsilon )+24 \epsilon -8}{128 \gamma  (\epsilon -1) (2 \epsilon -1) (3 \epsilon -1)}\,,\\
&\boldsymbol{\tilde h}_{111}=-\frac{(2 \epsilon +1) \left(\gamma ^3 \left(24 \epsilon ^2-24 \epsilon +5\right)-8 \gamma ^2 \left(6 \epsilon ^2-6 \epsilon +1\right)-24 \epsilon +8\right)}{128 \gamma  \left(\gamma ^2-1\right) (\epsilon -1) \epsilon  (2 \epsilon -1) (3 \epsilon -1)}\,,\\
\nonumber
     \end{split}
\end{align}
\begin{align}
\begin{split}
&\boldsymbol{\tilde h}_{112}=\frac{\gamma  \left(-\left(\gamma ^2 (\epsilon -1) (12 \epsilon -5)\right)-5 \epsilon +2\right)}{12 \left(\gamma ^2-1\right) (\epsilon -1)^2}\,,\boldsymbol{\tilde h}_{113}=\frac{\gamma  \left(\gamma ^2 (6 (3-2 \epsilon ) \epsilon -5)-6 \epsilon +2\right)}{96 \left(\gamma ^2-1\right) (\epsilon -1)^2}\,,\\& \boldsymbol{\tilde h}_{114}=\frac{1}{256(\gamma -1)^2\gamma (\gamma +1)^2(\epsilon -1)^2\left(6 \epsilon ^2-5 \epsilon +1\right)}\Bigg[-64 \gamma ^8 (\epsilon -1)^2 \epsilon ^2-32 \gamma ^7 (\epsilon -1)^2 \epsilon  \left(4 \epsilon ^2+4 \epsilon -3\right)\\&+\gamma ^6 \left(160 \epsilon ^5+1872 \epsilon ^4-5282 \epsilon ^3+4581 \epsilon ^2-1490 \epsilon +159\right)+16 \gamma ^5 \left(16 \epsilon ^5-300 \epsilon ^4+556 \epsilon ^3-347 \epsilon ^2+83 \epsilon -8\right)\\&-2 \gamma ^4 \left(80 \epsilon ^5-3816 \epsilon ^4+8202 \epsilon ^3-5819 \epsilon ^2+1453 \epsilon -120\right)-32 \gamma ^3 \left(4 \epsilon ^5-112 \epsilon ^4+213 \epsilon ^3-149 \epsilon ^2+39 \epsilon -4\right)\\&+\gamma ^2 \left(800 \epsilon ^4-2506 \epsilon ^3+1597 \epsilon ^2-184 \epsilon -27\right)+16 \gamma  \epsilon  \left(68 \epsilon ^3-144 \epsilon ^2+69 \epsilon -11\right)-4 \left(22 \epsilon ^2+7 \epsilon -3\right)\Bigg]\,,\\& \boldsymbol{\tilde h}_{115}=\frac{1}{1536 (\gamma -1) \gamma  (\gamma +1) (\epsilon -1)^2 \epsilon  (2 \epsilon -1) (3 \epsilon -1)}\Bigg[192 \gamma ^6 (\epsilon -1)^2 \epsilon ^2+96 \gamma ^5 (\epsilon -1)^2 \epsilon  \left(4 \epsilon ^2+4 \epsilon -3\right)\\&+4 \left(22 \epsilon ^2+7 \epsilon -3\right)-48 \gamma  \epsilon  \left(28 \epsilon ^3-48 \epsilon ^2+23 \epsilon -3\right)+\gamma ^2 \left(-1440 \epsilon ^4+2762 \epsilon ^3-1733 \epsilon ^2+292 \epsilon -1\right)\\&-48 \gamma ^3 \epsilon  \left(8 \epsilon ^4-36 \epsilon ^3+34 \epsilon ^2-3 \epsilon -3\right)+\gamma ^4\left(-944 \epsilon ^4+4078 \epsilon ^3-3551 \epsilon ^2+896 \epsilon -480 \epsilon +1\right)\Bigg]\,,\\&\boldsymbol{\tilde h}_{116}=-\frac{1}{1536 (\gamma -1)^2 \gamma  (\gamma +1)^2 (\epsilon -1)^2 \epsilon ^2 \left(6 \epsilon ^2-5 \epsilon +1\right)}\Bigg[\left(2 \epsilon +1\right)\big(192 \gamma ^6 (\epsilon -1)^2 \epsilon ^2\\&+96 \gamma ^5 \epsilon  \left(4 \epsilon ^4-28 \epsilon ^3+43 \epsilon ^2-22 \epsilon +3\right)-\gamma ^4 \left(480 \epsilon ^5-5360 \epsilon ^4+9646 \epsilon ^3-5663 \epsilon ^2+896 \epsilon +1\right)\\&-48 \gamma ^3 \epsilon  \left(8 \epsilon ^4-36 \epsilon ^3+46 \epsilon ^2-25 \epsilon +3\right)+\gamma ^2 \left(480 \epsilon ^4-1850 \epsilon ^3+1493 \epsilon ^2-292 \epsilon +1\right)\\&+48 \gamma  \epsilon  \left(20 \epsilon ^3-40 \epsilon ^2+19 \epsilon -3\right)-4 \left(22 \epsilon ^2+7 \epsilon -3\right)\big)\Bigg]\,,\\&\boldsymbol{\tilde h}_{117}=\frac{4 (\epsilon -3)-\gamma ^2 (\epsilon -1) \left(\gamma ^2 \left(320 \epsilon ^2-418 \epsilon +97\right)-336 \epsilon ^2+466 \epsilon -97\right)}{768 \left(\gamma ^2-1\right) (\epsilon -1)^2 (3 \epsilon -1)}\,,\\&
\boldsymbol{\tilde h}_{118}=\frac{\epsilon  (2 \epsilon +1) \left(\gamma ^2 (\epsilon -1) (2 \epsilon -1)-\epsilon  (2 \epsilon +5)+1\right)}{2 (\gamma ^2-1) (\epsilon -1) (2 \epsilon -1) (3 \epsilon -1)}\,,\boldsymbol{\tilde h}_{119}=\frac{(2 \epsilon +1)}{4-12 \epsilon }\,,\\&\boldsymbol{\tilde h}_{120}=\frac{(2 \epsilon +1) (\epsilon  (4 (\epsilon -1) \epsilon +5)-1)}{4 (\gamma^2 -1)  (\epsilon -1) \epsilon  (2 \epsilon -1) (3 \epsilon -1)}\,.
\nonumber
   \end{split}
\end{align}
\bibliography{ref}
\bibliographystyle{jhep}
\end{document}